\documentclass[a4paper,oneside,reqno]{amsart}

\usepackage[OT2,T1]{fontenc}
\usepackage[utf8]{inputenc}
\usepackage[english]{babel}
\usepackage[osf,sc]{mathpazo}
\usepackage{eucal}
\newcommand*{\Scale}[2][4]{\scalebox{#1}{$#2$}}
\normalfont

\usepackage{fullpage}
\usepackage[dvipsnames,table]{xcolor}
\usepackage{hyperref}
\usepackage{bookmark}
\usepackage{enumitem}
\usepackage{comment}

\hypersetup{
	colorlinks=true,
	linktocpage=true,
	pdfstartpage=1,
	pdfstartview=FitV,
	breaklinks=true,
	pdfpagemode=UseNone,
	pageanchor=true,
	plainpages=false,
	bookmarksnumbered,
	bookmarksopen=true,
	bookmarksopenlevel=1,
	hypertexnames=false,
	pdfhighlight=/O,
	urlcolor=BrickRed,
	linkcolor=RoyalBlue,
	citecolor=ForestGreen
}

\usepackage{graphicx}
\usepackage{float}
\usepackage{caption}
\usepackage{pgfplots}
\usepackage{tikz}
\usetikzlibrary{calc,decorations.markings}
\pgfplotsset{compat=1.16}
\usepackage{ytableau}
\newcommand{\cir}[2]{\begin{tikzpicture}
\node[draw,scale=0.5,circle,{#1},{#2}] at (0,0) {}; 
\end{tikzpicture}}

\usepackage[textsize=footnotesize]{todonotes}
\presetkeys{todonotes}{color=Apricot}{}
\usepackage{amsmath,amssymb,amsthm}
\usepackage{mathtools,thmtools} 
\usepackage{bm,braket,stmaryrd,dsfont,mathrsfs,esint}
\usepackage[scr=dutchcal]{mathalfa}
\numberwithin{equation}{section}

\usepackage[english]{varioref}
\usepackage[noabbrev]{cleveref}
\AddToHook{cmd/appendix/before}{\crefalias{section}{appendix}}

\crefname{thm}{theorem}{theorems}
\Crefname{thm}{Theorem}{Theorems}

\crefname{lem}{lemma}{lemmata}
\Crefname{lem}{Lemma}{Lemmata}

\crefname{prop}{proposition}{propositions}
\Crefname{prop}{Proposition}{Propositions}

\crefname{conj}{conjecture}{conjectures}
\Crefname{conj}{Conjecture}{Conjectures}

\crefname{defn}{definition}{definitions}
\Crefname{defn}{Definition}{Definitions}

\crefname{rem}{remark}{remarks}
\Crefname{rem}{Remark}{Remarks}

\crefname{ex}{example}{examples}
\Crefname{ex}{Example}{Examples}

\newcommand{\Ev}{\textnormal{Ev}}
\newcommand{\dd}{\mathrm{d}}
\newcommand{\ii}{\mathrm{i}}

\DeclareMathOperator*{\Res}{Res}
\newcommand{\Ram}{\textnormal{Ram}}
\newcommand{\N}{\mathbb{N}}
\newcommand{\Z}{\mathbb{Z}}
\newcommand{\Mu}{\textnormal{M}}

\newcommand{\dk}{\kappa}
\newcommand{\std}{\textnormal{std}}
\newcommand{\fiber}{\mathfrak{f}}

\makeatletter
\def\section{\@startsection{section}{2}%
  \z@{\linespacing\@plus.7\linespacing}{.3\linespacing}%
  {\centering \Large \scshape}}
\makeatother

\makeatletter
\def\subsection{\@startsection{subsection}{2}%
  \z@{\linespacing\@plus.7\linespacing}{.3\linespacing}%
  {\normalfont\bfseries}}
\makeatother

\makeatletter
\def\subsubsection{\@startsection{subsubsection}{2}%
  \z@{\linespacing\@plus.7\linespacing}{.3\linespacing}%
  {\normalfont\itshape}}
\makeatother

\usepackage{setspace}

\newtheorem{theorem}{Theorem}[section]

\newtheorem{proposition}[theorem]{Proposition}
\newtheorem{lemma}[theorem]{Lemma}
\newtheorem{corollary}[theorem]{Corollary}

\newtheorem{definition}[theorem]{Definition}
\newtheorem{remark}[theorem]{Remark}

\newcommand{\Du}{\textnormal{D}}
\newcommand{\Sol}{\Sigma}
\makeatletter
\newcommand{\vast}{\bBigg@{3}}
\newcommand{\Vast}{\bBigg@{4}}
\makeatother

\date{}

\makeatother

\begin{document}

\title{Partial Airy ideals and coloured blobbed topological recursion}

\author{Ga{\"e}tan Borot}%
\address{
	Humboldt-Universit{\"a}t zu Berlin, Institut f{\"u}r Mathematik und Institut f{\"u}r Physik, Unter den Linden 6, 10099 Berlin, Germany.}%
\email{gaetan.borot@hu-berlin.de}

\author{Vincent Bouchard}
\address{Department of Mathematical and Statistical Sciences, University of Alberta, University Commons 5-140
Edmonton, Alberta, Canada T6G 2N8.}
\email{vincent.bouchard@ualberta.ca}

\author{Swagata S. Datta}
\address{Department of Mathematical and Statistical Sciences, University of Alberta, University Commons 5-140
Edmonton, Alberta, Canada T6G 2N8.}
\email{ssdatta@ualberta.ca}

\begin{abstract}
We define and study the notion of partial Airy ideals, which are certain linear systems of PDEs that may contain less equations than variables. We show that they admit a unique solution of exponential type (the \emph{partition function}) modulo a choice of initial condition. We construct a large class of examples of partial Airy ideals attached to spectral curves using twisted free-field representations of $\mathcal{W}(\mathfrak{gl}_r)$-algebras at self-dual level. This generalizes the results of \cite{BBCCN24} by significantly enlarging the set of admissible spectral curves, provided they come with additional data that we call \emph{colour}. We show that the differential constraints on the partition function are equivalent to a set of equations on the associated correlators, which we call \emph{coloured loop equations}. We find a recursive residue formulation for the solutions of those equations, the \emph{coloured (blobbed) topological recursion}, which is a generalization of the currently known forms of topological recursion. It includes the blobbed topological recursion of \cite{BS16} as a special case, giving it a natural interpretation within the formalism of partial Airy ideals.
\end{abstract}

\maketitle

\newpage

\begingroup
\singlespacing
\tableofcontents
\endgroup

\section{Introduction}

\subsection{Weakening the notion of Airy ideal}

The idea of Airy ideals is to give a sufficient condition for systems of linear PDEs formed in the Rees Weyl algebra to admit a unique solution of exponential type
\[
Z(\bm{t}) = \exp\left(\sum_{\substack{g \in \frac{1}{2}\N,\,\, n \in \mathbb{N}  \\ 2g - 2 + n \geq 0}} \frac{\hbar^{2g - 2 + n}}{n!} F_{g,n}(\bm{t})\right),
\] 
where $F_{g,n}(\bm{t})$ is a homogeneous polyonomial of degree $n$ with $F_{g,0}=0$. This $Z$ is called the \emph{partition function} and $F = \log Z$ the \emph{free energy}. The existence and uniqueness of the partition function can be considered as a generalization to differential operators of the Frobenius integrability criterion for vector fields. Airy ideals were defined in \cite{KS18} to give an algebraic reformulation and generalization of the Eynard--Orantin topological recursion \cite{EO07}, see also \cite{ABCO24}. The topological recursion for curves with higher ramification proposed in \cite{BHLMR14,BE13} was retrieved and extended in this framework \cite{BBCCN24}, and several other extensions have been found since then \cite{BCHORS,BO21,BKS24,BBCC24,O21,CDO24,BCJ24,Bou}.

We propose and study a weaker axiomatic of \emph{partial Airy ideals}, where the system of linear PDEs does not include an evolution equation for certain \emph{spectator} variables. As a result, partition functions still exist but are not unique. Instead, they are uniquely determined by the value of $Z$ when all variables but the spectator variables are set to zero. In other words, for any choice of initial condition --- taking the form of a \emph{reduced partition function} $\hat{Z}$ depending only on spectator variables --- we have existence of uniqueness of a partition function $Z$ annihilated by a partial Airy ideal. This generalization is mild but allow the construction of many more interesting examples of partial Airy ideals (that are not Airy ideals).

While our work was under preparation, the notion of partial Airy ideals and the existence and uniqueness of partition functions modulo initial conditions first appeared in \cite{AKOO} in the context of superconformal topological recursion, but with less details. Our purpose is to study this notion systematically. Revisiting the construction of \cite{BBCCN24} of Airy ideals from twisted free-field representations of $\mathcal{W}(\mathfrak{gl}_r)$-algebras at self-dual level, we construct  partial Airy ideals associated to a large class of spectral curves together with extra data, which we call \emph{colour}. The colour specifies the spectator variables as well as some additional terms to put in $F_{g,1}$. If we add colours, many situations where the $1$-forms $\dd x$ and $\dd y$ have common zeros become admissible (such spectral curves were not admissible in \cite{BBCCN24}) and our work provides a natural definition of topological recursion for them (modulo the input of initial condition), which was so far unknown.

Encoding the partition function into meromorphic multi-differentials on the spectral curve called \emph{correlators}, we show that the partial Airy ideal differential constraints on the partition function are precisely equivalent to set of equations on the correlators that we call \emph{coloured loop equations}. Those generalize the usual loop equations. Moreover, in contrast to the construction of \cite{BBCCN24}, the correspondence here does not single out a unique solution satisfying a projection property. As such, it allows us to describe the space of solutions of these coloured loop equations. In fact, this is already interesting if we forget about colours, as it allows us to recover the spaces of solutions of the usual loop equations (and the blobbed topological recursion of \cite{BS16}) within the framework of partial Airy ideals.

We also show that we can write a recursive residue formulation for the solutions of the coloured loop equations. This recursive formulation generalizes the previously known forms of topological recursions in two directions. The first layer of generalization already manifests itself for trivial initial conditions and comes from the colouring: the recursion kernel does not involve anymore the bidifferential $\omega_{0,2}$ itself, but rather $\omega_{0,2}$ from which a part associated to spectator variables is removed. The second layer of generalization incorporates  initial conditions, i.e. the reduced partition function $\hat{Z}$. In the language of correlators, the choice of a reduced partition function becomes a choice of \emph{blobs}, which specify a certain part of the correlators. In the restricted case where the blobs are holomorphic, we retrieve the formula for blobbed topological recursion found in \cite{BS16}, but now extended to more general spectral curves. Our results therefore give a conceptual explanation for the blobbed topological recursion and a natural place for it in the realm of partial Airy ideals. When blobs are not holomorphic, our formula involves a different set of graphs and has a more complicated structure than all previously known ones. We call the general formula \emph{coloured blobbed topological recursion}. 

It would be interesting to understand the geometric meaning of the partition functions associated to the partial Airy ideals that we attach to coloured spectral curves. Are the partition functions descendant potentials for certain cohomological field theories? Are they $\tau$-functions for certain integrable hierarchies? Can we write an explicit formula for the $F_{g,n}$ as integrals over $\overline{\mathcal{M}}_{g,n}$? 

We do not have answers to these questions in general. However, \cite{BCGS} provides an answer for one very specific class of examples. In \cite{BCGS}, the authors study the $\Theta^{(r,s)}$ cohomological field theories, which are non-semisimple cohomological field theories obtained as the top degrees of the Chiodo cohomological field theories. They find a set of $\mathcal{W}$-constraints satisfied by the descendant potential of $\Theta^{(r,s)}$ for $s \in [r-1]$. As we explain, when $r$ and $s$ are coprime, these $\mathcal{W}$-constraints actually take the form of partial Airy ideals attached to the so-called $(r,s)$-spectral curves for very specific choices of colours. The descendant potential is then obtained as the unique partition function of the partial Airy ideal singled out by an appropriate reduced partition function determined in terms of some descendant integrals. It is also shown in \cite{BCGS} that this descendant potential is a $\tau$-function for the $r$-KdV integrable hierarchy. It would be interesting to see whether these results generalize to other choices of colours and initial conditions.

In a different direction, it may be interesting to revisit the rather stringent classification of the quadratic Airy ideals based on simple Lie algebras \cite{Hadasz} and determine whether partial Airy ideals offer more possibilities.

\subsection{Outline and main results}

The basic theory of partial Airy ideals is developed in \Cref{Sec2}. In particular, we prove the existence and uniqueness of the partition function up to initial conditions in \Cref{t:PAI_KS}. We explain in \Cref{Sec3} how to compute the partition function graphically as sums over certain sets of graphs. 

The rest of the article is devoted to the construction of classes of $\mathcal{W}$-constraints associated to spectral curves and taking the form of partial Airy ideals. In \Cref{Sec4} we carry out such a construction for spectral curves that are admissible in the sense of \cite{BBCCN24}. We establish its relation to the Airy ideals constructed in \textit{loc.cit.}, which is quite interesting and non-trivial at the level of partition functions (\Cref{ProP:Hi}). In \Cref{Sec5}, we switch to the language of correlators on the spectral curves and show that the differential constraints from the partial Airy ideals are equivalent to loop equations for the correlators, with no extra assumption on the solution. The partition functions thus retrieve general solutions of the usual loop equations. We then write a recursive residue formulation for the solutions: the highlight is \Cref{t:BS} which extends the blobbed topological recursion of \cite{BS16} to all admissible spectral curves. 

In \Cref{Sec6}, we construct a class of partial Airy ideals which do not have analogues in the world of Airy ideals. They are associated to spectral curves equipped with the additional data of a colour, which essentially consists in a choice of partition of the order of each ramification point and a choice of zero-mode shifts. Introducing colours makes a much larger class of spectral curves admissible for our construction. Colours come with a poset structure that we study in \Cref{secpose}. We also give general properties of those partial Airy ideals and their partition functions (\Cref{secprop}). Converting to correlators on spectral curves in \Cref{Sec7}, we show that the differential constraints are equivalent to a set of \emph{coloured loop equations} for the correlators. The partition functions thus correspond to general solutions of the coloured loop equations, and we show how those solutions can be recovered by a residue formula in the style of Chekhov--Eynard--Orantin. 

We finish in \Cref{Sec8} with examples. We explore two concrete examples: we discuss coloured topological recursion for the cuspidal spectral curve $4 y^2 + x^3 =0 $ coloured by the  partition $\nu=(1,1)$ and no zero-mode shifts (which we call \emph{Quintic Airy}), and the Airy-type spectral curve $4 y^2 + x = 0$ coloured by the non-canonical choice of partition $\nu=(2)$ and non-zero zero-mode shifts (which we call \emph{Queer Airy}). We also explain how the $\mathcal{W}$-constraints for the descendant potential of the $\Theta^{(r,s)}$-cohomological field theories obtained in \cite{BCGS} fit within our formalism.

\medskip

\noindent \textbf{Notation.}  The set of integers from $1$ to $n$ is denoted $[n]$. 

\medskip

\noindent \textbf{Acknowledgments.} We would like to thank Nitin K. Chidambaram and A. Giacchetto for valuable discussions. V.B. and S.D. acknowledge support from the National Science and Engineering Research Council of Canada. The University of Alberta respectfully acknowledges that they are situated on Treaty 6 territory, traditional lands of First Nations and M\'etis people.

\medskip

\noindent \textbf{Tool and computational resource disclosure.} No generative AI tools were used in the research, writing, or generation of any part of this work. All analysis, text, and conclusions are entirely the original work of the human authors.

\vspace{0.5cm}

\section{Partial Airy ideals and their partition functions}

\medskip
\label{Sec2}
We will follow the recent approach and conventions of \cite{BCJ24,Bou}.

\subsection{Rees Weyl algebra and completions}
\label{sec:S11}
 Let $I$ be a finite or countably infinite index set. We use the notation $\bm{t}_I \coloneqq \set{t_i\,\,|\,\,i \in I}$ for the set of variables, and $\bm{\partial}_I \coloneqq \set{\partial_i\,\,|\,\,i \in I}$ for the set of differential operators with $\partial_i = \frac{\partial}{\partial t_i}$.

\begin{definition}[Weyl algebra]\label{d:Weyl}
The \emph{Weyl algebra} $\mathcal{D}_I := \mathbb{C}[\bm{t}_I]\langle \bm{\partial}_I \rangle$ is the algebra of differential operators with polynomial coefficients. 
\end{definition}

The Weyl algebra $\mathcal{D}_I$ is a filtered algebra, but not a graded algebra. It admits many filtrations. The one that we will be interested in is called the Bernstein filtration.

\begin{definition}[Bernstein filtration]\label{d:bernstein}
The \emph{Bernstein filtration} on $\mathcal{D}_I$ is the exhaustive ascending filtration
 which is defined by assigning degree $1$ to all variables $ t_i $ and derivatives $ \partial_i $ and defining the subspaces $\mathcal{F}^d \mathcal{D}_I $ to contain all operators in $\mathcal{D}_I$ of degree $ \leq d $. Concretely, the elements of $\mathcal{F}^{d}\mathcal{D}_I$ take the form
\begin{equation}
 \sum_{m = 0}^{d}  \sum_{\mathbf{i} \in I^m} p_{\mathbf{i}}^{(d-m)}(\bm{t}_I) \partial_{i_1} \cdots \partial_{i_m},
\end{equation}
where the $ p_{\mathbf{i}}^{(k)}(\bm{t}_I)$ are polynomials of degree $\leq k$ and the sums contain finitely many terms.
 Here, $\mathcal{F}^0 \mathcal{D}_I = \mathbb{C}$.
\end{definition}

There is a natural way to construct a graded algebra out of any filtered algebra, which is sometimes called the ``Rees construction'' \cite{Rees}.

\begin{definition}[Rees Weyl algebra]\label{d:rees}
The \emph{Rees Weyl algebra} $\mathcal{D}_I^{\hbar}$ associated to $\mathcal{D}_I$ with the Bernstein filtration is
\begin{equation}\label{eq:RWA}
\mathcal{D}_I^{\hbar} = \prod_{d \geq 0} \hbar^d \mathcal{F}^d \mathcal{D}_I.
\end{equation}
\end{definition}
The $\hbar$-degree makes $\mathcal{D}_I^{\hbar}$ a graded algebra. 

\begin{remark}
For simplicity of notation, we define the Rees Weyl algebra in \eqref{eq:RWA} directly as a direct product, instead of first as a direct sum and then completing in $\hbar$ as in \cite{BCJ24,Bou}. In other words, we are already working with the $\hbar$-adic completion, in which infinite power series in $\hbar$ are allowed. This is necessary for the proof of \Cref{t:PAI_KS} and for \Cref{stablini}.
\end{remark}

More subtle aspects arise when $I$ is infinite, and we will need to work with further completions of the Rees Weyl algebra.
{}
\begin{definition}[$(V,D)$-completed Rees Weyl algebra]\label{d:comprees}
Let $V,D \subseteq I$ be empty or countably infinite subsets such that $V \cap D = \emptyset$. We define the \emph{$(V,D)$-completed Rees Weyl algebra} $\mathcal{D}^\hbar_{I;\llbracket V,D\rrbracket}$ to be the graded algebra of operators of the form
\begin{equation}\label{eq: operator final form}
    \sum_{d \geq 0} \hbar^d \sum_{m = 0}^{d}  \sum_{m'+n'=m}\sum_{\substack{\mathbf{j} \in V^{n'} \\ \mathbf{i} \in D^{m'}}} \left(\prod_{\ell = 1}^{n'} t_{j_\ell}\right) p^{(d-m)}_{d;\mathbf{j},\mathbf{i}}(\bm{t}_{I\setminus V},\bm{\partial}_{I\setminus D}) \left(\prod_{\ell = 1}^{m'} \partial_{i_\ell}\right),
\end{equation}
where the $p^{(d-m)}_{d;\mathbf{j},\mathbf{i}}(\bm{t}_{I\setminus V},\bm{\partial}_{I\setminus D})$ are normal-ordered polynomials of degree $\leq d-m$ in the variables $\bm{t}_{I\setminus V}$ and derivatives $\bm{\partial}_{I\setminus D}$. This means that we are allowing infinite sums in the variables $\bm{t}_V$ (when $V$ is infinite) and in the derivatives $\bm{\partial}_D$ (when $D$ is infinite).
\end{definition} 

\begin{remark} \label{rem:finites} We show in Appendix \ref{a:completion} that \Cref{d:comprees} can indeed be understood as a completion of the Rees Weyl algebra $\mathcal{D}^\hbar_I$. In particular, the general construction does not require $V,D \subset I$ to be disjoint, it only needs $V\cap D$ to be finite. But we show that the completion does not change if we add or subtract finite subsets to $V$ and $D$ (\Cref{lem:removfinit}), and hence we may as well take $V$ and $D$ to be disjoint from the start.
\end{remark}

When working with sets of operators in the $(V,D)$-completed Rees Weyl algebra, we will make use of a technical condition to avoid divergences.
\begin{definition}[Boundedness]\label{d:bounded}
We say that a set $\mathbf{H} \subset \mathcal{D}^\hbar_{I;\llbracket V,D\rrbracket}$ is bounded if, for all fixed choices of $d,m,\mathbf{j},\mathbf{i}$, there are only finitely many $H \in \mathbf{H}$ for which the normal-ordered polynomial  $p^{(d-m)}_{d;\mathbf{j},\mathbf{i}}$ appearing in the decomposition \eqref{eq: operator final form} of $H$ is non-zero.
\end{definition} 
If $\mathbf{H} \subset \mathcal{D}^\hbar_{I;\llbracket V,D\rrbracket}$ is a bounded set, infinite sums of the form $\sum_{H \in \mathbf{H}} C_H H$, where $C_H \in \mathcal{D}^{\hbar}_{I;\llbracket V,D \rrbracket}$, are well-defined elements of $\mathcal{D}^\hbar_{I;\llbracket V,D\rrbracket}$. When we talk about the ideal generated by a bounded set $\mathbf{H}$, we allow such infinite sums.

\subsection{Partial Airy Ideals}

We now define the notion of partial Airy ideals, generalizing Airy ideals. While this paper was being written, the concept of partial Airy ideals and the associated existence and uniqueness result was also proposed independently in a different context and with less details in \cite{AKOO}.

\begin{definition}[Partial Airy ideals]\label{d:PAI}
Let $I$ be an index set decomposed into $I = A \sqcup S$. Let $\mathcal{I} \subset \mathcal{D}^\hbar_{I;\llbracket S,A\rrbracket}$ be a left ideal in the $(S,A)$-completed Rees Weyl algebra. We say that $\mathcal{I}$ is a \emph{partial Airy ideal} with actor set $A$ and spectator set $S$ if there exists a generating set $\mathbf{H} = \set{H_a \in \mathcal{D}^\hbar_{I;\llbracket S,A\rrbracket}\,\,|\,\,a \in A}$ such that
\begin{enumerate}
\item $\mathbf{H}$ is bounded.
\item $\forall a \in A,\quad H_a = \hbar \partial_a - \hbar h_a^{(1)}(\bm{t}_I) + O(\hbar^2)$.\footnote{Since $H_a \in \mathcal{D}^\hbar_{I;\llbracket S,A\rrbracket}$, $h_a^{(1)}(\bm{t}_I)$ contains only terms of degree $\leq 1$ in the variables $\bm{t}_I$. However, it is not necessarily polynomial as it could involve infinite sums in the spectator variables $\bm{t}_S$.}
\item $[\mathcal{I},\mathcal{I}] \subseteq \hbar^2 \mathcal{I}$.
\end{enumerate}
It is \emph{stable} if $h_a^{(1)}(\bm{t}_I) = 0$ for all $a \in A$. It is an \emph{Airy ideal} if it is a partial Airy ideal with empty spectator set (i.e. $I = A$).
\end{definition}
 In other words, a partial Airy ideal is like an Airy ideal, but it does not include generators corresponding to spectator variables that would contain spectator derivatives $\partial_s$ in their $O(\hbar)$ terms. Yet, both actor and spectator variables can appear in $H_a$. In the  literature on Airy ideals preceding \cite{BBKN25}, the stability condition $h_a^{(1)}(\bm{t}_I) = 0$ was part of the definition. We prefer to make it a separate property which may or may not be satisfied.

The Kontsevich--Soibelman theorem \cite{KS18} for the existence and uniqueness of the partition function annihilated by an Airy ideal generalizes easily to partial Airy ideals.  Let us first define the notion of a partition function. 

\begin{definition}[Partition functions]\label{d:pf}
Let $I$ be an index set and $V,D \subseteq I$ such that $V \cap D = \emptyset$. A \emph{partition function} $Z$ is an object of the form 
\begin{equation}
\label{exp4}
Z = \exp\Bigg(\sum_{d \geq 0} \hbar^d f^{(d+2)}(\bm{t}_{I})\Bigg),
\end{equation}
where $\hbar^{d+2} f^{(d+2)}(\bm{t}_{I})$ is an element of $\mathcal{D}^\hbar_{I;\llbracket V,D\rrbracket}$ without derivatives for all $d \geq 0$. Its degree-$N$ truncation is
\[
Z^{(N)} = \exp\Bigg( \sum_{d = 0}^{N} \hbar^d f^{(d+2)}(\bm{t}_{I})\Bigg).
\]
The quantity inside the exponential \eqref{exp4} is called the \emph{free energy}.
\end{definition}
Unlike in the literature on Airy ideals, we do not assume $f^{(d+2)}(\bm{0}) = 0$. Later, the constant $f^{(d + 2)}(\bm{0})$ will be part of the initial conditions to get uniqueness of the partition function.

\begin{remark} The fact that $\hbar^{d+2}f^{(d + 2)}(\bm{t}_I)$ is an element of the completion $\mathcal{D}^{\hbar}_{I;\llbracket V,D \rrbracket}$ means that 1) it contains only terms of degree $\leq d+2$ in the variables $\bm{t}_I$ and 2) the coefficient of any given monomial in the variables $\bm{t}_V$ is a polynomial in finitely many variables among $\bm{t}_D$. In particular, $f^{(d+2)}(\bm{t}_I)_{|\bm{t}_V = 0}$ is a polynomial in finitely many variables among $\bm{t}_D$, but this may not be true of $f^{(d + 2)}(\bm{t}_I)$ itself, as it may involve infinite sums in the variables $\bm{t}_V$.
\end{remark}

Elements $H \in \mathcal{D}_{I;\llbracket V,D \rrbracket}^{\hbar}$ act on a partition function $Z$ as differential operators: we denote the outcome by $H \cdot Z$. Another way to think about this action is the following. While $Z$ is not in the Rees Weyl algebra $\mathcal{D}^{\hbar}_{I;\llbracket V,D \rrbracket}$, for any operator $H \in \mathcal{D}^{\hbar}_{I;\llbracket V,D \rrbracket}$ the conjugation $Z^{-1} H Z \in \mathcal{D}^{\hbar}_{I;\llbracket V,D \rrbracket}$ (here the product means product of differential operators). Indeed, conjugation is an exponential series in commutators, and the action of one commutator on a monomial decreases the number of derivatives and variables at least by $2$, and hence conjugation is well-defined and produces new operators in $\mathcal{D}^{\hbar}_{I;\llbracket V,D \rrbracket}$. For any $H \in \mathcal{D}^{\hbar}_{I;\llbracket V,D \rrbracket}$ we can replace the differential equation $H \cdot Z = 0$ by the equivalent equation $Z^{-1} H Z \cdot 1 = 0$.

\begin{remark}\label{r:weylaction}
In the following we will always use the notation $\cdot$ to denote the action of differential operators on functions, to distinguish with the product of operators in the Rees Weyl algebra, which we denote without a dot.
\end{remark}

We now state the generalization of the Kontsevich--Soibelman theorem:
 
\begin{theorem}[Partition functions of partial Airy ideals]\label{t:PAI_KS}
Let $I$ be an index set and $\mathcal{I} \subset \mathcal{D}^\hbar_{I;\llbracket S,A\rrbracket}$ a partial Airy ideal with actor set $A$ and spectator set $S = I \setminus A$. Let $\hat{Z}$ be a partition function in the variables $\bm{t}_S$, which we call \emph{reduced partition function}. Then there exists a unique partition function $Z$ in the variables $\bm{t}_I$ such that $\mathcal{I} \cdot Z = 0$ and   $Z_{|\bm{t}_A = 0} = \hat Z$.
\end{theorem}

The reduced partition function $\hat{Z}(\bm{t}_S) = Z(\bm{t}_I)_{|\bm{t}_A = 0}$ plays the role of \emph{initial condition}. Given this initial condition, there exists a unique solution to the differential constraints $\mathcal{I} \cdot Z = 0$. We call it \emph{the partition function} of the partial Airy ideal with initial condition $\hat{Z}$. The set of partition functions $\mathfrak{Z}$ satisfying the constraints $\mathcal{I} \cdot Z = 0$ is therefore  in bijection with the set $\hat{\mathfrak{Z}}$ of reduced partition functions. With this bijection, pointwise multiplication in $\hat{\mathfrak{Z}}$ transfers to a structure of abelian group on the solution set $\mathfrak{Z}$. If $Z,Z' \in \mathfrak{Z}$ we denote $Z * Z'$ this group operation; making $*$ explicit --- more than its definition --- is difficult in general. We will see in \Cref{Sec:44} a class of example where it can be done.

We could be more general in \Cref{t:PAI_KS} and allow initial conditions that do not necessarily take the form of \Cref{d:pf}, in the spirit of \cite{ABDKSblob}. For simplicity we do not address this generalization here. If $I$ is infinite, we could consider another generalization where one completes in disjoint sets of derivatives $D$ and variables $V$ such that $A \subseteq D$; \Cref{t:PAI_KS}  would remain valid. \Cref{d:PAI} corresponds to $A  = D$ and $S = V$, or equivalently $A \subseteq D$ with $D \setminus A$ finite in view of \Cref{rem:finites}. For simplicity and for lack of motivating examples, we do not pursue a study of cases with $D \setminus A$ infinite.

\begin{proof}[Proof of \Cref{t:PAI_KS}]
Let us fix a reduced partition function
\begin{equation}
\hat{Z} = \exp\Bigg( \sum_{d \geq 0} \hbar^d \hat{f}^{(d+2)}(\bm{t}_{S})\Bigg).
\end{equation}
We prove by induction on $N \geq 0$ the unique existence of a truncation $Z^{(N)}$ 
such that
\begin{equation}
\label{wesolve}
Z^{(N)}_{|\bm{t}_A = 0} = \hat{Z}^{(N)} \qquad \textnormal{and} \qquad \forall a \in A ,\quad (Z^{(N)})^{-1} H_a Z^{(N)} \cdot 1 = O(\hbar^{N + 2}).
\end{equation}
Note that the conjugation by $Z^{(N)}$ does not alter boundedness. 

{\bf Base case} ($N=0$): We want to show that there is a unique $Z^{(0)}$ such that 
\begin{equation}
\label{wesolvebase}
Z^{(0)}_{|\bm{t}_A = 0} = \hat{Z}^{(0)} \qquad \textnormal{and} \qquad \forall a \in A ,\quad (Z^{(0)})^{-1} H_a Z^{(0)} \cdot 1 = O(\hbar^{2}).
\end{equation}
By definition, $Z^{(0)} = \exp(f^{(2)}(\bm{t}_I))$. Since $H_a = \hbar \partial_a - \hbar h_a^{(1)}(\bm{t}_I) + O(\hbar^2)$, the equations in \eqref{wesolvebase} become
\begin{equation}
\label{baseeqn}
f^{(2)}(\bm{t}_I)_{|\bm{t}_A = 0} = \hat{f}^{(2)}(\bm{t}_S) \qquad \text{and} \qquad\forall a \in A, \quad \partial_a \cdot f^{(2)}(\bm{t}_I) = h_a^{(1)}(\bm{t}_I).
\end{equation}
These equations uniquely fix $f^{(2)}(\bm{t}_I)$.

{\bf Induction}: Let $N \geq 1$ and assume the claim holds for $Z^{(N - 1)}$. For each $a \in A$ define the operator
\[
G_a := (Z^{(N - 1)})^{-1} H_a Z^{(N - 1)}.
\]
By assumption, $G_a \cdot 1 = O(\hbar^{N + 1})$, therefore
\begin{equation}
\label{Gadefun}
G_a = \hbar \partial_a - \hbar^{N + 1}h_a^{(N + 1)}(\bm{t}_I) + \hbar^2 p_a(\bm{t}_I,\bm{\partial}_I) + O(\hbar^{N + 2}),
\end{equation}
where $\hbar^2 p_a(\bm{t}_I,\bm{\partial}_I) \in \mathcal{D}^\hbar_{I;\llbracket S,A\rrbracket}$ is such that $\hbar^2 p_a \cdot 1 = 0$, i.e. all normal-ordered terms in $p_a$ contain at least one derivative on the right) and $h_a^{(N+1)}(\bm{t}_I)$ does not contain derivatives. All we are saying here is that there are no derivative-free terms of $\hbar$-degree less than $N + 1$. By definition of partial Airy ideals we know that
\[
[H_a,H_b] = \hbar^2 \sum_{c \in A} C_{abc} H_c\qquad \textnormal{for some}\,\,C_{abc} \in \mathcal{D}^\hbar_{I;\llbracket S,A\rrbracket}.
\]
The sum may contain infinitely many terms but is well-defined because $\mathbf{H}$ is bounded. Conjugating with $Z^{(N - 1)}$, we get
\[
[G_a,G_b] = \hbar^2\sum_{c \in A} D_{abc} G_c, \qquad D_{abc} := (Z^{(N - 1)})^{-1}C_{abc} Z^{(N - 1)}.
\]
Acting on the function $1$, the right-hand side produces $O(\hbar^{N+3})$ since by the induction assumption, $G_c \cdot 1 = O(\hbar^{N+1})$. Taking into account \eqref{Gadefun}, the left-hand side produces
\[
G_a \cdot \big(\hbar^{N + 1}h_b^{(N + 1)}(\bm{t}_I)\big) - G_b \cdot (\hbar^{N + 1} h_a^{(N + 1)}(\bm{t}_I)\big) = O(\hbar^{N + 3}).
\]
The left-hand side has a term of order $\hbar^{N + 2}$. We thus we obtain a constraint which takes the form
\begin{equation}
\partial_a \cdot h_b^{(N + 1)}(\bm{t}_I) = \partial_b \cdot h_a^{(N + 1)}(\bm{t}_I).
\end{equation} 
By Poincaré lemma there exists a unique $f^{(N + 2)}(\bm{t}_I)$ such that
\begin{equation}
\label{theconstrain}
f^{(N + 2)}(\bm{t}_I)_{|\bm{t}_A = 0} = \hat{f}^{(N + 2)}(\bm{t}_S)\qquad \textnormal{and} \qquad \forall a \in A,\quad h_a^{(N + 1)}(\bm{t}_I) = \partial_a \cdot f^{(N + 2)}(\bm{t}_I).
\end{equation}
The boundedness of $\set{G_a | a \in A}$ tells us that any given monomial in spectator variables can only occur in $G_a$ for finitely many $a \in A$. This property transfers to the potential $f^{(N+2)}$: any given monomial in spectator variables can only appear with finitely many actor variables in $f^{(N + 2)}$. Therefore $\hbar^{N+2} f^{(N + 2)} \in \mathcal{D}^{\hbar}_{I;\llbracket S,A \rrbracket}$. Finally, set
\[
Z^{(N)} := Z^{(N - 1)}\exp\big(\hbar^{N} f^{(N + 2)}(\bm{t}_I)\big).
\]
Then the equations \eqref{theconstrain} are equivalent to the desired equations \eqref{wesolve} for $Z^{(N)}$. 
\end{proof}

The proof was exactly the same as the proof of Kontsevich--Soibelman theorem \cite{KS18}, but with the integration constant in the application of Poincaré lemma being uniquely fixed in terms of the initial condition instead of being set to zero, and with subtleties related to completions if $I$ is infinite. For Airy ideals, the initial condition is simply an overall scale factor $\hat{Z} = \exp\big(\sum_{d \geq 0} \hbar^d c_d\big)$ for some constants $c_d$. In previous references on Airy ideals, the reduced partition function was set to $\hat{Z} = 1$ to get a unique solution, that is, it was additionally required that $c_d = f^{(d+2)}(\bm{0}) =0$ for all $d$.

\subsection{Stabilizing and reinitializing the operators}

\label{stablini}

To study the partition function, it can be convenient to 1) get rid of the $\hbar$-degree zero term in the free energy, i.e. of $f^{(2)}(\bm{t}_I)$; (2) to transfer the initial conditions to the operators and then only work with a constant initial condition. Let us explain how this can be done.

Let $\mathbf{H}$ be a partial Airy ideal, $\hat{Z}$ a reduced partition function and $Z$ be the partition function of $\mathbf{H}$ corresponding to this initial condition. Recall from the base case $N = 0$ of the proof of \Cref{t:PAI_KS} that $h_a^{(1)}(\bm{t}_I)$ is an affine function of $\bm{t}_I$ and it gives access to the quadratic function $f^{(2)}(\bm{t}_I)$ via \eqref{baseeqn}, which we reproduce here for convenience:
\begin{equation}
\label{000eqn}
f^{(2)}(\bm{t}_I)_{|\bm{t}_A = 0} = \hat{f}^{(2)}(\bm{t}_S) \qquad \text{and} \qquad\forall a \in A, \quad \partial_a \cdot f^{(2)}(\bm{t}_I) = h_a^{(1)}(\bm{t}_I).
\end{equation}
We define the partition function $\check{Z}$ by the decomposition
\begin{equation}
Z = \exp\big(f^{(2)}(\bm{t}_I) - \hat{f}^{(2)}(\bm{t}_S)\big)\hat{Z} \check{Z}.
\end{equation}
It is designed such that $\check{Z} = \exp(O(\hbar))$ and $\check{Z}_{|\bm{t}_A = 0} = 1$. It is easy to see that it satisfies the differential equations
\begin{equation}
\forall a \in A, \qquad \check{H}_a\cdot \check{Z} = 0
\end{equation}
with the operators 
\begin{equation}
\label{defchecks}
\check{H}_a \coloneqq \big(\hat{Z}\exp(f^{(2)}(\bm{t}_I) - \hat{f}^{(2)}(\bm{t}_S))\big)^{-1}H_a \hat{Z}\exp(f^{(2)}(\bm{t}_I) - \hat{f}^{(2)}(\bm{t}_S)).
\end{equation}

\begin{lemma}
\label{lem:renorm} The set $\check{\mathbf{H}} = \set{\check{H}_a | a \in A}$ generates a stable partial Airy ideal in $\mathcal{D}^\hbar_{I;\llbracket S,A\rrbracket}$ and $\check{Z}$ is its partition function with trivial initial condition.
\end{lemma}
\begin{proof} Boundedness is not affected because the only infinite sums that can occur in $\hat{Z}$ involve spectator variables, which we always complete in. The ideal condition is preserved by conjugation. It remains to check that $\check{H}_a = \hbar \partial_a + O(\hbar^2)$ for all $a \in A$. As the conjugation cannot decrease the $\hbar$-degree, the terms of $\hbar$-degree $\leq 1$ in $\check{H}_a$ can only come from the conjugation of $\hbar\partial_a -  \hbar h_a^{(1)}(\bm{t}_I)$. But the latter commutes with $\hat{Z} \exp(-\hat{f}^{(2)}(\bm{t}_S))$ in absence of spectator derivatives, and the conjugation of $\hbar \partial_a$ with $\exp(f^{(2)}(\bm{t}_I))$ kills the $h_a^{(1)}$-term due to \eqref{000eqn}.
\end{proof} 

Note that $\check{Z}$ can still depend on spectator variables, but they always come with at least one actor variable. In general, $\check{\mathbf{H}}$ can contain arbitrary powers of $\hbar$, even if $\mathbf{H}$ did not. There is one situation when this does not happen: if $\hat{Z}$ has $\hbar$-degree $0$, then we are just conjugating by $\exp(f^{(2)}(\bm{t}_I))$ and this preserves the $\hbar$-degree.

\section{Graphical calculus}

\label{Sec3}

In this section we explain how we can calculate the partition function of a partial Airy ideal as a sum over graphs.
 
\subsection{Graphical representation of the recursion}
\label{Sec31}

We will consider graphs with three types of vertices: $H$-vertices, $\hat{F}$-vertices and $F$-vertices, encoding respectively generators of the partial Airy ideal, reduced free energies and free energies.

First, we represent elements of $\mathcal{D}_I^{\hbar}$ of the form $\hbar^{1+d} t_{j_1} \cdots t_{j_n} \partial_{i_1} \cdots \partial_{i_m}$ as a vertex with $n$ incoming half-edges carrying indices $j_1,\ldots,j_n \in I$, with $m$ outgoing half-edges carrying indices $i_1,\ldots,i_m \in I$, and carrying a genus $g \in \frac{1}{2}\N$ defined by
\[
2g - 2 + (1  + n) + m = d.
\] 
The \emph{$\hbar$-degree} of the vertex is $d$, and its \emph{topology} is $(g,1+n,m)$.

Given a partial Airy ideal with generators $\set{H_a|a \in A}$, we can decompose
\begin{equation}
\label{Hdecompos}
H_a = \hbar \partial_a - \sum_{g,m,n} \sum_{\substack{\mathbf{j} \in I^n \\ \mathbf{i} \in I^m}} \frac{\hbar^{2g + n + m}}{n!m!} H_{g,1+n,m}\begin{bsmallmatrix} a,\mathbf{j} \\ \mathbf{i}\end{bsmallmatrix}\, t_{j_1} \cdots t_{j_n} \partial_{i_1} \cdots \partial_{i_m}.
\end{equation} 
The scalars $H_{g,1+n,m}\begin{bsmallmatrix} a,\mathbf{j} \\ \mathbf{i} \end{bsmallmatrix}$ define weights for the $H$-vertices. These vertices have one incoming half-edge which plays a special role (it carries the index $a$), while the weight is invariant under permutation of the other incoming half-edges. It is also invariant under permutation of the outgoing half-edges. $H$-vertices cannot have topology $(0,1,1)$: this would correspond to the term $\hbar\partial_a$ which is written separately in the sum, or to monomial $\hbar\partial_s$ with $s \in S$ which are forbidden by the definition of partial Airy ideals. The only $H$-vertices of $\hbar$-degree $0$ are $H_{0,2,0}$ and $H_{\frac{1}{2},1,0}$, corresponding respectively to the linear term and the constant terms in $h_a^{(1)}(\bm{t}_I)$.

For free energies our convention is slightly different. We represent a monomial of the form $\hbar^d t_{j_1}\cdots t_{j_n}$ appearing in $\hbar^d f^{(d + 2)}(\bm{t}_I)$ as a vertex with $n$ incoming half-edges. The genus $g \in \frac{1}{2}\N$ is defined by
\[
2g - 2 + n  = d.
\]
The \emph{$\hbar$-degree} of the vertex is $d$, and its \emph{topology} is $(g,n)$. We decompose
\begin{equation}
\label{Zdecompo}
\hat{Z} = \exp\Bigg(\sum_{g,n} \sum_{\mathbf{j} \in I^n} \frac{\hbar^{2g - 2 + n}}{n!} \hat{F}_{g,n}[\mathbf{j}]\, t_{j_1} \cdots t_{j_n} \Bigg).
\end{equation}
where we consider $\hat{F}_{g,n}[\textbf{j}]$ to be zero if $j_c \in A$ for some $c \in [n]$. The scalars $\hat{F}_{g,n}[\textbf{j}]$ define weights for the $\hat{F}$-vertices. These vertices cannot have topology $(0,0)$, $(\frac{1}{2},0)$ and $(0,1)$, because their $\hbar$-degree is nonnegative. The only $\hat{F}$-vertices of $\hbar$-degree $0$ are $\hat{F}_{0,2}$, $\hat{F}_{\frac{1}{2},1}$ and $\hat{F}_{1,0}$.  We define $F$-vertices by representing in a similar way the partition function $Z = \exp(F)$. We will ignore $\hat{F}$ and $F$-vertices without leaves, as they completely decouple and $F_{g,0} = \hat{F}_{g,0}$ for all $g \geq 0$.
 
If we form a graph on vertices of the kind we have described above, edges are oriented and carry an index in $I$: they are obtained by connecting an outgoing half-edge to an incoming half-edge with matching indices. A \emph{leaf} (incoming or outgoing) is a half-edge which is not paired with another one. All leaves are labeled and carry an index in $I$. The \emph{weight of a graph} is the product of the weights of vertices, and we will separately divided it by the number of automorphisms (taking indices into account). The \emph{$\hbar$-degree of a graph} is the sum of $\hbar$-degrees of its vertices. The \emph{genus of a graph} is the sum of genera of the vertices plus the first Betti number of the graph. In our drawings, thick edges indicate a bundle of $s \geq 1$ edges,  while thin edges indicate a (possibly empty) bundle of $s \geq 0$ edges. We will consider sums over various sets of graphs: this means that we sum over all graphs together with decorations of their edges by indices in $I$ while the leaf indices are fixed.

We put an orange dot to indicate the special incoming half-edge of $H$-vertices. We highlight (half-)edges in orange when we constrain them to carry an index in the actor set $A$, in purple when we constrain them to carry an index in the spectator set $S$. (Half-)edges without constraints are drawn in black. For consistency, we take the convention that the weight of a graph vanishes if we put an actor index on a purple half-edge  or a spectator index on an orange half-edge.

\begin{figure}[h!]
\begin{center}
\includegraphics[width=0.5\textwidth]{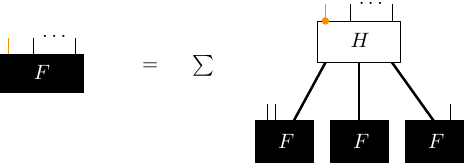}
\caption{\label{fig:recgraph} The recursion equivalent to $H_a \cdot Z = 0$, where $a \in A$ is the index carried by the orange incoming leaf.}
\end{center} 
\end{figure}

The equation $Z^{-1} H_a  Z \cdot 1 = 0$ is depicted in \Cref{fig:recgraph}. It is a finite sum over connected two-level graphs: the top level has a single $H$-vertex; the bottom level (possibly empty) is made of $F$-vertices, each of them connected by at least one edge to the top vertex. All outgoing half-edges of the top vertex must be connected to the bottom vertices. Incoming leaves can be distributed on the top vertex and the bottom vertices. An automorphism of the two-level graph is a permutation of its edges leaving the graph invariant.

\begin{figure}[h!]
\begin{center}
\includegraphics[width=0.9\textwidth]{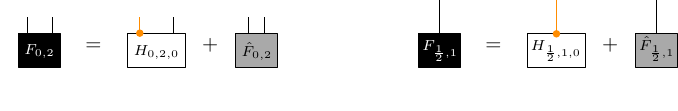}
\vspace{-0.7cm}
\caption{\label{fig:02graph} $F$-vertices of $\hbar$-degree $0$.}
\end{center}
\end{figure}

For $\hbar$-degree $0$ we find \Cref{fig:02graph}.

\subsection{Stabilized and reinitialized recursion}

\begin{figure}[h!]
\begin{center}
\includegraphics[width=0.65\textwidth]{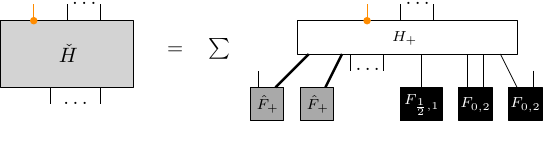}
\caption{\label{fig:renorm}  $H_+$ and $\hat{F}_+$ indicates vertices of positive $\hbar$-degree: the contributions of $H$- and $\hat{F}$-vertices of $\hbar$-degree $0$ are already included in $F_{0,2}$ and $F_{\frac{1}{2},1}$.}
\end{center}
\end{figure}

To emphasize the recursive nature of  \Cref{fig:recgraph} and absorb the contributions of the initial condition $\hat{F}$, we can work with the operators $\check{H}_a$ of \Cref{stablini}.  We decompose them as in \eqref{Hdecompos} to define $\check{H}$-vertices which only have positive $\hbar$-degree. Like $H$-vertices, $\check{H}$-vertices have a special incoming half-edge, that we indicate with an orange dot. The conjugation relation \eqref{defchecks} expresses $\check{H}$-vertices as a finite sum over connected two-level graphs (\Cref{fig:renorm}): the top level is a single $H$-vertex; the bottom level (possibly empty) is made of $F$-vertices of $\hbar$-degree zero and $\hat{F}$-vertices of positive $\hbar$-degree.
 
\begin{figure}[h!]
\begin{center}
\includegraphics[width=0.35\textwidth]{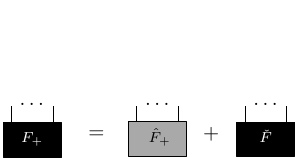}
\caption{\label{fig:FFFF} The full free energy $F$.}
\end{center}
\end{figure}

\begin{figure}[h!]
\begin{center}
\includegraphics[width=0.42\textwidth]{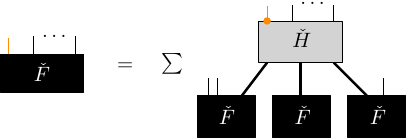}
\caption{\label{fig:recgraph2} The graphical recursion for $\check{F}$ contains only vertices of positive $\hbar$-degree.}
\end{center}
\end{figure}

\begin{figure}[h!]
\begin{center}
\includegraphics[width=0.8\textwidth]{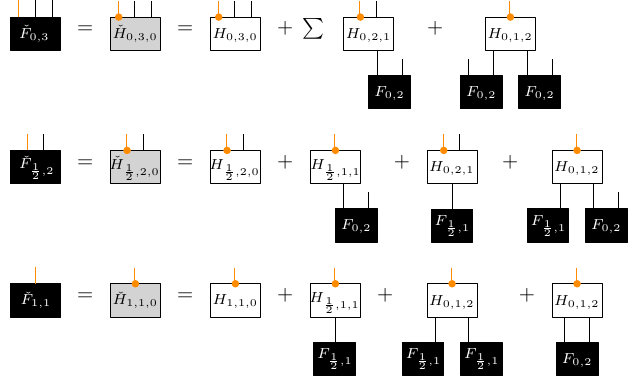}
\caption{\label{fig:recinicheck} Base cases for the recursion of \Cref{fig:recgraph2}, where one should plug \Cref{fig:02graph} for $F_{0,2}$ and $F_{\frac{1}{2},1}$.}
\end{center}
\end{figure}

By \Cref{lem:renorm}, we can obtain the free energies for $2g - 2 + n > 0$ as $F_{g,n} = \check{F}_{g,n} + \hat{F}_{g,n}$ (\Cref{fig:FFFF}), where the weight of a $\check{F}$-vertex vanishes unless it has at least one orange leaf. The relation $\check{H}_a \cdot \check{Z} = 0$  translates into the graphical recursion of \Cref{fig:recgraph2}. It is a sum over connected two-level graphs: the top level is a $\check{H}$-vertex; the bottom level (possibly empty) is made of $\check{F}$-vertices, each of them connected by at least one edge to the top vertex. Due to the nature of $\check{F}$-vertices, we can always consider that one of their leaf is orange and then use \Cref{fig:recgraph2}.  Since all vertices in \Cref{fig:recgraph2} have positive $\hbar$-degree, this is manifestly a recursion on the $\hbar$-degree $2g - 2 + n > 0$ to compute $\check{F}_{g,n}$ knowing $\check{H}$. The base cases for this recursion correspond to the $\hbar$-degree $1$ vertices $\check{F}_{0,3}$, $\check{F}_{\frac{1}{2},2}$ and $\check{F}_{1,1}$ (\Cref{fig:recinicheck}).

We now describe the formula that is obtained by iterating the recursion of \Cref{fig:recgraph2} until reaching the base cases materialized by $\check{H}$-vertices of $\hbar$-degree $1$ without any outgoing half-edge. We need to be careful in keeping track of which edge is turned orange to apply one step of the recursion. We can do so by making a distinction between orange and purple leaves.

Recall that a rooted tree $\mathfrak{t}$ is a tree with a marked vertex. Considering all leaves to be incoming, flowing out of the root vertex gives a canonical orientation to all (half-)edges in $\mathfrak{t}$. Two vertices $v$ and $v'$ are \emph{apparented} if they are connected by an oriented path in $\mathfrak{t}$. 

\begin{definition}
\label{def:GGG} Let $g \in \frac{1}{2}\N$, $n' \in \N^*$ and $n'' \in \N$. We define the set $\check{\mathbb{G}}_{g,n',n''}$ of pairs $(\Gamma,\mathfrak{t})$ such that
\begin{enumerate}
\item $\Gamma$ is a connected graph with $n'$ labeled orange leaves, $n''$ labeled purple leaves, and no outgoing leaves.
\item Vertices carry a genus in $\frac{1}{2}\N$ and have positive $\hbar$-degree.
\item The sum of $\hbar$-degrees of the vertices is $2g - 2 + n' + n''$.
\item $\mathfrak{t}$ is a rooted spanning tree of $\Gamma$; it contains all leaves and the root is incident to the first orange leaf. 
\item Edges that are not in $\mathfrak{t}$ can only connect apparented vertices in $\mathfrak{t}$.
\end{enumerate}
$\Gamma$ inherits an orientation from $\mathfrak{t}$. Each vertex $v$ is a $\check{H}$-vertex, with the following rule to decide which incoming half-edge is special: if $v$ has an orange leaf, the special one is the orange leaf with minimal label; if there are no orange leaves, the special one receives the unique edge in $\mathfrak{t}$ from the vertex just above $v$ (these special edges are coloured orange, while all the other edges remain black). We add the rule
\begin{itemize}
\item[(6)] $\mathfrak{t}$ contains all special (orange) edges.
\end{itemize}

Given $\mathbf{j} \in A^{n'} \times S^{n''}$, we denote $\check{\mathbb{G}}_{g,n',n''}[\mathbf{j}]$ the set of graphs in $\check{\mathbb{G}}_{g,n',n''}$ equipped with decorations of edges by indices in $I$ (black) or $A$ (orange), and decorations of leaves specified by $\mathbf{j}$. An automorphism of a graph $\bm{\Gamma} \in \check{\mathbb{G}}_{g,n',n''}[\mathbf{j}]$ is a permutation of edges having common index decoration and common source vertex and connecting to isomorphic components if we remove this source vertex (being in $\mathfrak{t}$ or not does not play a role for the definition of automorphisms).
\end{definition}

\begin{figure}[h!]
\begin{center}
\includegraphics[width=0.3\textwidth]{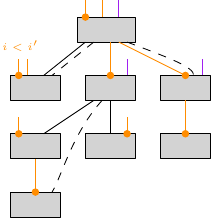}
\caption{\label{fig:sumgraph} An example of a graph from \Cref{def:GGG}. The edges that are not in $\mathfrak{t}$ are dashed. The number of automorphisms is $s_1  s_1'  s_{2}$, where: $s_1 = 2$ if the two edges relating the root and the leftmost vertex of the second level carry the same index, and $s_1 = 1$ otherwise; $s_1' = 2$ if the two edges relating the root and the rightmost vertex of the second level carry the same index, and $s_1' = 1$ otherwise; $s_2 = 2$ if the two edges issued from the middle vertex of the second level carry the same index (even though they arrive to different vertices), and $s_2 = 1$ otherwise.}
\end{center}
\end{figure}

See \Cref{fig:sumgraph} for an illustration of this definition. Due to (2)-(3), the set $\check{\mathbb{G}}_{g,n',n''}$ is finite. For each fixed $\mathbf{j}$ the set of graphs in $\check{\mathbb{G}}_{g,n',n''}[\mathbf{j}]$ with non-zero weight is finite because $\check{\mathbf{H}}$ is bounded in the completion $\mathcal{D}_{I;\llbracket S,A \rrbracket}^{\hbar}$.

\begin{proposition}[First graphical represetation for free energies]
\label{p:firstsumg} For $2g - 2 + n' + n'' > 0$ with $n' \geq 1$ and $n'' \geq 0$, we have for all $\mathbf{j} \in A^{n'} \times S^{n''}$
\[
\check{F}_{g,n' + n''}[\mathbf{j}] = \sum_{\bm{\Gamma} \in \check{\mathbb{G}}_{g,n',n''}[\mathbf{j}]} \frac{w_{\check{H}}(\bm{\Gamma})}{|\textnormal{Aut}(\bm{\Gamma})|},
\]
where $w_{\check{H}}(\bm{\Gamma})$ is the weight associated to $\bm{\Gamma}$ by considering all vertices to have $\check{H}$-weights, and $\textnormal{Aut}(\bm{\Gamma})$ is the automorphism group of the decorated graph $\bm{\Gamma}$.
\end{proposition}
\begin{proof} Removing the root vertex gives a decomposition of the set $\check{\mathbb{G}}_{g,n',n''}[\mathbf{j}]$ which matches the recursion of \Cref{fig:recgraph2}.
\end{proof}

\subsection{A solution operator}

Instead of hiding the reduced free energies $\hat{F}$ in the $\check{H}$-vertices, we can reorganize the graphs of \Cref{p:firstsumg} to emphasize how to pass from the reduced free energies to the free energies. For this purpose, we define a new set of graphs.

\begin{definition}
\label{def:GGGh}
Let $g \in \frac{1}{2}\N$, $n' \in \N^*$, and $n'',m\in \N$. Let $\bm{M} = \set{M_1,\ldots,M_p}$ be a partition of $[m]$ into pairwise disjoint non-empty subsets. We indicate this with the notation $\bm{M} \vdash [m]$. We define the set $\mathbb{G}_{g,n',n'';\bm{M}}$ of pairs $(\Gamma,\mathfrak{t})$ such that
\begin{itemize}
\item[(1)'] $\Gamma$ is a connected graph with $n'$ labeled incoming orange leaves, $n''$ labeled incoming purple leaves, and $m$ labeled outgoing purple leaves.
\item[(2)'] Vertices carry a genus in $\frac{1}{2}\N$ and have positive $\hbar$-degree, except for vertices of topology $(0,2,0)$ and $(\frac{1}{2},1,0)$.
\item[(3)'] The sum of $\hbar$-degrees of the vertices is $2g - 2 + n' + n'' + m$.
\item[(4),(5),(6)] like in \Cref{def:GGG}.
\item[(7)] Pairs of outgoing leaves included in some $M_c$ for $c \in [p]$ must be incident to apparented vertices in $\mathfrak{t}$.
\end{itemize}
We consider each vertex as a $H$-vertex, and the rule from \Cref{def:GGG} to decide which are the special  (and therefore orange) incoming (half-)edges. 

Given $\mathbf{j} \in A^{n'} \times S^{n''}$ and $\mathbf{i} \in S^{m}$ we denote $\mathbb{G}_{g,n',n'';\bm{M}}\begin{bsmallmatrix} \mathbf{j} \\ \mathbf{i} \end{bsmallmatrix}$ the set of graphs in $\mathbb{G}_{g,n',n'';\bm{M}}$ equipped with decorations of edges by indices in $I$ (black) or $A$ (orange), decoration of incoming leaves specified by $\mathbf{j}$, and decorations of outgoing leaves specified by $\mathbf{i}$. Automorphisms are  permutations of edges like in \Cref{def:GGG}, of purple outgoing leaves preserving the decorated graph and the cluster $\bm{M}$, and permutation of clusters $M_1,\ldots,M_p$ preserving the decorated graph.
\end{definition} 

\begin{figure}[h!]
\begin{center}
\includegraphics[width=0.4\textwidth]{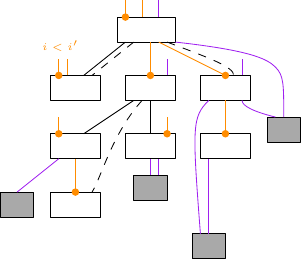}
\caption{\label{fig:sumgraph2} Example of a graph from \Cref{def:GGGh} (the H-vertices appear in white and build the subgraph $\bm{\Gamma}$). It contributes to \Cref{prop:gencolu} by adding $\hat{F}$-vertices (in grey) that can only connect to apparented H-vertices.}
\end{center}
\end{figure}

\begin{proposition}[Second graphical representation for free energies]
\label{prop:gencolu}For $2g - 2 + n' + n'' > 0$ with $n' \in \N^*$ and $n'' \in \N$, we have for all $\mathbf{j} \in A^{n'} \times S^{n''}$
\begin{equation}
\label{Fgun2}
F_{g,n'+n''}[\mathbf{j}] = \sum_{\substack{m \geq 0 \\ \mathbf{i} \in S^{m}}} \sum_{\substack{1 \leq p \leq m \\ \mathbf{g} \in (\frac{1}{2}\N)^p}} \,\,\sum_{\substack{\bm{M} \vdash [m] \\ |\bm{M}| = p}} \,\sum_{\sqcup_{c} N''_c \subseteq N''}\,\sum_{\bm{\Gamma} \in \mathbb{G}_{g_0,n',n''_0;\bm{M}}\begin{bsmallmatrix} \mathbf{j} \\ \mathbf{i}\end{bsmallmatrix}} \, \frac{w_{H}(\bm{\Gamma})}{|\textnormal{Aut}(\bm{\Gamma},\bm{M})|}  \prod_{c = 1}^{p} \hat{F}_{g_c,|M_c| + |N_{c}''|}[\mathbf{i}_{M_c},\mathbf{j}_{N''_c}],
\end{equation}
where we have set
\begin{equation}
\label{rule33}
N'' \coloneqq \set{n'+1,\ldots,n' + n''},\qquad  g_0 \coloneqq g - \sum_{c = 1}^p (g_c + |M_c| - 1),\qquad n''_0 = n'' - \sum_{c = 1}^{p} |N''_p|.
\end{equation}
The weight $w_{H}(\bm{\Gamma})$ is associated to $\bm{\Gamma}$ by considering all vertices to have $H$-weights, and $\textnormal{Aut}(\bm{\Gamma},\bm{M})$ is the group of automorphisms of the decorated graph $\bm{\Gamma}$ as in \Cref{def:GGGh}. See \Cref{fig:sumgraph2} for an example of term in \eqref{Fgun2}. 
\end{proposition} 
\begin{proof}
We return to \Cref{p:firstsumg} where we replace $\check{H}$-vertices by their expression of \Cref{fig:renorm}. In the latter formula, $\hat{F}$-vertices can appear in two ways: either fully attached to a $H$-vertex by purple edges, or partially attached to a $H$-vertex by purple edges and having purple incoming leaves. If we look at an individual $\hat{F}$-vertex now placed in a graph $\check{\bm{\Gamma}}$ of type $\check{\mathbb{G}}$ (\Cref{def:GGG}), each of those incoming purple half-edge either remains a purple incoming leaf or connects to the outgoing half-edge of a $H$-vertex above it (\emph{clustering rule}). Removing all $\hat{F}$-vertices (let us call $p$ their number) and labeling the $m \geq 0$ outgoing purple edges that were connected to them, we get a graph $\bm{\Gamma}$ of type $\mathbb{G}$ (\Cref{def:GGGh}), where the role of the partition $\bm{M} \vdash [m]$ is to remember the clustering rule. This graph has $H$-vertices only, its weight is $w_{H}(\bm{\Gamma})$.  We get all possible graphs $\bm{\Gamma}$ in this way. To retrieve $\check{\bm{\Gamma}}$ we need to take into account all possible ways to take $p$ different $\hat{F}$-vertices, possibly having incoming half-edges labeled by $N_c'' \subseteq N''$ that remain leaves in $\check{\bm{\Gamma}}$, and to connect the remaining $m$ incoming purple half-edges to the corresponding outgoing purple half-edges of $\bm{\Gamma}$ respecting the clustering rule relative to $\set{M_1,\ldots,M_p}$. The rules \eqref{rule33} are explained as follows. If we want a total of $n''$ incoming purple leaves  in $\check{\bm{\Gamma}}$ we need $n''_0$ of them in $\bm{\Gamma}$. Since connecting the $c$-th $\hat{F}$-vertex creates $|M_c| - 1$ independent cycles in $\check{\bm{\Gamma}}$, the rule that the total $\hbar$-degree is $2g - 2 + n' +n''$ forces the genus of $\bm{\Gamma}$ to be $g_0$.
\end{proof}

In general, \Cref{prop:gencolu} cannot be expressed at the level of the partition function by the linear action of a global differential operator on $\hat{Z}$. We can attach a differential operator to each graph $\bm{\Gamma}$ of type $\mathbb{G}$, where the outgoing leaves correspond to derivatives. But, not all derivatives in this operator are acting on the variables of a collection of $m$ free energies: they can only act following the clustering rule.

\vspace{0.5cm}

\section{Partial Airy ideals attached to spectral curves}

\label{Sec4}

In this section we review the notion of spectral curves and construct examples of partial Airy ideals that are naturally associated to a certain class of spectral curves called admissible. The main idea behind this construction is to obtain compatible systems of PDEs based on a free-field representation of the $\mathcal{W}(\mathfrak{gl}_r)$-algebra at self-dual level, twisted using the data of the spectral curve. The partition functions are then solutions of these \emph{$\mathcal{W}$-constraints}. As we will see in the next section, these $\mathcal{W}$-constraints can be reformulated as loop equations, and topological recursion was invented precisely to solve these loop equations.

In fact, what we do is a variant of the construction of stable Airy ideals for admissible spectral curves carried out in \cite{BBCCN24}. The main difference with \cite{BBCCN24} is that we add variables indexed by all integers, not only positive ones. This is reflected in the free-field representation that we use (see \Cref{l:rephei}, which can be compared to \cite[Equation 2.5]{BBCCN24}). We establish in \Cref{Sec:44} a correspondence between the two constructions, which is non-trivial at the level of partition functions.

For pedagogical reasons, it is useful to go through this construction before we come to the new setting of coloured spectral curves in \Cref{Sec6}, even though it is a special case of the latter.

\subsection{Spectral curves}

Spectral curves can be defined in many different ways, depending on the context of application. In this paper we follow the local definition of \cite{BBKN25}, which includes the choice of standard local coordinates, but we do not include the choice of fundamental bidifferential of the second kind (or Bergman kernel) as part of the data of a spectral curve. We will see in \Cref{Sec:44} that such bidifferentials can be naturally incorporated in the data of the initial condition $\hat{Z}$. 

\begin{definition}[Spectral curve]\label{d:sc}
  A \emph{local spectral curve} $ \mathcal{S} = ( C, x, \omega_{0,1}) $ is given by the following data:
  \begin{itemize}
  \item $C$ is a finite disjoint union of small disks $ C = \bigsqcup_{\alpha} C_{\alpha}$.\footnote{By disk, we mean an open domain of $\mathbb{C}$ homeomorphic to a disk and containing $0$.}
  \item $x : C \rightarrow \mathbb{C}$ is a map specified component-wise by
  \begin{align}
  x \colon& C_{\alpha} \rightarrow \mathbb{C}  \nonumber\\
  & z \longrightarrow z^{r_\alpha} + x_\alpha,
  \end{align}
  where $ x_\alpha \in \mathbb{C}$ and $r_\alpha \geq 2$ is an integer.
  \item $\omega_{0,1}$ is a holomorphic $1$-form on $C$ whose series expansion in $C_\alpha$ takes the form:
  \begin{equation}
    \omega_{0,1}^{(\alpha)} (z) 
    \coloneqq
    \sum_{k \geq s_\alpha} F_{0,1}\begin{bsmallmatrix}\alpha \\ -k \end{bsmallmatrix} z^{k-1} \dd z
  \end{equation}
  for some integer $s_{\alpha} > 0$ and such that $\tau_{\alpha} \coloneqq -F_{0,1}\begin{bsmallmatrix} \alpha \\ - s_{\alpha} \end{bsmallmatrix} \neq 0$.
  \end{itemize}
  \end{definition}

The origin in each $C_{\alpha}$ is a ramification point for the map $x$. We call $\Ram$ the set of ramification points and we consider indices $\alpha$ as elements of $\Ram$. In particular, we will write $\Res_{z = \alpha}$ for the residue at the ramification point in $C_{\alpha}$.

We next define the condition of admissibility, which is a particular requirement on the local behavior of $\omega_{0,1}$ near the ramification points.

\begin{definition}[Admissibility]\label{d:admissible}
We say that a spectral curve is \emph{admissible} if for any $\alpha \in \Ram$ we have $s_{\alpha} \in [r_{\alpha} + 1]$ and $r_{\alpha} = \pm 1\,\,\textnormal{mod}\,\,s_{\alpha}$. In particular, this implies that $s_\alpha$ and $r_\alpha$ are coprime.
\end{definition}
 
We could have been a little more general here, by considering $x$ taking values in the Riemann sphere, allowing $r_{\alpha} \in \mathbb{Z}^*$, allowing $\omega_{0,1}$ to be meromorphic (see the definition of admissibility in \cite[Definition 2.6]{BBCKS}), and allowing an additional non-zero term $F_{0,1}\begin{bsmallmatrix} \alpha \\ -s_{\alpha} - 1 \end{bsmallmatrix}$ when $s_{\alpha} + 1$ is divisible by $r_{\alpha}$. It would not really add anything interesting to the story (for instance, some $\alpha$ would not contribute to the topological recursion) and we do not discuss these extensions. The present setting will anyway be a particular case of the new general setting we propose in \Cref{Sec6}.

To construct partial Airy ideals associated to admissible spectral curves, we will work with the following completed algebras (see \Cref{d:comprees}, \Cref{a:completion} and in particular \Cref{r:aboutcomp}).

\begin{definition}[Completions]
\label{d:WeylZcomplete}
We introduce the completed Weyl algebras
\[
\widehat{\mathcal{D}}_{\mathbb{Z}} \coloneqq \mathcal{D}_{\mathbb{Z};\llbracket \mathbb{Z}_{<0},\N\rrbracket}\qquad \textnormal{and more generally}\quad \widehat{\mathcal{D}}_{\Ram\times \mathbb{Z}}  \coloneqq \mathcal{D}_{\Ram\times \mathbb{Z};\llbracket \Ram\times\mathbb{Z}_{<0}, \Ram \times \N\rrbracket}
\]
and the corresponding completed Rees Weyl algebras $\widehat{\mathcal{D}}_{\Z}^{\hbar}$ and $\widehat{\mathcal{D}}_{\Ram\times\Z}^{\hbar}$.
\end{definition} 
In words, we introduce a set of coordinates indexed by integers for each ramification point, and we complete in the variables for negative integers and in the derivatives for nonnegative integers.

\subsection{Partial Airy ideals for a single ramification point}
\label{Sec41}

We now construct partial Airy ideals attached to admissible spectral curves. 
For simplicity, we start by constructing partial Airy ideals associated to admissible spectral curves $\mathcal{S} = (C,x,\omega_{0,1})$ with a single ramification point. In this case, $C$ is a disk, we write $r := r_1$, $s := s_1$ and we remove indices $\alpha$ everywhere. We also introduce $\theta := \exp(\frac{2\ii\pi s}{r})$.

The first step in the construction is only sensitive to the order $r$ of the ramification point and is identical to \cite{BBCCN24} (see also \cite[Section 2]{BBKN25}). The starting point is the $\mathcal{W}(\mathfrak{gl}_r)$-algebra at self-dual level. Without getting into the details, what is important to know here is that the $\mathcal{W}(\mathfrak{gl}_r)$-algebra is strongly freely generated by a set of fields $W^{(i)}(z)$, $i \in [r]$, with mode expansions
\begin{equation}
W^{(i)}(z) = \sum_{k \in \mathbb{Z}} W^{(i)}_k z^{-k-i}.
\end{equation}
The modes $W^{(i)}_k$ of the strong generators generate an associative algebra after completion, which we denote $\mathcal{U}_r$. This algebra has a representation in terms of a  $\Z_r$-twisted free-field \cite{FL88}, i.e. can be described as a subalgebra of the universal enveloping algebra $\mathcal{H}$ of the Lie algebra generated by $\set{J_k\,\,|\,\,k \in \mathbb{Z}}$ with commutation relations
\begin{equation}
\label{Heisco}
[J_m, J_n] = m \delta_{m+n,0}
\end{equation}
and completed in the nonnegative modes. Concretely, this representation is the algebra morphism $\mu : \mathcal{U}_r \rightarrow \mathcal{H}$ with\footnote{This is \cite[Proposition 4.5 and Corollary 4.7]{BBCCN24} except that our $W_k^{(i)}$ do not include the normalization factor $r^{1-i}$, and powers of $\hbar$ since our commutation relation \eqref{Heisco} do not contain any.}
\begin{equation}
\label{Miur}
 \mu( W^{(i)}_k)
   \coloneqq
    \frac{1}{r^i} \sum_{j=0}^{\lfloor i/2 \rfloor} \frac{i!}{2^j j! (i-2j)!} \sum_{\substack{m_{2j+1}, \ldots, m_i \in \Z \\ \sum_l m_l = rk}} \Psi^{(i,j)}_r (m_{2j+1}, \ldots, m_i) : \prod_{l=2j+1}^i J_{m_l} : \,,
\end{equation}
  where
  \begin{equation}
 \label{Psiiki}   \Psi^{(i,j)}_r (m_{2j+1}, \ldots, m_i) \coloneqq \frac{1}{i!} \sum_{\substack{n_1, \ldots, n_i = 0\\ n_l \neq n_k}}^{r-1} \prod_{l = 1}^j \frac{\theta^{n_{2l-1} + n_{2l}}}{(\theta^{n_{2l-1}} - \theta^{n_{2l}})^2} \prod_{l=2j+1}^i \theta^{-n_l m_l} \, .
  \end{equation}
The normal ordering $: \cdots :$ moves the $J$s with positive indices to the right. For $j=\frac{i}{2}$ the condition $\sum_l m_l = rk$ is understood as $\delta_{k,0}$. 

For computations it is useful to record the following evaluations of \eqref{Psiiki}.
   
  \begin{lemma} \cite[Appendix A]{BBCCN24}
\label{evallem} For $r \geq 2$ and $i = 2$, we have:
\[
 \Psi_r^{(2,0)}(m_1,m_2) = \frac{r}{2}(r\delta_{r|m_1}\delta_{r|m_2} - \delta_{r|m_1 + m_2}),\qquad  \Psi_r^{(2,1)}(\emptyset) = - \frac{r(r^2 - 1)}{24}.
\]
 For $j = 0$ and $i \in [r]$, we have:
  \[
  \Psi_r^{(i,0)}(m_1,\ldots,m_i) = \frac{1}{i!} \sum_{\mathbf{L} \vdash [i]} r^{|\mathbf{L}|} (-1)^{i - |\mathbf{L}|} \prod_{L \in \mathbf{L}} (|L| - 1)! \,\delta_{r|\sum_{l \in L} m_l},
  \]
  where the sum ranges over set partitions $\mathbf{L} = \set{L_1,\ldots,L_p} \vdash [i]$, and we call $p = |\mathbf{L}|$.
  \end{lemma}
  
  The route taken in  \cite{BBCCN24} consists in realizing the $W_k^{(i)}$ as differential operators in variables $\bm{t}_{\N^*}$, that is mapping them to elements of $\mathcal{D}_{\N^*;\llbracket \N^*,\emptyset \rrbracket}$. This can be achieved simply by mapping $J_0$ to a scalar, $J_{-k}$ to $|k|t_{|k|}$ and $J_{k}$ to $\partial_k$ for $k > 0$. 

Here we will rather  realize the $W_k^{(i)}$ as differential operators in variables $\bm{t}_{\mathbb{Z}}$, that is mapping them to elements of $\widehat{\mathcal{D}}_{\Z}$. We are using here the Weyl algebra completed in negative variables and nonnegative derivatives (\Cref{d:WeylZcomplete}).

\begin{lemma}[Free-field representations]\label{l:rephei}
The map $\sigma: \mathcal{H} \rightarrow \widehat{\mathcal{D}}_{\mathbb{Z}}$  defined by $\sigma(J_k) = \partial_k + \delta_{k < 0}  |k| t_{|k|}$ is an algebra morphism. 

The map $\rho = \sigma \circ \mu: \mathcal{U}_r\rightarrow \widehat{\mathcal{D}}_{\mathbb{Z}}$ given by
 \begin{equation}\label{eq:wwr}
   \rho( W^{(i)}_k)
    =
    \frac{1}{r^i} \sum_{j=0}^{\lfloor i/2 \rfloor} \frac{i!}{2^j j! (i-2j)!} \sum_{\substack{m_{2j+1}, \ldots, m_i \in \Z \\ \sum m_l = rk}} \Psi^{(i,j)}_r (m_{2j+1}, \ldots, m_i) : \prod_{l=2j+1}^i  \left(  \partial_{m_l} + |m_l| t_{|m_l|} \delta_{m_l < 0}\right) : 
  \end{equation}
is an algebra morphism.  Here, the normal ordering $: \cdots :$ moves all derivatives to the right.\footnote{The normal ordering in \eqref{Miur} moves factors with positive $m_l$ to the right. With the representation $\sigma$ those factors contain the positive derivatives. Then, the factors with negative $m_l$ are on the left, and they only involve the (mutually commuting) negative derivatives and positive variables. Thus, we do not affect the outcome to move those negative derivatives to the right. The net effect is that we are allowed to move all derivatives to the right, as announced here.}
\end{lemma}
\begin{proof}
Let $A_{k,l}$ be scalars indexed by $k,l \in \Z$. If we set $\sigma(J_k) = \partial_{k} + \sum_{l \in \Z} A_{k,l}t_l$ we compute
\[
\forall m,n \in \Z\qquad [\sigma(J_{m}),\sigma(J_{n})] = A_{n,m} - A_{m,n}.
\]
For $A_{k,l} = l \delta_{k < 0} \delta_{k + l,0} = -k \delta_{l > 0} \delta_{k + l,0}$ we find $A_{n,m} - A_{m,n} = m\delta_{m + n,0}$. This shows that $\mu$ is an algebra morphism, and so remains the composition with $\sigma$. The operator written in \eqref{eq:wwr} does belong to the completion $\widehat{\mathcal{D}}_{\Z}$: for each fixed $i$, $k$ and $b_1,\ldots,b_q < 0$, the monomial $\partial_{b_1}\cdots \partial_{b_q}$ can  only appear among the  finitely many terms satisfying $m_l \in \set{b_1,\ldots,b_q}$ or $0 \leq m_l \leq rk + \sum_{l} |b_l|$.
\end{proof}

\begin{remark}\label{r:bergrep}
The coefficients of $\bm{t}_{\mathbb{Z}}$ in $\sigma(J_k)$ correspond to the coefficients of the expansion of the standard bidifferential (see \Cref{d:sofc}) as $|z_1| < |z_2|$ both approach the ramification point at $0$:
\[
\omega_{0,2}^{\std}(z_1,z_2) \coloneqq \frac{\dd z_1 \dd z_2}{(z_1 - z_2)^2} = \sum_{k < 0} |k|z_1^{|k| - 1}z_2^{k - 1} \dd z_1 \dd z_2.
\]
We could have been more general here and define $\sigma(J_k)$ to correspond to a more general choice of fundamental bidifferential, but it is simpler to work with this standard $\sigma$ and move the choice of bidifferential into the initial condition.
\end{remark}

In a second step, we make the operators $\rho(W^{(i)}_k)$ homogeneous by multiplying by $\hbar^i$ to get operators living in the completed Rees Weyl algebra $\widehat{\mathcal{D}}^\hbar_\mathbb{Z}$, and incorporate the data of $\omega_{0,1}$ from the spectral curve by a \emph{dilaton shift}.

\begin{definition}[Dilaton shift]{}
\label{d:Hik}
We set $\overline{W}^{(i)}_k := \hbar^i \rho(W^{(i)}_k)$ and  $H^{(i)}_k := Z_{(0,1)}^{-1} \overline{W}^{(i)}_k Z_{(0,1)}$, where
\[
Z_{(0,1)} = \exp\Bigg(  \frac{1}{\hbar}\sum_{m \geq s} F_{0,1}[-m] t_{-m}  \Bigg).
\]
\end{definition}
The net effect of the conjugation by $Z_{(0,1)}$ is to replace $\sigma(J_m)$ by
\[
Z_{(0,1)}^{-1} \hbar \sigma(J_m) Z_{(0,1)} = \hbar \partial_m + \hbar |m|t_{|m|}\delta_{m<0} + F_{0,1}[m],
\]
which de facto breaks the homogeneity of $\overline{W}^{(i)}_k$. Since $s > 0$, we can also see the dilaton shift as equivalent to the transformation
\begin{equation}
\label{dailtonshit}
\forall m \geq 1, \qquad t_{m} \longrightarrow t_{m} + \frac{F_{0,1}[-m]}{\hbar m}.
\end{equation}

\begin{theorem}\label{t:PAISC1}
Let $\mathcal{S} = (C,x,\omega_{0,1})$ be an admissible spectral curve with a single ramification point. Let $\mathcal{I}_\mathcal{S} \subset \widehat{\mathcal{D}}^\hbar_{\mathbb{Z}}$ be the left ideal generated by the set of operators
\[
\mathbf{H} \coloneqq \set{H^{(i)}_k \,\,|\,\,i \in [r]\quad \textnormal{and}\quad k \geq \dk^{(i)}},\qquad \textnormal{where}\quad \dk^{(i)} \coloneqq - \left\lfloor \frac{s(i-1)}{r} \right\rfloor.
\]
Then $\mathcal{I}_{\mathcal{S}}$ is a stable partial Airy ideal on the set $I \coloneqq \mathbb{Z}$ with actor set $A \coloneqq \N$. 

Furthermore, for any initial condition, its associated partition function $Z$ is independent of $t_{rm}$ for all $m \geq 0$. Also, if for some $m < 0$ the initial condition $\hat{Z}$ does not depend on $t_{m}$, then $Z$ also does not depend on $t_{m}$.
\end{theorem}
\begin{proof}
It is clear that $\bm{H}$ is bounded. The ideal condition $[\mathcal{I}_\mathcal{S}, \mathcal{I}_{\mathcal{S}}] \subseteq \hbar^2 \mathcal{I}_{\mathcal{S}}$ is proved in  \cite[Theorem 4.9]{BBCCN24}. The dilaton shift \eqref{dailtonshit} is the same as the one performed in \cite{BBCCN24}, so we obtain the same result as in Proposition 4.13 there, which yields a generating set $\set{G_j\,\,|\,\,j \in \N}$ such that
\[
G_j = \hbar \rho(J_j) + O(\hbar^2) = \hbar \partial_j + O(\hbar^2).
\]

Furthermore, the set $\bm{H}$ always contain $H_m^{(1)}$ for $m \geq 0 $, which is proportional to $\hbar \partial_{rm}$. Therefore, the partition function is independent of $t_{rm}$.

Finally, the operators in \Cref{d:Hik} do not involve $\bm{t}_{-\N}$ but can involve the derivatives $\bm{\partial}_{-\N}$. Suppose that for some fixed $m<0$, $\hat{Z}$ does not depend on $t_m$. Proceeding by induction as in the proof of \Cref{t:PAI_KS}, we  then see that for all $N \geq 0$ there exists a unique truncation $Z^{(N)}$ satisfying the properties \eqref{wesolve} and such that $Z^{(N)}$ does not depend on $t_m$. As a result, we conclude that $Z$ cannot depend on $t_m$.
\end{proof}

\subsection{The case of several ramification points}

For spectral curves $\mathcal{S}=(C,x,\omega_{0,1})$ with an arbitrary set of ramification points, the construction is essentially the same, with one copy of the operators for each component $C_\alpha$ with $\alpha \in \Ram$. Combining the formulae of the previous section, we consider the operators
\begin{equation}\label{eq:overline}
\overline{W}_{\alpha,k}^{(i)} \coloneqq 
    \frac{\hbar^i}{r_\alpha^i} \sum_{j=0}^{\lfloor i/2 \rfloor} \frac{i!}{2^j j! (i-2j)!} \sum_{\substack{m_{2j+1}, \ldots, m_i \in \Z \\ \sum_l m_l = r_\alpha k}} \Psi^{(i,j)}_{r_\alpha} (m_{2j+1}, \ldots, m_i) : \prod_{l=2j+1}^i  \left( \partial_{\alpha,m_l} + |m_l| t_{\alpha,|m_l|} \delta_{m_l < 0}\right) :.
\end{equation}
and incorporate a dilaton shift as usual
\begin{equation}\label{eq:HikN}
H^{(i)}_{\alpha,k} := Z_{(0,1)}^{-1} \overline{W}_{\alpha,k}^{(i)} Z_{(0,1)} \qquad \text{with}\qquad Z_{(0,1)} = \exp\left(  \frac{1}{\hbar}\sum_{\substack{\alpha \in \Ram \\ m \geq s_\alpha} } F_{0,1}\begin{bsmallmatrix} \alpha \\ -m \end{bsmallmatrix} t_{\alpha,-m}\right). 
\end{equation}
Concretely, $H^{(i)}_{\alpha,k}$ is obtained from \eqref{eq:overline} after substituting
\begin{equation}
\label{theshift}
\forall (\alpha,m) \in \Ram \times \N^*,\qquad t_{\alpha,m} \longrightarrow t_{\alpha,m} + \frac{1}{\hbar m} F_{0,1}\begin{bsmallmatrix} \alpha \\ -m \end{bsmallmatrix}.
\end{equation}
Since these operators are just independent copies of those of \Cref{d:Hik} for each $\alpha \in \Ram$, the following result is a corollary of \Cref{t:PAISC1}.
\begin{theorem}\label{t:PAISC}
Let $\mathcal{S} = (C,x,\omega_{0,1})$ be an admissible spectral curve. Let $\mathcal{I}_\mathcal{S} \subset \widehat{\mathcal{D}}^\hbar_{\Ram \times \Z}$ be the left ideal generated by the set of operators
\[
\mathbf{H} \coloneqq \set{H_{\alpha,k}^{(i)}\,\, |\,\, \alpha \in \Ram, \quad i \in [r_{\alpha}],\quad \textnormal{and} \quad k \geq \kappa^{(i)}_{\alpha}},\qquad \textnormal{where} \quad \kappa_{\alpha}^{(i)} \coloneqq - \left\lfloor \frac{s_{\alpha}(i-1)}{r_{\alpha}}\right\rfloor.
\]
Then $\mathcal{I}_{\mathcal{S}}$ is a stable partial Airy ideal on the set $I \coloneqq \Ram \times \mathbb{Z}$ with actor set $A \coloneqq \Ram \times \N$.

Furthermore, for any initial condition, its associated partition function $Z$ does not depend on the variables $\set{t_{\alpha,r_{\alpha}m} | \alpha \in \Ram}$ for any $m \geq 0$. Also, if for some $\alpha \in \Ram$ and $m < 0$ the initial condition $\hat{Z}$ does not depend on $t_{\alpha,m}$, then $Z$ also does not depend on $t_{\alpha,m}$.
\end{theorem}

Although there is an independent set of operators associated to each $\alpha \in \Ram$, the partition function may not factorize as $Z = \prod_{\alpha} Z_{\alpha}$: coupling can occur through a non-factorized initial condition $\hat{Z}$.

\subsection{Relation with the Airy ideals of \cite{BBCCN24}}

\label{Sec:44}

The partial Airy ideals constructed in \Cref{t:PAISC} differ from the Airy ideals originally constructed in \cite[Theorem 4.9]{BBCCN24}. The main difference comes in the free-field representation from \Cref{l:rephei}, which lives in the Rees Weyl algebra for the variables $\bm{t}_{\Ram \times \mathbb{Z}}$; on the other hand, in \cite{BBCCN24}, the representation used (see \cite[eq.(2.5)]{BBCCN24}) lives in the Rees Weyl algebra for only the  positive variables $\bm{t}_{\Ram \times \mathbb{N}^*}$. With the free-field representation of \Cref{l:rephei}, we get more general solutions depending on a choice of initial condition $\hat{Z}$ (which is a function of the spectator variables $\bm{t}_{\alpha, -\mathbb{N}^*}$).

What we show in this section is the following:
\begin{enumerate}
\item To every partial Airy ideal in \Cref{t:PAISC}, we construct an associated Airy ideal in the actor variables $\bm{t}_A$ of the form of \cite{BBCCN24}. We denote by $\mathring{Z}$ its unique partition function such that $\mathring{Z}_{|\bm{t}_A = 0} = 1$; note that $\mathring{Z}$ only depends on the $\bm{t}_A$. We call $\mathring{Z}$ the \emph{active partition function}.
\item Given a constant initial condition $\hat{Z}$ (i.e. independent of $\bm{t}_S$), the full partition function $Z$ of the partial Airy ideal of \Cref{t:PAISC} does not depend on $\bm{t}_S$ and is given by $Z = \mathring{Z} \hat{Z}$.
\item For a general choice of initial condition $\hat{Z}$, the situation is more intricate. We show that there exists a \emph{mixing operator} $\textnormal{M}$ such that $Z = \textnormal{M} \cdot \mathring{Z} \hat{Z}$. Crucially, $\textnormal{M}$ is constructed in terms of the data of $\mathring{Z}$ and a duality operator $\textnormal{D}$. As a corollary, we get an explicit \emph{solution operator} $\Sigma$ that takes as input the initial condition $\hat{Z}$ and returns the full partition function $Z$.
\end{enumerate}
These statements provide an explicit connection between the construction here and that of \cite{BBCCN24} for all choices of initial condition $\hat{Z}$.

\begin{lemma}\label{l:tobbccn}
Consider a partial Airy ideal from \Cref{t:PAISC} with generators $H^{(i)}_{\alpha,k}$. For all $\alpha$ and $k$, define
\begin{equation}
\mathring{H}_{\alpha,k}^{(i)} = H_{\alpha,k}^{(i)}{}_{|\bm{\partial}_{S} = 0},
\end{equation}
where we set the derivatives with respect to the spectator variables to zero in normally-ordered expressions. Then, for each $\alpha \in \Ram$, the left ideal generated by the set of operators\footnote{By $\widehat{\mathcal{D}}^\hbar_{\{\alpha\} \times \mathbb{N}}$ we mean the Rees Weyl algebra for the set $\{\alpha\} \times \mathbb{N}$ where we complete in all derivatives. Actually, in \cite{BBCCN24} the representation involves the set $\{\alpha\} \times \mathbb{N}^*$, because they set $H^{(1)}_{\alpha,0} = \hbar J_{\alpha,0} = 0$. But one could have instead set $H^{(1)}_{\alpha,0} = \hbar J_{\alpha,0} = \hbar \partial_{\alpha,0}$ and obtained an Airy ideal on the set $\{\alpha\} \times \N$. Since $H^{(1)}_{\alpha,0}$ is always in the ideal, one concludes right away that $Z$ is independent of $t_{\alpha,0}$, so introducing $(t_{\alpha,0},\partial_{\alpha,0})$ would have changed nothing in the construction of \cite{BBCCN24}.}
\[
\mathring{\bm{H}}_{\alpha} \coloneqq \set{\mathring{H}_{\alpha,k}^{(i)}\,\, |\,\,i \in [r_{\alpha}] \quad \textnormal{and} \quad k \geq \kappa^{(i)}_{\alpha}} \subset \widehat{\mathcal{D}}^{\hbar}_{\{\alpha\} \times \N}
\]
is an Airy ideal of the form of \cite[Theorem 4.9]{BBCCN24}.
\end{lemma}
\begin{proof}
This is clear, since we can go from the free-field representation of \Cref{l:rephei} to the free-field representation \cite[Equation 2.5]{BBCCN24} simply by setting $\partial_{\bm{t}_S}=0$ for all spectator variables.
\end{proof}

\begin{definition}[Active partition function] 
\label{def:activep} Let $\mathring{Z}$ be the unique partition function of the Airy ideal $\sqcup_{\alpha} \mathring{\bm{H}}_{\alpha}$ satisfying $\mathring{Z}_{|\bm{t}_A = 0}=1$.
\end{definition}

Next, we consider the very simple case where we choose a constant initial condition $\hat{Z}$. 

\begin{lemma}\label{l:PQconst}
Let $Z$ be the partition function of a partial Airy ideal from \Cref{t:PAISC} with an initial condition $\hat{Z}$ such that $\hat{Z}_{|\bm{t}_S=0} = \hat{Z}$. Then $Z = Z_{|\bm{t}_S=0} =\mathring{Z} \hat{Z}$.
\end{lemma}
\begin{proof}
If $\hat{Z}$ does not depend on the spectator variables, from \Cref{t:PAISC} we know that $Z$ does not depend on the spectator variables either, and hence $Z = Z_{|\bm{t}_S=0}$. Then, for all $\alpha$ and $k$, $H^{(i)}_{\alpha,k} \cdot Z = \mathring{H}^{(i)}_{\alpha, k} \cdot Z = 0$. Thus $Z = A \mathring{Z}$ for some constant $A$. Since $Z_{|\bm{t}_A=0} = \hat{Z}$ and $\mathring{Z}_{|\bm{t}_A = 0}=1$, we conclude that $Z = \hat{Z} \mathring{Z}$. 
\end{proof}

Thus, when $\hat{Z}$ is constant, the full partition function $Z$ is simply the product of the reduced partition function $\hat{Z}$ and the active partition function $\mathring{Z}$. This is not true however for more general choices of initial conditions. What we can show instead is that there exists a mixing operator which reconstructs $Z$ from the product $\hat{Z} \mathring{Z}$. 

We first introduce a duality operator.\footnote{In this section we will deal with operators that are not quite in the Rees Weyl algebra $\widehat{\mathcal{D}}^{\hbar}_{\Ram \times \Z}$. However, like partition functions, they can be included in a larger set where the product with elements of $\widehat{\mathcal{D}}^{\hbar}_{\Ram \times \Z}$ or the conjugation with partition functions is well defined.}

\begin{definition}[Duality operator]\label{d:duality}
We define the \emph{duality operator}:
\begin{equation}
\label{DuDu}
\Du \coloneqq \exp\left(\sum_{\substack{\alpha \in \Ram \\ m \in \Z^*}} \frac{\partial_{\alpha,m}\partial_{\alpha,-m}}{2|m|}\right).
\end{equation}
Conjugating with $\Du$ amounts to doing the substitutions
\[
\forall (\alpha,m) \in \Ram \times \Z^*, \qquad mt_{\alpha,m} \longrightarrow mt_{\alpha,m} + \partial_{\alpha,-m}.
\]
\end{definition}

The duality operators gives us an explicit relation between the $H^{(i)}_{\alpha,k}$ and the $\mathring{H}_{\alpha,k}^{(i)}$:

\begin{lemma}\label{l:HHring}
Consider a partial Airy ideal from \Cref{t:PAISC} with generators $H^{(i)}_{\alpha,k}$, and define the $\mathring{H}_{\alpha,k}^{(i)}$ as in \ref{l:tobbccn}. Then
\begin{equation}\label{eq:HHring}
H_{\alpha,k}^{(i)} = \Du \mathring{H}_{\alpha,k}^{(i)} \Du^{-1}.
\end{equation}
\end{lemma}

\begin{proof}
The only difference between $H^{(i)}_{\alpha,k}$ and $\mathring{H}_{\alpha,k}^{(i)}$ is that $H_{\alpha,k}^{(i)}$ has a factor $\partial_{\alpha,-m} + mt_{\alpha,m}$ each time $\mathring{H}_{\alpha,k}^{(i)}$ contains a factor $mt_{\alpha,m}$ for $(\alpha,m) \in \Ram \times \N^*$. Since $\mathring{H}_{\alpha,k}^{(i)}$ does not contain negative variables, this difference can be reproduced by conjugating with the duality operator from \Cref{d:duality}. This gives \eqref{eq:HHring}.
\end{proof}

Next we define a dagger operator.

\begin{definition}[Dagger operator]\label{d:dagger}
We define the algebra morphism $\dagger$ which exchanges $\partial_{\alpha,m} \leftrightarrow mt_{\alpha,-m}$ for $(\alpha,m) \in \Ram \times \Z^*$ and sends $t_0$ and $\partial_0$ to zero.
\end{definition}

The dagger and duality operators can be related as follows:
\begin{lemma}\label{l:daggerduality}
Let $Y$ be a partition function that depends only on the positive variables $\bm{t}_{\Ram \times \N^*}$. Then 
\begin{equation}
\label{daggeru}
(\Du Y \Du^{-1})_{|\bm{t}_{\Ram \times \N^*} = 0} = Y^{\dagger},\qquad \Du \cdot Y = Y.
\end{equation}
A similar statement is true if $Y$ depends only on the negative variables $\bm{t}_{\Ram \times -\N^*}$.
\end{lemma}

\begin{proof}
Since $Y$ only depends on the positive variables, conjugating by $\Du$ does not introduce any normal-ordering issues, and we can bring all derivatives to the right. Then it is clear that setting all variables $\bm{t}_{\Ram \times \N^*}$ to zero simply reproduces the action of the dagger operation, which is \eqref{daggeru}. Furthermore, since $Y$ does not have simultaneously negative and positive variables, we have $\Du \cdot Y = Y$. The same holds if $Y$ depends only on the negative variables.
\end{proof}

Finally we define a mixing operator.

\begin{definition}[Mixing operator]\label{d:mixing}
Let $Z$ be a partition function as in \Cref{t:PAISC} and $\mathring{Z}$ its active partition function from \Cref{l:tobbccn}. We define $\Mu = \Du (\mathring{Z}^\dagger)^{-1}$.
\end{definition}

We can now state the main result of this section.

\begin{proposition}
\label{ProP:Hi}
Let $\mathcal{S} = (C,x,\omega_{0,1})$ be an admissible spectral curve. Construct a partial Airy ideal as in \Cref{t:PAISC} and let $Z$ be the associated partition function with initial condition $\hat{Z} = \hat{Z}(\bm{t}_S)$. Let $\mathring{Z} =\mathring{Z}(\bm{t}_A)$ be the active partition function of \Cref{def:activep}. Then
\begin{equation}
Z = \Mu \cdot (\hat{Z} \mathring{Z}),
\end{equation}
\end{proposition}
\begin{proof}
Define $Y = \Mu \cdot (\hat{Z} \mathring{Z})$. We want to show that $Y= Z$. To prove this claim, we need to show that check that their specialization to $\bm{t}_A = 0$ agree and that they are both annihilated by the differential operators in $\bm{H}$. For clarity we introduce the notation $\Phi \coloneqq (\mathring{Z}^\dagger)^{-1} \cdot \hat{Z}$, which is a partition function depending only on spectator variables. For the initial condition, we have
\begin{equation*}
\begin{array}{ccll}
Y_{|\bm{t}_A=0} & = &  \left( (\Du (\mathring{Z}^\dagger)^{-1}) \cdot \hat{Z} \mathring{Z}\right)_{|\bm{t}_A=0} & \quad \\[5pt]
& = &  \left(\Du \mathring{Z} \cdot ( (\mathring{Z}^\dagger)^{-1} \cdot \hat{Z}) \right)_{|\bm{t}_A=0}  & \quad \text{since $\mathring{Z}$ depends only on $\bm{t}_A$} \\[5pt]
& = & (\Du \mathring{Z}\Du^{-1} \cdot \Phi)_{|\bm{t}_A=0} & \quad \text{since $\Phi= (\mathring{Z}^\dagger)^{-1} \cdot \hat{Z}$ only depends on $\bm{t}_S$, hence $\Du^{-1} \cdot \Phi = \Phi$} \\[5pt]
& = & (\Du \mathring{Z} \Du^{-1})_{|\bm{t}_A=0}\cdot \Phi & \quad \text{since $\Phi$ only depends on $\bm{t}_S$}\\[5pt]
&= & \mathring{Z}^\dagger \cdot \Phi & \quad \text{by \Cref{l:daggerduality}}\\[5pt]
&= & \mathring{Z}^\dagger (\mathring{Z}^\dagger)^{-1} \cdot \hat{Z} & \quad \text{since $\Phi = (\mathring{Z}^{\dagger})^{-1} \cdot \hat{Z}$} \\[5pt]
&= &\hat{Z}. & 
\end{array}
\end{equation*}
For the differential equation, we have for $\alpha \in \Ram$, $i \in [r_{\alpha}]$ and $k \geq \kappa^{(i)}_{\alpha}$
\begin{equation*}
\begin{array}{ccll}
H^{(i)}_{\alpha,k} \cdot Y & = &H^{(i)}_{\alpha,k}\Du (\mathring{Z}^\dagger)^{-1} \cdot \hat{Z} \mathring{Z} & \\[5pt]
& = &H^{(i)}_{\alpha,k}\Du \mathring{Z} \cdot \big((\mathring{Z}^\dagger)^{-1} \cdot\hat{Z} \big) & \quad \text{since $\mathring{Z}$  only depends on $\bm{t}_A$}\\[5pt]
&=&\Du \mathring{H}^{(i)}_{\alpha,k}\mathring{Z}  \cdot \Phi  & \quad \text{by \Cref{l:HHring}}\\[5pt]
&=&\Du \mathring{H}^{(i)}_{\alpha,k}\Phi \cdot \mathring{Z} & \\[5pt]
&=& \Du \Phi \mathring{H}^{(i)}_{\alpha,k} \cdot \mathring{Z} & \quad \text{since $\Phi$ only depends on $\bm{t}_S$ and $\mathring{H}^{(i)}_{\alpha,k}$ does not involve $\bm{\partial}_S$}\\[5pt]
&=& 0 & \quad \text{by definition of $\mathring{Z}$.}
\end{array}
\end{equation*}
\end{proof}
 
\Cref{ProP:Hi} can also be read in the following way.

\begin{corollary}\label{c:solution}
Define the \emph{solution operator} $\Sigma \coloneqq \Mu \mathring{Z}= \Du (\mathring{Z}^\dagger)^{-1} \mathring{Z} = \Du \mathring{Z} (\mathring{Z}^\dagger)^{-1} $. Then $Z = \Sigma \cdot \hat{Z}$. 
\end{corollary}

This tells us how to reconstruct the full partition function $Z$ by acting with the solution operator $\Sigma$ on the initial condition $\hat{Z}$. Furthermore, the solution operator $\Sigma$ is built from the duality operator $\Du$ using the active partition function $\mathring{Z}$ as a pivot.

We can extract from \Cref{ProP:Hi} a second rewriting.

\begin{corollary}\label{c:dressed}
Define the \emph{dressed initial condition} $\Phi = (\mathring{Z}^\dagger)^{-1} \cdot \hat{Z}$ and the \emph{dressing operator} $\Phi^{\dagger}$. Then $Z_{|\bm{t}_S=0} = \Phi^\dagger \cdot \mathring{Z}$. 
\end{corollary}
\begin{proof}
We have  $Z = \Mu \cdot \hat{Z} \mathring{Z} = \Du \mathring{Z} \cdot \big((\mathring{Z}^\dagger)^{-1} \cdot \hat{Z}\big) = \Du \mathring{Z} \cdot \Phi = \Du \Phi \cdot \mathring{Z} = \Du \Phi \Du^{-1} \cdot \mathring{Z}$  since $\mathring{Z}$ only depends on $\bm{t}_A$. Using again this property and \Cref{l:daggerduality}, we find $Z_{|\bm{t}_S=0} = (\Du \Phi\Du^{-1})_{|\bm{t}_S=0} \cdot \mathring{Z} =  \Phi^{\dagger} \cdot \mathring{Z}$.
\end{proof}

This tells us how to construct the full partition function $Z$ starting from the active partition function $\mathring{Z}$: we need to act on $\mathring{Z}$ with the dressing operator $\Phi^\dagger$ obtained from the initial condition $\hat{Z}$.

\subsubsection{The abelian group structure on the set of partition functions}

\Cref{c:solution} gives an explicit solution operator to go from the initial condition $\hat{Z}$ to the partition function $Z$. Remarkably, this solution operator is acting linearly on the initial condition $\hat{Z}$ --- this was not true in general with \Cref{prop:gencolu}. This also means that we have a direct access to the abelian group structure on the set $\mathfrak{Z}$ of partition functions associated to $\mathcal{I}_{\mathcal{S}}$: the unit is $\mathring{Z}$ and
\[
\forall Z_1,Z_2 \in \mathfrak{Z} \qquad Z_1 * Z_2 = \Sol \cdot (\Sol^{-1}Z_1.\Sol^{-1}Z_2),
\]
where $.$ is the usual product of functions. 

\subsubsection{The active partition functions and choices of polarization}

By \Cref{c:dressed}, we see that $Z$ can be fully reconstructed from its active partition function $\mathring{Z}$. In other words, the partial Airy ideals of \Cref{t:PAISC} and their partition functions can be fully reconstructed in terms of the Airy ideals of \cite{BBCCN24} and their partition functions. 

More precisely, the restriction of $Z$ to actor variables, i.e. $Z_{|\bm{t}_S = 0}$, is up to a constant factor the partition function of a conjugated stable Airy ideal generated by
\begin{equation}
\label{conjguih}
\Phi^{\dagger}\mathring{\mathbf{H}}(\Phi^{\dagger})^{-1}.
\end{equation}
The initial condition $\hat{Z}$ corresponding to $Z$ specifies the relevant conjugation via a dressing transformation. 

This may sound a little disappointing: why should we care about partial Airy ideals coming from spectral curves if we could simply work with Airy ideals instead? We see  a few reasons to care. First, the construction here in terms of partial Airy ideals seems a lot cleaner than the action of the solution and dressing operators, which are quite tricky. Furthermore, as we will see in the next section, the partial Airy ideals are directly equivalent to the loop equations for correlators, without any assumption on the solution such as the projection property, while the old Airy ideals only reproduced the particular solution of the loop equations that satisfies the projection property. Second, a new (and non-trivial) aspect is the way the information about $\hat{Z}$ is encoded in the partition function $Z$. We will explain in \Cref{Sec5} the relevance of this information in the usual formulation of topological recursion on spectral curves. In particular, we will discover in \Cref{t:BS} that \Cref{ProP:Hi} allows incorporating the blobbed topological recursion of \cite{BS16} in the formalism of partial Airy ideals --- something that could not be achieved with Airy ideals --- and in fact extend it to all admissible spectral curves. Third, we will construct in \Cref{Sec6} partial Airy ideals associated to more general spectral curves equipped with additional data, whose understanding will not be reduced to Airy ideals technology for lack of duality between spectator and actor sets.

The astute reader may also wonder what happened to the choice of bidifferential. In \cite[Proposition 4.14 and Theorem 5.30]{BBCCN24} the data of a spectral curve contains a choice of bidifferential $\omega_{0,2}$, while our construction does not. The coefficients of the expansion of this bidifferential near ramification points were used to produce a conjugated Airy ideal like \eqref{conjguih} with
\[
\Phi \coloneqq \exp\left(\frac{1}{2} \sum_{\substack{\alpha_1,\alpha_2 \in \Ram \\ m_1,m_2 \geq 1}} \phi_{0,2}\begin{bsmallmatrix} \alpha_1 & \alpha_2 \\ -m_1 & -m_2 \end{bsmallmatrix} t_{\alpha_1,-m_1} t_{\alpha_2,-m_2} \right).
\]
Such conjugations correspond to a change of polarization in the description of the Weyl algebra as a quantization of the algebra of functions. For us, this choice appears as a special case corresponding to $\Phi$ of $\hbar$-degree $0$, and therefore are matched by the specific choice of initial condition $\hat{Z} = \mathring{Z}^{\dagger} \cdot \Phi$.

The main message of the construction presented here is that the choice of bidifferential finds a natural place in the choice of initial condition $\hat{Z}$, so that there is no need to include it as part of the data of the spectral curve.

\subsubsection{More general cases}

We observe that the proof of \Cref{ProP:Hi} would hold for arbitrary partial Airy ideals with generator sets $\mathbf{H}$ and active generator set $\mathring{\mathbf{H}}$ when the following conditions are met:
\begin{itemize}
\item $\mathring{\mathbf{H}}$ does not involve spectator variables or derivatives, i.e can be seen as an Airy ideal in the actor variables;
\item up to discarding actor variables and derivatives that decouple in $\mathring{\mathbf{H}}$ (like $t_{\alpha,0}$ in our example), there is a pairing between actors and spectators, giving rise to a duality operator as in \Cref{d:duality};
\item $\mathring{\mathbf{H}}$ and $\mathbf{H}$ are related by conjugation with the duality operator as in \Cref{l:HHring}.
\end{itemize}

\vspace{0.5cm}

\section{Equivalence between partial Airy ideals attached to spectral curves and loop equations}

\label{Sec5}

In the previous section we constructed partial Airy ideals attached to spectral curves (\Cref{t:PAISC}). Those take the form of $\mathcal{W}$-constraints. We can reformulate these differential constraints in a different way. Instead of encoding the free energies $F_{g,n}$ in a partition function $Z$, we can encode them in a system of symmetric meromorphic multi-differentials $\omega_{g,n}$ (called correlators) on (powers of) the spectral curve. The differential constraints satisfied by $Z$ then become a system of equations satisfied by the correlators $\omega_{g,n}$, known as \emph{loop equations}. More precisely, loop equations control the order of the poles of certain combinations of correlators at the ramification points.

The goal of this section is to establish this correspondence and deduce a bijection between the set of solutions of loop equations and the set of partition functions of the partial Airy ideals of \Cref{t:PAISC}. This improves on the results of \cite{BBCCN24}, which only considered the special solution of the loop equations satisfying a particular property known as the projection property. 

The translation between the partition function $Z$ and the correlators $\omega_{g,n}$ is very explicit. To go from $Z$ to $\omega_{g,n}$, we take the $(g,n)$ term in the free energy
\[
\sum_{\substack{\alpha_1,\ldots,\alpha_n \in \Ram \\ k_1,\ldots,k_n \in \mathbb{Z}}} F_{g,n}[k_1,\ldots,k_n] \prod_{i=1}^n t_{\alpha_i, k_i}
\]
and we do the replacement $t_{\alpha_i,k_i} \longrightarrow z_i^{-k_i-1} \dd z_i$, where the $z_i$ are copies of the coordinates on the components $C_{\alpha_i}$ of the spectral curve from \Cref{d:sc}. The result is the Laurent series expansion of a symmetric multi-differential on $C^n$ with only poles (in each coordinate) at the origin of the components $C_\alpha$ of the spectral curve. We note that meromorphicity --- the fact there are finitely many negative powers of $z$, but possibly infinitely many positive powers --- matches our convention of completing the Rees Weyl algebra only in negative variables. Through this translation, we see that the partition function is encoded into a system of correlators, and that the reduced partition function (which only involves negative variables) specifies the holomorphic part of these correlators, which are called \emph{blobs}.

Of all the solutions of the loop equations (correspondingly, of all partitions functions associated to the partial Airy ideal), there is a particular one that satisfies the projection property. In this case, one can write an explicit residue formula that recursively reconstructs its correlators. This is the well-known topological recursion from \cite{EO07}, extended to higher ramification in \cite{BHLMR14,BE13}. For a general solution of the loop equations, one can also write a residue formula, which is known as blobbed topological recursion \cite{BS16}. We comment on this in \Cref{Sec:55} and extend the results of \cite{BS16} to the more general case considered here.

\subsection{Blobs and correlators}

We start by defining an abstract notion of blobs and correlators on a spectral curve. By $n$-differential on $C^n$ for $n = 0$, we just mean a scalar.
 
\begin{definition}[Holomorphic blobs] \label{d:blobs}
Let $\mathcal{S}= ( C, x, \omega_{0,1})$ be a spectral curve. A \emph{system of holomorphic blobs} on $\mathcal{S}$  is a collection
\[
\bm{\phi} = \Set{\phi_{g,n}\,\,|\,\,g  \in  \tfrac{1}{2} \N \,\,\textnormal{and}\,\,n \in \N\,\,\textnormal{such that}\,\,2g - 2 + n \geq 0},
\]
where each $\phi_{g,n}$ is a holomorphic symmetric $n$-differential on $C^n$.
\end{definition}
We stress holomorphicity here because it will be absent in the more general situation of \Cref{Sec71}. Other authors \cite{ABDKSblob} have included holomorphic blobs in the data of the spectral curve. Yet, from the point of view of partial Airy ideals, it is meaningful to distinguish between the ingredients of the spectral curve and the blobs. The spectral curve is needed to define the differential operators in the partial Airy ideal (see \Cref{Sec4}), while the blobs rather appear as initial conditions to solve the differential equations. Note that we also include differentials indexed by half-integer $g$ to be as general as possible. 

Given a system of holomorphic blobs, their power series expansions have the form
\begin{equation}\label{eq:expblobs}
\phi_{g,n}(z_1,\ldots,z_n) =\sum_{ m_1, \ldots, m_n \geq 1}\phi_{g,n} \begin{bsmallmatrix} \alpha_1 & \cdots & \alpha_n \\ -m_1 & \cdots & -m_n \end{bsmallmatrix} \prod_{i=1}^n z_i^{m_i-1} \dd  z_i.
\end{equation}
when $z_i$ approaches the ramification point in $C_{\alpha_i}$ for each $i \in [n]$.

\begin{definition}[Correlators]\label{d:sofc}
Let $\mathcal{S} = ( C, x, \omega_{0,1})$ be a spectral curve. A \emph{system of correlators} on $\mathcal{S}$ is a collection $\bm{\omega} = \set{\omega_{g,n}\,\,|\,\,g \in \frac{1}{2} \N\,\,\textnormal{and}\,\,n \in \N}$, where
\begin{itemize} 
	\item $\omega_{0,0} = \omega_{\frac{1}{2},0} = 0$.
	\item $\omega_{0,1}$ is specified in the data of the spectral curve;
	\item $\omega_{0,2}$ is a \emph{fundamental bidifferential of the second kind},  i.e. a meromorphic and symmetric bidifferential on $C^2$ with a double pole on the diagonal $\Delta \subset C \times C$ with biresidue $1$, and which can have poles at $(\Ram \times C) \sqcup (C \times \Ram) \subset C \times C$.
	\item For $(g,n) = (\frac{1}{2},1)$ or $2g-2+n > 0$, $\omega_{g,n}(\bm{z}_{[n]})$ is a meromorphic symmetric $n$-differential on $C^n$ which can have poles only at $z_i \in \Ram$ for each $i \in [n]$.
\end{itemize}
We call the correlators $\omega_{g,n}$ with $2g-2+n \leq 0$ \emph{unstable}, and those with $2g-2+n > 0$ \emph{stable}.
\end{definition}

Given a system of correlators, their Laurent series expansions have the form
\begin{equation}\label{eq:expcorr}
\omega_{g,n}(z_1,\ldots,z_n) = \sum_{m_1, \ldots, m_n \in \mathbb{Z}} F_{g,n} \begin{bsmallmatrix} \alpha_1 & \cdots & \alpha_n \\ -m_1 & \cdots & -m_n \end{bsmallmatrix} \prod_{i=1}^n z_i^{m_i-1} \dd z_i +\delta_{g,0} \delta_{n,2}\omega_{0,2}^{\std}(z_1,z_2),
\end{equation}
where $z_i$ approaches the ramification point in $C_{\alpha_i}$ for each $i \in [n]$. In \eqref{eq:expcorr}, we have used the \emph{standard bidifferential} $\omega_{0,2}^{\std}$, which is the unique fundamental bidifferential of the second kind which takes the following form on $C_{\alpha_1} \times C_{\alpha_2}$:
  \begin{equation}\label{eq:sbd}
  \omega_{0,2}^{\std} (z_1, z_2) = \frac{\dd z_1 \dd z_2}{(z_1-z_2)^2} \delta_{\alpha_1, \alpha_2}.
  \end{equation}
By definition, we have the vanishing rule $F_{0,1}\begin{bsmallmatrix} \alpha \\ m \end{bsmallmatrix} =  0$ for $m > -s_\alpha$ and $\alpha \in \Ram$, and for each $(g,n)$, there exists $N_{g,n}$ such that $F_{g,n}\begin{bsmallmatrix} \alpha_1 & \cdots & \alpha_n \\ m_1 & \cdots & m_n \end{bsmallmatrix} = 0$ if $m_i > N_{g,n}$ for some $i \in [n]$. 

We relate the notions of correlators and blobs as follows.

\begin{definition}[Compatibility]\label{d:compatibility}
We say that a system of correlators $\bm{\omega}$ is compatible with a system of holomorphic blobs $\bm{\phi}$  if $\omega_{g,0} = \phi_{g,0}$ for all $g \geq 1$, and
\[
\omega_{\frac{1}{2},1}(z) = \phi_{\frac{1}{2},1}(z), \qquad  \omega_{0,2}(z_1,z_2) = \omega_{0,2}^{\textnormal{std}}(z_1,z_2) + \phi_{0,2}(z_1,z_2),
\]
and for all $g \in \frac{1}{2}\N$ and $n \in \N^*$ such that $2g-2+n > 0$, for all $\alpha_1,\ldots,\alpha_n \in \Ram$ and $m_1,\ldots,m_n \geq 1$, we have
\begin{equation}
F_{g,n} \begin{bsmallmatrix} \alpha_1 & \ldots & \alpha_n \\ -m_1 & \ldots & -m_n \end{bsmallmatrix} = \phi_{g,n} \begin{bsmallmatrix} \alpha_1 & \ldots & \alpha_n \\ -m_1 & \ldots & -m_n \end{bsmallmatrix}.
\end{equation}
\end{definition}

In other words, a system of correlators is compatible with a system of holomorphic blobs if the blobs give the holomorphic part of the correlators in the stable case, and the blobs prescribe the correlators in the unstable case or for $n = 0$. In particular, the bidifferential $\omega_{0,2}$ can only have a double pole on the diagonal.

\subsection{Loop equations}
\label{Sec33}

We are interested in systems of correlators that solve a particular set of equations known as loop equations. They may seem to come out of nowhere, but they appear in this form in many applications to enumerative geometry, and thus it is important to study their solution set. To formulate the loop equations, we need certain combinations of correlators, which we now define.

\begin{definition}[Local fibers]\label{d:fz}
For each $\alpha \in \Ram$,   the fiber of the map $x : C_{\alpha} \rightarrow \mathbb{C}$ above $x(z)$ is
\[
\fiber_{\alpha}(z) \coloneqq \set{ \theta_{\alpha}^k z \,\,|\,\,k\in [r_\alpha]},\qquad \textnormal{where}\qquad \theta_{\alpha} \coloneqq e^{\frac{2{\rm i}\pi}{r_\alpha}}.
\]
We also define $\fiber_{\alpha}'(z) = \fiber_{\alpha}(z) \setminus \set{z}$. 
\end{definition}

  \begin{definition}[$\mathcal{W}_{g,n}^{(i)}$ and $\mathcal{E}_{\alpha;g,n}^{(i)}$]\label{d:EW}
  Let $\bm{\omega}$ be a system of correlators on a spectral curve $\mathcal{S}$.  For any $i \in \N^*$, we define the objects:
  \begin{equation}\label{eq:dW}
     \mathcal{W}_{g,n}^{(i)} (\bm{z}_{[i]} ; \bm{w}_{[n]}) \coloneqq \sum_{\substack{\mathbf{P} \vdash [i] \\ \bigsqcup_{S \in \mathbf{P}} N_S = [n] \\ \sum_{S \in \mathbf{P}} (g_S -1) = g-i}} \prod_{S \in \mathbf{P}} \omega_{g_S, |S| + |N_S|}(\bm{z}_S,\bm{w}_{N_S}) \,, 
     \end{equation}
  where the sum is 1) over set partitions $\mathbf{P}$ of $[i]$, 2) over all possible decompositions of $[n]$ into possibly empty disjoint subsets $N_S$ where $S$ runs over all parts of $\mathbf{P}$, 3) over all families of nonnegative half-integers $(g_S)_{S \in P}$ such that $\sum_{S \in P} (g_S - 1) = g-i$. We define $\mathcal{W}_{g,n}^{(i)\,\prime}$ by the same sum but excluding the terms containing a factor of $\omega_{0,1}$.

For each $\alpha \in \Ram$ and $i \in [r_{\alpha}]$ we also define the objects
    \begin{equation}
    \label{eq:57}
    \mathcal{E}_{\alpha;g,n}^{(i)}(z; \bm{z}_{[n]}) = \sum_{\substack{\Xi \subseteq \fiber_{\alpha}(z) \\ |\Xi| = i}} \mathcal{W}_{g,n}^{(i)}(\Xi; \bm{z}_{[n]}).
  \end{equation}
  \end{definition}

\begin{definition}[Loop equations]\label{d:le}
We say that a system of correlators $\bm{\omega}$ satisfies the \emph{loop equations} if, for all $\alpha \in \Ram$, $i \in [r_\alpha]$ and $2g-2+(1+n) > 0$,
  \begin{equation}
    \mathcal{E}_{\alpha;g,n}^{(i)} (z; \bm{z}_{[n]}) = O\big( z^{r_\alpha(1 - \dk_{\alpha}^{(i)})} \big) \left(\frac{\dd z}{z}\right)^i \qquad \text{as $z \rightarrow 0$}.
  \end{equation}
  where we recall from \Cref{t:PAISC} the notation
  \[
  \dk_{\alpha}^{(i)} \coloneqq - \left\lfloor \frac{s_{\alpha}(i - 1)}{r_{\alpha}} \right\rfloor.
  \]
\end{definition}

\subsection{Correspondence with partial Airy ideals}
\label{Sec51} 

 \textit{A priori}, it is not clear whether there exist systems of correlators that satisfy the loop equations. The situation becomes more transparent after we prove the equivalence between loop equations and the partial Airy ideals constructed in \Cref{t:PAISC}.

\begin{theorem}\label{t:LEPAI}
Let $\mathcal{S}= ( C, x, \omega_{0,1})$ be an admissible spectral curve and $\bm{\omega}$ a system of correlators on $\mathcal{S}$. 
Using the expansion \eqref{eq:expcorr} for the correlators, we construct a partition function
\begin{equation}\label{eq:Zpf}
  Z \coloneqq \exp \left( \sum_{\substack{g,n \\ 2g - 2 + n \geq 0}} \frac{\hbar^{2g-2+n}}{n!} \sum_{\substack{\alpha_1, \ldots, \alpha_n \in \Ram \\ m_1, \ldots, m_n \in \mathbb{Z}}} F_{g,n} \begin{bsmallmatrix} \alpha_1 & \ldots & \alpha_n \\ m_1 & \ldots & m_n \end{bsmallmatrix}  \prod_{j = 1}^{n} t_{\alpha_j,m_j} \right).
\end{equation} 
Then $\bm{\omega}$ satisfies the loop equations from \Cref{d:le} if and only if $\mathcal{I}_\mathcal{S} \cdot Z = 0$, where $\mathcal{I}_\mathcal{S}$ is the partial Airy ideal attached to $\mathcal{S}$ via \Cref{t:PAISC}.
\end{theorem}

\begin{corollary}
\label{t:EU}
Let $\mathcal{S}$ be an admissible spectral curve, and fix a system of blobs $\bm{\phi}$ on $\mathcal{S}$. Then there exists a unique system of correlators $\bm{\omega}$ that satisfies the loop equations and is compatible with $\bm{\phi}$ in the sense of \Cref{d:compatibility}. All solutions of the loop equations are obtained in this way.
\end{corollary}

The upshot is that the loop equations for correlators are equivalent to a set of $\mathcal{W}$-constraints for a partition function $Z$ that involves both positive and negative variables. In this approach, the singular term $\omega_{0,2}^{\std}$ does not contribute to the partition function; it is in fact hidden within the Heisenberg representation $\rho$ of \Cref{l:rephei} (see \Cref{r:bergrep} and the proof of \Cref{l:simplephi}). The partial Airy ideals of \Cref{t:PAISC} are designed to bypass the singular $\omega_{0,2}^{\std}$. Alternatively, we could think of the loop equations as equivalent to a set of $\mathcal{W}$-constraints using the representation $J_k \rightarrow \hbar \partial_k$, and $J_{-k} \rightarrow \hbar t_{k}$ for $k > 0$, but then we would need to use a modified partition function including a contribution from the singular term $\omega_{0,2}^{\std}$ in a normal-ordered way (see \Cref{Sec:44}).

To prove \Cref{t:LEPAI}, we first need a few definitions and lemmata.

\begin{definition}[Fields]\label{d:fields}
For $\alpha \in \Ram$, let $\set{J_{\alpha,m}\,\,|\,\,m\in \mathbb{Z}}$ be a copy of the Heisenberg modes. We define
\begin{equation*}
\begin{split}
J^{-}_{\alpha}(z) \coloneqq  \sum_{m \geq 1} J_{\alpha,-m} z^{m-1}\dd z, & \qquad\qquad \nabla^{-}_{\alpha}(z)  \coloneqq  \sum_{m \geq 1}  \partial_{\alpha,-m} z^{m-1}\dd z, \\
J^{+}_{\alpha}(z)  \coloneqq \sum_{m  \geq 0} J_{\alpha,m} z^{-m-1}\dd z,& \qquad\qquad \nabla_{\alpha}^{+}(z) \coloneqq  \sum_{m \geq 0} \partial_{\alpha,m} z^{-m-1}\dd z, \\
J_\alpha(z) \coloneqq J^{-}_{\alpha}(z)+J^{+}_{\alpha}(z),& \qquad\qquad \nabla_{\alpha}(z)   \coloneqq  \nabla_{\alpha}^{-}(z)+\nabla_{\alpha}^{+}(z).
\end{split}
\end{equation*}
For $i \in [r_\alpha]$, we define
\begin{equation}\label{eq:fieldsW}
\overline{W}^{(i)}_{\alpha}(z) \coloneqq \sum_{k \in \mathbb{Z}} \frac{\overline{W}_{\alpha,k}^{(i)}}{(x(z) - x_{\alpha})^k} \left(\frac{\dd x(z)}{x(z)-x_\alpha}\right)^i = \sum_{k \in \mathbb{Z}} r_\alpha^i\overline{W}_{\alpha,k}^{(i)} \frac{(\dd z)^i}{z^{r_\alpha k +i}},
\end{equation}
where the operators $\overline{W}^{(i)}_{\alpha,k}$ were defined in \eqref{eq:overline}.
\end{definition}  

\begin{definition}[From $\bm{t}$ to $\bm{z}$]\label{d:projRWA}
Let $I \coloneqq \Ram \times \Z$ and $f$ be of the form
\begin{equation}
f = \sum_{g,n} \frac{\hbar^{2g+n}}{n!} f_{g,n}(\bm{t}_I),
\end{equation}
where $f_{g,n}(\bm{t}_I)$ is homogeneous of degree $n$ in the variables $\bm{t}_{I}$. We define the operation
\begin{equation}
\Ev_{g,n}[f](\bm{z}_{[n]}) \coloneqq \sum_{\alpha_1, \ldots, \alpha_n \in \Ram} \Bigg(\prod_{j=1}^n \nabla_{\alpha_j}(z_j) \Bigg) f_{g,n}(\bm{t}_I),
\end{equation}
which picks the coefficient of $\hbar^{2g + n}$ and homogeneous terms of degree $n$ in the variables $\bm{t}_{I}$, and then replaces $t_{\alpha_j,m_j}$ by $z_j^{-m_j-1}\dd z_j$ for $j \in [n]$, $\alpha_j \in \Ram$ and $m_j \in \Z$.
\end{definition}

With this rewriting we have the following straightforward lemma.
\begin{lemma}\label{l:Zres}
Let $\mathcal{S}$ be a spectral curve, $\bm{\omega}$ a system of correlators, and $Z$ the associated partition function \eqref{eq:Zpf}. Define
\begin{equation}
\label{Zbar}\overline{Z}:= Z_{(0,1)} Z = \exp \left( \sum_{\substack{g,n \\ 2g - 2 + n \geq -1}} \frac{\hbar^{2g-2+n}}{n!} \sum_{\substack{\alpha_1, \ldots, \alpha_n \in\Ram \\ m_1, \ldots, m_n \in \mathbb{Z}}} F_{g,n} \begin{bsmallmatrix} \alpha_1 & \ldots & \alpha_n \\ m_1 & \ldots & m_n \end{bsmallmatrix}  \prod_{j = 1}^{n} t_{\alpha_j,m_j} \right).
\end{equation}
Then $\mathcal{I}_\mathcal{S} \cdot Z = 0$ if and only if, for any $2g - 2 + (1+n) > 0$ and $k \geq  \dk^{(i)}_{\alpha}$ we have
\begin{equation}\label{eq:overWres}
\Res_{z=0} \frac{z^{r_\alpha k + i - 1}}{(\dd z)^{i-1}} \Ev_{g,n} \left[  \overline{Z}^{-1} \overline{W}^{(i)}_{\alpha}(z) \overline{Z} \cdot 1\right](\bm{z}_{[n]})= 0.
\end{equation}
\end{lemma}

\begin{proof}
By definition of the ideal $\mathcal{I}_{\mathcal{S}}$ in \Cref{t:PAISC}, the constraints $\mathcal{I}_\mathcal{S} \cdot Z = 0$ are equivalent to the constraints $ Z^{-1} H^{(i)}_{\alpha,k}  Z \cdot 1 = 0$ for all $\alpha \in \Ram$, $i \in [r_\alpha]$ and $k \geq \dk_{\alpha}^{(i)}$, and $H^{(i)}_{\alpha,k} = Z_{(0,1)}^{-1} \overline{W}^{(i)}_{\alpha,k}  Z_{(0,1)}$. Including the factor $Z_{(0,1)}$ in the partition function, we can rewrite the constraints as $\overline{Z}^{-1}  \overline{W}^{(i)}_{\alpha,k}  \overline{Z} \cdot 1 = 0$. From \eqref{eq:fieldsW} one can see that this is equivalent to imposing the residue constraints
\begin{equation}
\forall k \geq \kappa_{\alpha}^{(i)}\qquad \Res_{z=0} \frac{z^{r_\alpha k + i - 1}}{(\dd z)^{i-1}} (\overline{Z}^{-1}  \overline{W}^{(i)}_{\alpha}(z)  \overline{Z} \cdot 1) = 0.
\end{equation}
Finally, acting with $\Ev_{g,n}$ splits the constraints into separate constraints for different powers of $\hbar$ and homogeneous terms in the variables $\bm{t}_I$, replacing the latter with monomials in the variables $\bm{z}_{[n]}$. This is a linear operation which commutes with the residue at $z = 0$.
\end{proof}

Now we would like to rewrite \eqref{eq:overWres} in terms of the system of correlators $\bm{\omega}$.

\begin{lemma}\label{l:simplephi}
Let $\mathcal{S}$ be a spectral curve, $\bm{\omega}$ a system of correlators, $Z$ the associated partition function, and $\overline{Z} = Z_{(0,1)} Z$ like in \eqref{Zbar}. Then
\begin{equation}
\Ev_{g,n} \left[  \overline{Z}^{-1}  \overline{W}^{(i)}_{\alpha}(z)  \overline{Z} \cdot 1 \right](\bm{z}_{[n]}) = \mathcal{E}_{\alpha;g,n}^{(i)}(z;\bm{z}_{[n]}) \qquad \text{as $z \rightarrow 0$.}
\end{equation}
\end{lemma}

\begin{proof}
We proceed as in \cite[Section 4.1]{BKS24}. The starting point is the following rewriting of the fields \eqref{eq:fieldsW}
\begin{equation}\label{eq:Wcomb}
\overline{W}^{(i)}_{\alpha}(z) = \hbar^i \sum_{\substack{\Xi \subseteq \fiber_{\alpha}(z) \\ |\Xi| = i}} \sum_{\substack{ \mathbf{P} \in \textnormal{Pair}[\fiber_{\alpha}(z)] \\  (\sqcup \mathbf{P}) \sqcup \Xi^- \sqcup \Xi^+ = \Xi}} \prod_{\set{z', z''} \in \mathbf{P}} \omega_{0,2}^{\std}(z',z'') \prod_{z' \in \Xi^-} J^{-}_{\alpha} (z') \prod_{z' \in \Xi^{+}} J^+_{\alpha}(z').
\end{equation}
Here, if $K$ is a set, we have called $\textnormal{Pair}[K]$ the set of partitions of a subset of $K$ into disjoint pairs. If $\mathbf{P} \in \textnormal{Pair}[K]$ we have called $\sqcup \mathbf{P}$ the union of the pairs in $\mathbf{P}$.  The sum in \eqref{eq:Wcomb} ranges over decomposition of $\Xi$ into disjoint subsets $\Xi^-$, $\Xi^+$ and $\sqcup \mathbf{P}$ for some pair partition $\mathbf{P}$ of $\fiber_{\alpha}(z)$. We shall transform the loop equations to eventually recognize the expression \eqref{eq:Wcomb}.

Let $\mathcal{E}^{(i)\,\#}_{\alpha;g,n} (z;\bm{z}_{[n]})$ be the same combinatorial expression as $\mathcal{E}_{\alpha;g,n}^{(i)} (z;\bm{z}_{[n]})$ (see \Cref{d:EW}) but with the substitution 
\begin{equation}
\omega_{0,2}(z_1,z_2) \longrightarrow \omega_{0,2}(z_1,z_2) - \omega_{0,2}^{\std}(z_1,z_2) = \sum_{m_1,m_2 \geq 1} F_{0,2} \begin{bsmallmatrix} \alpha_1 & \alpha_2 \\ -m_1 &  -m_2 \end{bsmallmatrix}  z_1^{m_1-1} z_2^{m_2-1} \dd z_1 \dd z_2
\end{equation}
when $z_i \in C_{\alpha_i}$.  After some algebra, one can show that
\begin{equation}\label{eq:startingpoint}
\Ev_{g,n} \left[  \overline{Z}^{-1}\Bigg(\hbar^i \sum_{\substack{\Xi \subseteq \fiber_{\alpha}(z) \\ |\Xi| = i}} \prod_{z' \in \Xi} \nabla_{\alpha}(z')\Bigg)\overline{Z} \cdot 1 \right](\bm{z}_{[n]}) =\mathcal{E}^{(i)\,\#}_{\alpha;g,n}(z;\bm{z}_{[n]}),
\end{equation}
which is simply the action of the derivatives on $\overline{Z}$. What we need to do is re-insert the contributions from $\omega_{0,2}^{\std}$ in $\mathcal{E}^{\#}$ to get $\mathcal{E}$.

There are two types of contributing factors: those with $\omega_{0,2}^{\std}(z',z_j)$ where $z' \in \fiber_{\alpha}(z)$, and those with $\omega_{0,2}^{\std}(z',z'')$ where $\set{z',z''} \subseteq \fiber_{\alpha}(z)$. For the first type of terms, we notice that
\begin{equation}
\omega_{0,2}^{\std}(z,z_j) = \sum_{m \geq 1} m z^{m-1} z_j^{-m-1}\, \dd z\,\dd z_j
\end{equation}
as $|z_j| > |z|$ approach $0$.  To incorporate such terms in \eqref{eq:startingpoint}, on the left-hand side we would need to include a term of the form\footnote{Note that $Z_{\textnormal{std}}^{\dagger} = \Du$ is the duality operator of \eqref{DuDu}.}
\begin{equation}
Z_{\std} \coloneqq \exp\left(\sum_{\substack{\alpha \in \Ram \\ m  \geq 1}}  m t_{\alpha,-m} t_{\alpha,m} \right)
\end{equation}
in $\overline{Z}$, but in such a way that only derivatives with respect to the variables $t_{\alpha,-m}$ act on it. We can achieve this using normal ordering. That is, we can rewrite the left-hand side of \eqref{eq:startingpoint} as
\begin{equation}
\Ev_{g,n} \left[  \overline{Z}^{-1} \Bigg(\hbar^i \sum_{\substack{\Xi \subseteq \fiber_{\alpha}(z) \\ |\Xi| = i}}\,\sum_{ \Xi^{-} \sqcup \Xi^{+} = \Xi} \bigg( Z_{\std}^{-1} \prod_{z' \in \Xi^-} \nabla^-_{\alpha}(z') Z_\std \bigg) \prod_{z' \in \Xi^{+}} \nabla^+_{\alpha}(z') \Bigg) \overline{Z} \cdot 1 \right](\bm{z}_{[n]}). 
\end{equation}
It is easy to see that the conjugation by $Z_{\std}$ amounts to substituting
\begin{equation}
\forall m \geq 1 \qquad \partial_{\alpha,-m} \longrightarrow \partial_{\alpha,-m} + m t_{\alpha,m}.
\end{equation}
This is precisely what the representation of \Cref{l:rephei} does. Therefore, we can forget about $Z_\std$ and simply replace the fields $\nabla_{\alpha}(z)$ by $J_{\alpha}(z)$. We get:
\begin{equation}\label{eq:startingpoint2} 
\Ev_{g,n} \left[ \overline{Z}^{-1} \Bigg(\hbar^i \sum_{\substack{\Xi \subseteq \fiber_\alpha(z) \\ |\Xi| = i}}\,\sum_{ \Xi^- \sqcup \Xi^+ = \Xi} \,\prod_{z' \in \Xi^-} J^-_{\alpha}(z') \prod_{z' \in \Xi^+} J^{+}_{\alpha}(z') \Bigg) \overline{Z} \cdot 1 \right](\bm{z}_{[n]}) = \widetilde{\mathcal{E}}_{\alpha;g,n}^{(i)}(z;\bm{z}_{[n]}),  
\end{equation}
where the right-hand side is again the same combination of correlators but now with contributions from terms of the form $\omega_{0,2}^{\std}(z',z_j)$ with $z' \in \fiber_\alpha(z)$ included.

To incorporate the second type of terms, that is $\omega_{0,2}^{\std}(z',z'')$ with $\set{z',z''} \subseteq \fiber_\alpha(z)$, all we need to do is extract pairs of elements in $\Xi \subseteq \fiber_{\alpha}(z)$, and then add the contributions from $\omega_{0,2}^{\std}(z',z'')$ explicitly to the operators. The result is precisely \eqref{eq:Wcomb}.
\end{proof}

We are now ready to prove \Cref{t:LEPAI} and \Cref{t:EU}.

\begin{proof}[Proof of \Cref{t:LEPAI}]
First, the loop equations from \Cref{d:le} are satisfied if and only if the following residue constraints hold, that is for all $\alpha \in \Ram$,  $i \in [r_\alpha]$, $k \geq \dk_\alpha^{(i)}$ and $2g-2+(1+n) > 0$ we have
\begin{equation}
 \Res_{z=0} \frac{z^{r_\alpha k + i - 1}}{(\dd z)^{i-1}} \mathcal{E}_{\alpha;g,n}^{(i)}(z; \bm{z}_{[n]}) = 0.
\end{equation}
Here we used the fact that $\mathcal{E}_{\alpha;g,n}^{(i)}(z; \bm{z}_{[n]})$ is a $i$-differential in the variable $z$. Using \Cref{l:simplephi}, we can rewrite these residue constraints as
\begin{equation}
 \Res_{z=0} \frac{z^{r_\alpha k + i - 1}}{(\dd z)^{i-1}}\Ev_{g,n} \left[  \overline{Z}^{-1}  \overline{W}^{(i)}_{\alpha}(z)  \overline{Z} \cdot 1 \right](\bm{z}_{[n]}) = 0.
\end{equation}
We conclude the proof by resorting to \Cref{l:Zres}.
\end{proof}

\begin{proof}[Proof of \Cref{t:EU}]
Let $\bm{\omega}$ be a system of correlators that is compatible with $\bm{\phi}$. Expand the $\omega_{g,n}$ and $\phi_{g,n}$ as in \eqref{eq:expcorr} and \eqref{eq:expblobs}.  As stated in \Cref{d:compatibility}, the compatibility of the correlators with the blobs is equivalent to requiring that 
\begin{equation}\label{eq:compp}
\forall \alpha_1,\ldots,\alpha_n \in \Ram \quad \forall m_1,\ldots,m_n \geq 1\qquad F_{g,n} \begin{bsmallmatrix} \alpha_1 & \ldots & \alpha_n \\ -m_1 & \ldots & -m_n \end{bsmallmatrix} = \phi_{g,n} \begin{bsmallmatrix} \alpha_1 & \ldots & \alpha_n \\ -m_1 & \ldots & -m_n \end{bsmallmatrix}.
\end{equation}
We want to show that there is a unique such system of correlators satisfying \eqref{eq:compp} and the loop equations. By \Cref{t:LEPAI}, this is equivalent to showing that there is a unique partition function $Z$ of the form \eqref{eq:Zpf} with the identification \eqref{eq:compp} such that $\mathcal{I}_{\mathcal{S}} \cdot Z = 0$.

We can reformulate this statement as follows. The total index set of variables is $I = \Ram \times \mathbb{Z}$, and we split it in spectator set $S = \Ram \times \mathbb{Z}_{<0}$ and actor set $A = \Ram \times \N$. Construct a reduced partition function
\begin{equation}\label{eq:rp}
  \hat{Z} \coloneqq \exp \left( \sum_{g,n} \frac{\hbar^{2g-2+n}}{n!} \sum_{\substack{\alpha_1, \ldots, \alpha_n \in \Ram \\ m_1, \ldots, m_n  \geq 1} } \phi_{g,n} \begin{bsmallmatrix} \alpha_1 & \ldots & \alpha_n \\ -m_1 & \ldots & -m_n \end{bsmallmatrix} \prod_{j = 1}^{n} t_{\alpha_j,-m_j} \right),
 \end{equation} 
 which precisely specifies the blobs.
It only depends on the spectator variables in $\bm{t}_{S}$. What we want to show is that there is a unique partition function $Z$ of the form \eqref{eq:Zpf} such that $\mathcal{I}_{\mathcal{S}} \cdot Z = 0$ and $Z_{|\bm{t}_A = 0} = \hat{Z}$. But we showed in \Cref{t:PAISC} that $\mathcal{I}_\mathcal{S}$ is a partial Airy ideal: the claim is a consequence of the general \Cref{t:PAI_KS}.
\end{proof}

\subsection{Blobbed topological recursion}

\label{Sec:55}

There is one particular solution of the loop equations that has been studied extensively in the past: this is the solution that satisfies the so-called "projection property".

 \begin{definition}[Projection property]\label{d:projprop}
A system of correlators $\bm{\omega}$ has the \emph{projection property} if $\omega_{1,\frac{1}{2}} = 0$, the bidifferential $\omega_{0,2}$ has no pole outside the diagonal, and for all $2g-2+(1+n) > 0$ we have
\begin{equation}\label{eq:pp}
\omega_{g,1+n}(z_0,\bm{z}_{[n]}) = \sum_{\alpha \in \Ram} \Res_{z=\alpha} \left( \int^z_{\alpha} \omega_{0,2}(\cdot, z_0)\right)\omega_{g,1+n}(z, \bm{z}_{[n]}).
\end{equation}
The result does not depend on the choice of path of integration between $\alpha$ and $z$ since $C_{\alpha}$ is a disk and $\omega_{0,2}$ has no residue.
 \end{definition}
  The assumption $\omega_{\frac{1}{2},1} = 0$ is equivalent to requiring that $\omega_{\frac{1}{2},1}$ is holomorphic and satisfies the projection property \eqref{eq:pp}.

 \begin{theorem}\cite{BBCCN24,BBCKS}
 \label{t:BBCCN}
Let $\mathcal{S}$ be an admissible spectral curve. There ex{}ists a unique system of correlators $\bm{\omega}$ satisfying the loop equations and the projection property. Furthermore, these correlators can be reconstructed via the topological recursion of \cite{BE13}. This means that we have the recursion on $2g - 2 + (1+n) > 0$
\begin{equation}
\label{usualTR} 
\omega_{g,1+n}(z_0,\bm{z}_{[n]}) = \sum_{\alpha \in \Ram} \Res_{z = \alpha}  \sum_{i = 2}^{r_{\alpha}}\, \sum_{\substack{\Xi \subseteq \fiber_{\alpha}'(z) \\ |\Xi| = i}}  K_{\alpha}^{(i)}(z_0,z,\Xi) \mathcal{W}^{(i)\,\prime}_{\alpha;g,n}(z,\Xi;\bm{z}_{[n]}),
\end{equation} 
with the recursion kernels given for $(z_0,z) \in C \times C_{\alpha}$ and  $\Xi \in \textnormal{Sym}^{i - 1} C_{\alpha}$ by
\[ 
K_{\alpha}^{(i)}(z_0,z,\Xi) \coloneqq \frac{-\int_{\alpha}^{z} \omega_{0,2}(\cdot,z_0)}{\prod_{z' \in \Xi} (\omega_{0,1}(z') - \omega_{0,1}(z))}.  
\]
 \end{theorem}
Note that the admissibility assumption on the spectral curve is essential for the existence and uniqueness of solutions, as explained in \cite{BBCCN24}.

It is natural to ask which system of holomorphic blobs $\bm{\phi}$ is compatible with the system of correlators $\bm{\omega}$ given by \Cref{t:BBCCN}, in other words, what is the purely holomorphic part of the correlators constructed by \Cref{t:BBCCN}  when expanded in the $z$-coordinates. If $\omega_{0,2}=\omega_{0,2}^{\std}$, then by the projection property \eqref{eq:pp} we see that the purely holomorphic part of the correlators vanish, and thus all blobs are zero, $\phi_{g,n} = 0$. If $\omega_{0,2} \neq \omega_{0,2}^{\std}$, that is $\phi_{0,2} \neq 0$, then the blobs will also be non-zero.

The general solution of the loop equations --- not necessarily satisfying the projection property --- has been studied in \cite{BS16} when the genus is integer-valued and for \emph{regular} spectral curves, that is when $(r_{\alpha},s_{\alpha}) = (2,3)$ for all $\alpha \in \Ram$. A residue formula that recursively reconstructs these solutions was obtained and called \emph{blobbed topological recursion}. Our definition of blobs correspond to the "KdV blobs" of \cite[Section 3]{BS16} and the formula for $Z_{|\bm{t}_S = 0}$ that we found in \Cref{ProP:Hi} retrieves the structure of \cite[Equation 3.6]{BS16}.  We can extend these results to all admissible spectral curves and reformulate \Cref{t:EU} in terms of the blobbed topological recursion.

\begin{theorem}[Blobbed topological recursion]\label{t:BS} 
Let $\mathcal{S}$ be an admissible spectral curve, and fix a system of holomorphic blobs $\bm{\phi}$ on $\mathcal{S}$. The unique system of correlators $\bm{\omega}$ satisfying loop equations and compatible with $\bm{\phi}$ can be reconstructed via the blobbed topological recursion. This means that, in terms of the unique system of correlators $\mathring{\bm{\omega}}$ satisfying the loop equations and compatible with zero blobs, we have the formula for $2g - 2 + n > 0$ 
\begin{equation}
\label{bipsum}
\omega_{g,n}(\bm{z}_{[n]}) = \sum_{\Gamma \in \textnormal{Bip}_{g,n}} \frac{1}{|\textnormal{Aut}(\Gamma)|}  \bigg\langle \prod_{e \in V_{\cir{}{}}(\Gamma)} \mathring{\omega}_{g(v),|E(v)|}(\bm{w}_{E(v)}) , \prod_{e \in V_{\cir{fill=gray}{}}(\Gamma)} \phi_{g(v),|E(v)|}(\bm{w}_{E(v)})\bigg\rangle.
\end{equation}
The sum ranges over the set $\textnormal{Bip}_{g,n}$ of bipartite graphs $\Gamma$ with $n$ labeled leaves such that 
\begin{itemize}
\item each vertex $v$ carries a genus $g(v) \in \frac{1}{2}\N$;
\item white vertices have at least a leaf and satisfy the stability condition $2g(v) - 2 + |E(v)| > 0$;
\item grey vertices satisfy the semi-stability condition $2g(v) - 2 + |E(v)| \geq 0$;
\item $g = b_1(\Gamma) + \sum_{v} g(v)$.
\end{itemize}
We called $V_{\cir{}{}}(\Gamma)$ (resp. $V_{\cir{fill=gray}{}}(\Gamma)$) the set of white (resp. grey) vertices of $\Gamma$, and $E(v)$ the set of edges and leaves incident to the vertex $v$. Each edge or leaf $e$ carries a variable $w_e \in C$. If $e$ is the $i$-th leaf this variable is $w_e \coloneqq z_i$, otherwise it is integrated out via the pairing  of residueless meromorphic $1$-forms in $C$
\begin{equation}
\label{pairing}\big\langle \omega,\varphi \big\rangle = \sum_{\alpha \in \Ram} \Res_{w_e = \alpha} \bigg(\omega(w_e) \int^{w_e}_{\alpha}\varphi\bigg).
\end{equation}
Since the white vertices are assumed stable, $\mathring{\omega}_{g(v),|E(v)|}$ has no poles outside the ramification divisor, so the order in which we integrate out the variables carried by internal edges is irrelevant.
\end{theorem}
 
\begin{corollary}[Blobbed topological recursion, II]
\label{c:BS}
In the situation of \Cref{t:BS}, we have
\begin{equation}
\label{bipsum2}
\omega_{g,n}(\bm{z}_{[n]}) = \sum_{\Gamma \in \textnormal{Bip}'_{g,n}} \frac{1}{|\textnormal{Aut}(\Gamma)|}  \bigg\langle \prod_{e \in V_{\cir{}{}}(\Gamma)} \omega^0_{g(v),|E(v)|}(\bm{w}_{E(v)}) , \prod_{e \in V_{\cir{fill=gray}{}}(\Gamma)} \phi_{g(v),|E(v)|}(\bm{w}_{E(v)})\bigg\rangle.
\end{equation}
where $\textnormal{Bip}'_{g,n} \subset \textnormal{Bip}_{g,n}$ are those bipartite graphs without $(0,2)$-grey vertices, and $\boldsymbol{\omega}^0$ is the unique system of correlators satisfying the loop equations and the projection property (cf. \Cref{t:BBCCN}). 
\end{corollary}

\begin{figure}[h!]
\begin{center}
\includegraphics[width=0.5\textwidth]{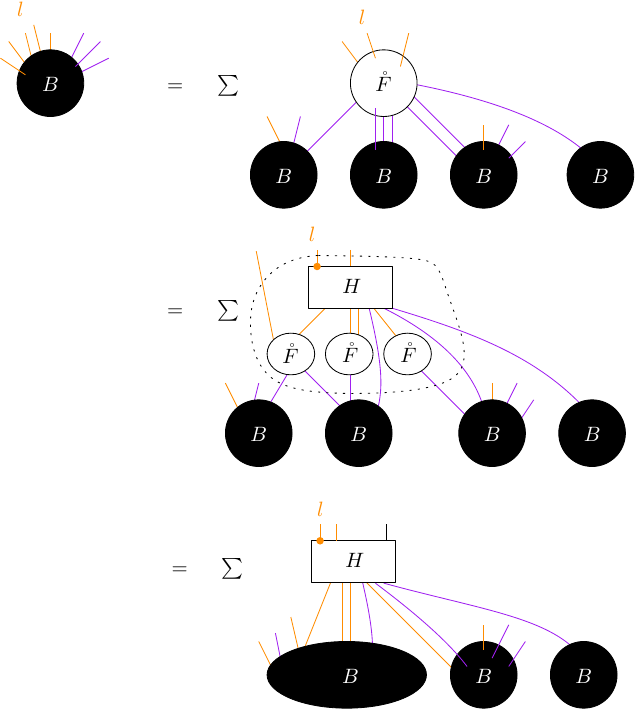}
\caption{\label{fig:recgraph3} Re-organizing the sum over bipartite graphs to get a recursion (we only draw representative examples of graphs).}
\end{center}
\end{figure}

\begin{proof}[Proof of \Cref{t:BS}]
We work with the graphical recursion of \Cref{Sec31}, recalling that orange edges correspond to actor indices and purple edges to spectator indices. The partial Airy ideals constructed in \Cref{Sec4} have several remarkable features: 
\begin{enumerate}
\item Actor variables are positive (we can ignore indices $(\alpha,0)$ since the partition function cannot depend on $t_{\alpha,0}$), spectator variables are negative.
\item The operators do not depend on negative variables, thus $H$-vertices do not have incoming purple half-edges.
\item The negative derivatives appear in the operators only through the combination $\partial_{\alpha,-m} + mt_{\alpha,m}$ (here $m > 0$). Thus, the weight of a $H$-vertex with an outgoing purple half-edge $h$ carrying the index $(\alpha,-m)$ for some $(\alpha,m) \in \Ram \times \mathbb{N}^*$ is the same as $\frac{1}{m}$ times the weight of the $H$-vertex where $h$ is replaced by an incoming orange half-edge carrying the index $(\alpha,m)$.
\item For $2g - 2 + n = 0$, a vertex $F_{g,n}$ has zero weight unless its incoming half-edges are all purple (in that case its weight is $\hat{F}_{g,n}$).
\end{enumerate}
Properties (2)-(3) mean that we can make all half-edges of $H$-vertices orange. And, when computing weights expressed in terms of correlators (see \Cref{d:projRWA}), Property (3) means that an edge obtained by connecting one purple half-edge turned orange to a purple half-edge is realized by the pairing \eqref{pairing}. Property (4) is incorporated in the compatibility of systems of correlators with holomorphic blobs in \Cref{d:compatibility}.

Even taking these properties into account, the graphical recursion of \Cref{Sec31} does not quite give the sum of bipartite graphs we are looking for. It does give a sum over graphs made of $H$-vertices and $\hat{F}$-vertices. We can cluster $H$-vertices having only orange (half-)edges in connected subgraphs equipped with a rooted spanning tree. But the $\hat{F}$-vertices (grey vertices) must then be connected to apparented vertices of a given cluster. Compare with $\textnormal{Bip}_{g,n}$: substituting the white vertices with the sum over graphs on $H$-vertices resulting from \Cref{Sec31} does give similar clusters having only orange edges, but a given grey vertex can connect to several clusters and even within one cluster it is not forbidden to connect to vertices that are not apparented.

From this perspective it is rather surprising that a formula like \eqref{bipsum} can be true. There is no a priori contradiction though. First, this naive argumentation ignored symmetry factors. Second, $\mathring{\omega}$ is symmetric, so the non-symmetric recursion of \Cref{Sec31} can be written starting from different incoming leaves, leading to non-trivial rewriting rules. Instead of trying to prove \eqref{bipsum} by a direct resummation of graphs from \Cref{Sec3} and rewriting rules, we do it by checking that both sides satisfy the same recursion on $2g - 2 + n > 0$ with the same initial condition.

Instead of working with $z_i$-variables and $\omega_{g,n}$, we work with $\bm{t}$-variables and $F_{g,n}$ (\Cref{d:projRWA}), so that \Cref{Sec31} is directly applicable. We call bipartite graph any graph belonging to $\textnormal{Bip}_{g,n}$ for some $g,n$.  Call  $B_{g,n}$ the expression in $\bm{t}$-variables corresponding to the right-hand side of \eqref{bipsum}, and $\mathring{F}_{g,n}$ the free energies for trivial initial condition. We draw $B_{g,n}$ as a black oval vertex, and $\mathring{F}_{g,n}$ as white oval vertex; in the present context it is convenient to decide that the orange leaves are incoming and the purple one are outgoing; all those leaves are labeled.

\Cref{fig:Euler1} shows that \eqref{bipsum} is true for $2g - 2 + n = 1$.

\begin{figure}[h!]
\begin{center}
\includegraphics[width=0.8\textwidth]{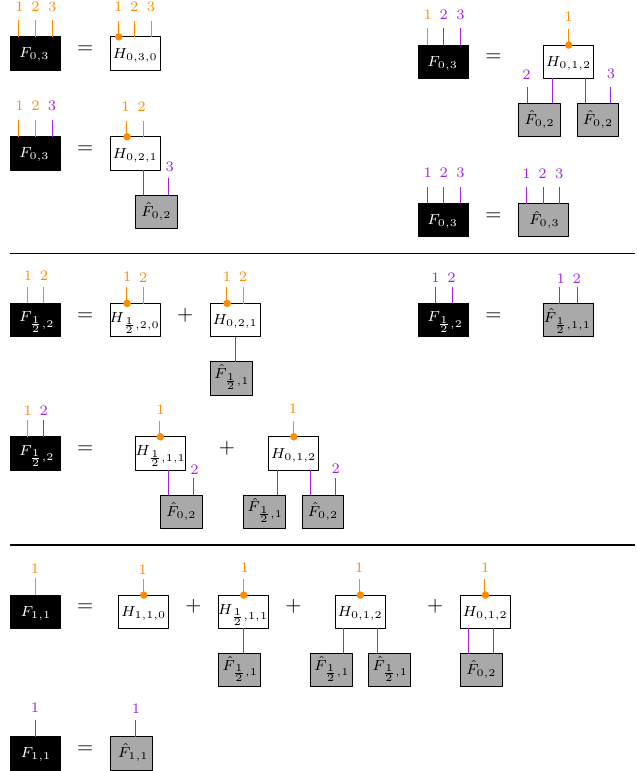}
\caption{\label{fig:Euler1} Graphical representation of $F_{g,n}$ for $2g - 2 + n = 1$ for all possible leaf colors (up to leaf relabeling), taking into account the properties (1)--(4). If we set $\hat{F} = 0$ we get  $\mathring{F}_{g,n} = H_{g,n,0}$ for $2g - 2 + n = 1$. In general, the vertices $H_{g',n',m'}$ appearing here all have $2g' - 2 + n' + m' = 1$ and the $m$ outgoing edges are purple and connected to $\hat{F}$-vertices. Using (3), we can replace them by vertices of weight $H_{g,n+m,0} = \hat{F}_{g,n+m}$ (i.e. white vertices) that are connected to vertices of weight $\hat{F}$ (i.e. grey vertices) using the pairing \eqref{pairing} (when  all weights are expressed in terms of correlators). This constructs bipartite graphs and yields $F_{g,n} = B_{g,n}$ for $2g - 2 + n = 1$.}
\end{center}
\end{figure}

The contribution to $B_{g,n}$ of the bipartite graph $*$ consisting of a single grey vertex is $\hat{F}_{g,n}$. In any other $\Gamma \in \textnormal{Bip}_{g,n}$, we look at the white vertex $v_0$ carrying the orange leaf with minimal label $l$ (\Cref{fig:recgraph3}). It sits in $\Gamma$ by being connected to a collection of grey vertices.

\medskip

\noindent \textsc{Step 1.} If we remove $v_0$, we get connected components $\Gamma_1,\ldots,\Gamma_a$ which are bipartite graphs, whose topologies assemble (together with the topology of $v_0$) to $(g,n)$. All tuples of bipartite graphs satisfying this topological constraint can occur when we vary $\Gamma$. The automorphisms that were lost when removing $v_0$ are those which permute the edges that we cut in doing so. Note that the components containing at least one leaf can be distinguished by the leaf labels.

\medskip

\noindent \textsc{Step 2.} The vertex $v_0$ has weight $\mathring{F}$. We decompose it --- starting from $l$ --- using the recursion of \Cref{fig:recgraph} specialized to the case where there are no $\hat{F}$-vertices. It replaces $v_0$ by a sum over connected two-level graphs. On the top level there is a $H$-vertex of genus $g_0$ with special incoming leaf $l$. On the second (possibly empty) level there are $\mathring{F}$-vertices $v_1',\ldots,v'_b$ connected to the $H$-vertex above by respectively $n_1,\ldots,n_b$ orange edges. Such two-level graphs come with a symmetry factor accounting for the fact that the edges out of $v_0$ are unlabeled. The purple edges which connected grey vertices to $v_0$ can land either on the $H$-vertex or on the vertices $v_1',\ldots,v_b'$. Likewise, the leaves of $v_0$ besides $l$ are distributed in some way as leaves on the $H$-vertex or $v_1',\ldots,v_b'$. 

\medskip

\noindent \textsc{Step 3.} If we remove the $H$-vertex, we get connected components $\Gamma''_1,\ldots,\Gamma''_c$ which are bipartite graphs, that can be connected to the $H$-vertex by orange or purple edges (and not only by purple edges as it was the case in Step 1).  In this picture the natural symmetry factor can be larger than in Step 2, but it is compensated with more possibilities for labeling of edges within each component $\Gamma''_i$, thus effectively retrieving the symmetry factor of Step 2. Arguing like in Step 1, this decomposition is compatible with symmetry factors. The topologies of the components assemble (together with the topology of $v_0'$) to $(g,n)$. All tuples of bipartite graphs satisfying this topological constraint can occur when we vary $\Gamma$.

\medskip

Since the weight is fully factorized, we can reorganize the sum over $\Gamma \in \textnormal{Bip}_{g,n} \setminus \set{*}$ by first computing the sum over individual components $\Gamma''_i$ (this retrieves a weight $B$), and then sum over all topological ways they can be attached to a $H$-vertex with special leaf $l$. The result is a recursion for $B_{g,n}$ which has exactly the same form as the recursion of \Cref{fig:recgraph} for $F_{g,n}$. As the base cases $2g - 2 + n = 1$ agree this shows that $F_{g,n} = B_{g,n}$. 
\end{proof}

\begin{proof}[Proof of \Cref{c:BS}]
Given a graph in $\textnormal{Bip}_{g,n}$, we can isolate the clusters made of white vertices and genus-zero bivalent grey vertices (having weight $\phi_{0,2}$). The latter are of two kinds: either they have one leaf and one half-edge connected to a white vertex, or two half-edges are connected to a white vertex. If we replace them respectively by a purple leaf and by an "edge", the clusters are stable graphs with vertices of weight $\bm{\omega}^0$ and "edges" of weight $\phi_{0,2}$, having either orange leaves or purple leaves (to be interpreted as an orange half-edge connected to a $\phi_{0,2}$-vertex). Sums over stable graphs correspond, at the level of partition functions, to action of the exponential of a quadratic expression in derivatives:  this corresponds to a change of polarization and  retrieves the solution of loop equations satisfying the projection property \cite{BBCCN24}. The special treatment of the leaves corresponds to decomposition of $\omega_{g,n}$ on the basis of 1-forms in the image of the projection coming from $\omega_{0,2} = \omega_{0,2}^{\textnormal{std}} + \phi_{0,2}$.

Note that the weight of $\Gamma$ is a product of the weight of the clusters and the weight of grey vertices, that the automorphisms of $\Gamma$ are products of automorphisms internal to each cluster, and automorphisms coming from permutations of edges between clusters and grey vertices. In this way, when we sum over all possible stable graphs realizing the clusters, we realize that the contribution of $\Gamma$ (including symmetry factors) is completely factorized over clusters and grey vertices.

Given a graph $\Gamma \in \textnormal{Bip}_{g,n}$, contracting all bivalent grey vertices of genus $0$ and their incident edges (taking into account the additional genera creating at vertices) gives a graph $\Gamma' \in \textnormal{Bip}'_{g,n}$. The fiber of this forgetful map $\textnormal{Bip}_{g,n} \rightarrow \textnormal{Bip}'_{g,n}$ is obtained by replacing each white vertex with an arbitrary stable graph as we described above. Since the weight of $\Gamma$ factorizes, we get a sum over bipartite graphs $\Gamma'$ with weight $\bm{\omega}^0$ for white vertices and unchanged weight $\bm{\phi}$ for grey vertices: this is \eqref{bipsum2}.
\end{proof}

\vspace{0.5cm}

\section{Partial Airy ideals for coloured spectral curves}

\label{Sec6}
In the previous sections, we constructed partial Airy ideals attached to admissible spectral curves. We showed that the resulting $\mathcal{W}$-constraints are precisely equivalent to the loop equations for correlators. This is a mild generalization of \cite{BBCCN24}, as our construction does not impose the projection property on correlators. As such, we obtained the general solution space of the loop equations, which allowed us to incorporate the blobbed topological recursion within the framework of partial Airy ideals. For zero blobs, the partition function that we obtained were the same as the partition functions constructed in \cite{BBCCN24} from Airy ideals with standard polarization.

In this section we show that we can get rid of the most stringent part of the admissibility condition of \Cref{d:admissible}, that is, the somewhat mysterious condition that $r_{\alpha} = \pm 1\,\,\textnormal{mod}\,\,s_{\alpha}$, and still get partial Airy ideals. This is a novel possibility that has no analogue in the formalism of Airy ideals.

\subsection{Preliminary: ideals and partitions}

We start by reviewing the known ideals $\mathcal{I}$ in the ($\hbar$-rescaled) algebra of modes of $\mathcal{W}(\mathfrak{gl}_r)$ at self-dual level which satisfy $[\mathcal{I},\mathcal{I}] \subseteq \hbar^2 \mathcal{I}$. We consider only left ideals that contain all nonnegative modes.

\begin{definition}[Partitions]\label{d:lambdaj}
Let $\nu = (\nu_1, \nu_2, \ldots, \nu_\ell)$ be a partition of $r$, i.e. a sequence of integers $\nu_1 \geq \nu_2 \geq \ldots \geq \nu_\ell \geq 1$ such that $\sum_{j=1}^\ell \nu_j = r$. Its length is $\ell(\nu) = \ell$. For $i \in [r]$, we define
\begin{equation}
\nu[i] \coloneqq \min \set{ j \in [\ell] | \nu_1 + \ldots + \nu_j \geq i} \qquad \kappa^{(i)}(\nu) := 1 - \nu[i].
\end{equation}
We also define the sets
\[
I_r(\nu) := \set{(i,k) \in [r] \times \Z\,\,| \,\,k \geq \kappa^{(i)}(\nu)}.
\]
\end{definition}

\begin{figure}[h!]
\begin{center}
\includegraphics[width=0.1\textwidth]{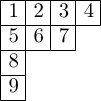}
\caption{Example of a partition of size $9$ with $\nu[1] = \cdots = \nu[4] = 1$, $\nu[5] = \nu[6] = \nu[7] = 2$, $\nu[8] = 3$, $\nu[9] = 4$. Numbers on the first line correspond to $\kappa^{(i)}(\nu)= 0$, on the second line to $\kappa^{(i)}(\nu) = -1$, etc. This $\kappa^{(i)}(\nu)$ indicates the minimal mode of conformal weight $i$ that is present in the ideal $\mathcal{I}_r(\nu)$.}
\end{center}
\end{figure}

The following map, which appeared in \cite[Definition 5.14]{BCGS} --- see also the proof of \cite[Theorem 4.9]{BBCCN24} --- will play an important role.

\begin{definition}[The map $\Pi_{r,s}$]\label{d:map}
Let $r \geq 2$ and $s \geq 1$ be two integers. We define the map
\begin{equation} 
\begin{array}{rrcl} \Pi_{r,s}: & [r] \times \Z & \longrightarrow & \mathbb{Z} \\
& (i,k) & \longmapsto & rk + s(i-1).
\end{array}
\end{equation}
Given a partition $\nu$ of $r$, we call $A_{r,s}(\nu) \coloneqq \Pi_{r,s}(I_r(\nu))$ and $S_{r,s}(\nu) \coloneqq \Z \setminus A_{r,s}(\nu)$. 
\end{definition}

\begin{lemma}[Ideals]
\label{l:partitionideal} Let $\nu$ be a partition of $r \geq 2$. The left ideal $\mathcal{J}_r(\nu) \subset \widehat{\mathcal{D}}_{\Z}^{\hbar}$ generated by
\[
\set{\hbar^i \rho(W_k^{(i)}) | (i,k) \in I_{r}(\nu)}
\]
satisfies $[\mathcal{J}_{r}(\nu),\mathcal{J}_{r}(\nu)] \subseteq \hbar^2\mathcal{J}_{r}(\nu)$. 
\end{lemma}
\begin{proof} This is the content of \cite[Theorem  3.16]{BBCCN24}, to which we apply $\rho$ from \Cref{l:rephei} and homogeneization.
\end{proof}

The ideals used in \Cref{t:PAISC1} correspond to specific partitions, which are maximal in a certain sense.

\begin{definition}[Canonical partitions]
\label{def:canopartition} Let $s \in [r + 1]$ such that $r = \pm 1\,\,\textnormal{mod}\,\,s$ (this comes from the admissibility condition in \Cref{d:admissible}). The canonical partition of $r$ associated to $(r,s)$ is defined to be $(1,\ldots,1)$ if $s = r + 1$, or 
\[
\nu_1 = \ldots = \nu_{r''} = r'+ 1,\qquad \nu_{r'' + 1} = \cdots = \nu_{s} = r'
\]
if $s < r$, where we decomposed $r= r's + r''$ with $r'' \in \set{1,s - 1}$ and $r' \geq 0$.
\end{definition}

\begin{table}[h!]
	\centering
		{\def\arraystretch{2.2}
			\begin{tabular}{|c|@{}p{2pt}@{}|c|c|c|c|}
				\cline{1-1}\cline{3-6}
				$r,s$ && $s=1$  & $r=r's + 1$ & $r=r's + s - 1$ & $s=r+1$ \\
				\cline{1-1}\cline{3-6}
				$\nu$ &&
				\begin{minipage}{0.14\textwidth}
					\centering
					\vspace{0.7em}
				$\overbrace{\Scale[0.6]{\begin{ytableau}%
					\, & \, & \, & \none[\dots] & \, & \,\\
					\none \\
					\none \\
					\none \\
					\none \\
					\none \\
					\end{ytableau}}}^{r\text{ boxes}}$
				\vspace{0.7em}
				\end{minipage} &
				\begin{minipage}{0.14\textwidth}
			\centering
			\vspace{0.7em}
			$\overbrace{\Scale[0.6]{\begin{ytableau}%
					\, & \, & \none[\dots] & \, & \, \\
					\, & \, & \none[\dots] & \, \\
					\, & \, & \none[\dots] & \, \\
					\none[\vdots] & \none[\vdots] & \none[\ddots] & \none[\vdots] \\
					\, & \, & \none[\dots] & \, \\
					\, & \, & \none[\dots] & \, \\
					\end{ytableau}}}^{(r' + 1)\text{ columns}}$
				\vspace{0.7em}
				\end{minipage} & 
				\begin{minipage}{0.14\textwidth}
					\centering
					\vspace{0.7em}
					$\overbrace{\Scale[0.6]{\begin{ytableau}%
					\, & \, & \none[\dots] & \, & \, \\
					\, & \, & \none[\dots] & \, & \, \\
					\, & \, & \none[\dots] & \, & \, \\
					\none[\vdots] & \none[\vdots] & \none[\ddots] & \none[\vdots] & \none[\vdots] \\
					\, & \, & \none[\dots] & \, & \, \\
					\, & \, & \none[\dots] & \, \\
						\end{ytableau}}}^{(r' + 1)\text{ columns}}$
					\vspace{0.7em}
				\end{minipage} &
				\begin{minipage}{0.14\textwidth}
					\centering
					\vspace{0.7em}
					$\Scale[0.6]{\begin{ytableau}%
					\none \\
					\none \\
					\, \\
					\, \\
					\, \\
					\none[\vdots] \\
					\, \\
					\, \\
					\end{ytableau}}$
					\vspace{0.7em}
				\end{minipage} \\
				\cline{1-1}\cline{3-6}
			\end{tabular}}
		\vspace{0.3cm}
		\caption{\label{tab:coxeter_partition_diagrams} Canonical partitions (reproduced from \cite[Table 1]{BKS24}).}
\end{table}

\begin{lemma}
\label{lem:N} Let $\nu$ be a partition of $r \geq 2$. Assume that $s$ is coprime to $r$ and that $s \geq \ell(\nu)$.
\begin{enumerate}
\item $0 \in A_{r,s}(\nu)$
\item The set $\N \setminus A_{r,s}(\nu)$ is finite.
\item $A_{r,s}(\nu) \subseteq \N$, with equality if and only if $r = \pm 1\,\,\textnormal{mod}\,\,s$ and $\nu$ is the canonical partition associated to $(r,s)$.
\item For all $i \in [r]$ we have $\kappa^{(i)}(\nu) \geq - \big\lfloor \frac{(i - 1)s}{r} \big\rfloor$, with simultaneous equalities if and only if $r = \pm 1\,\,\textnormal{mod}\,\,s$ and $\nu$ is the canonical partition associated to $(r,s)$.
\end{enumerate}
\end{lemma}

\begin{proof}
(1) Since $\nu[i=1] = 1$, we always have $\kappa^{(1)}(\nu) = 0$ and thus $0 \in A_{r,s}(\nu)$.

(2) For each $i \in [r]$, by definition, $\Pi_{r,s}(i,\kappa^{(i)}(\nu))$ is the minimal element of $A_{r,s}(\nu)$ congruent to $(i - 1)s$ modulo $r$, and $A_{r,s}(\nu)$ contains all integers larger than $\Pi_{r,s}(i,\kappa^{(i)}(\nu))$ in the same congruence class. Remember that $r,s$ are coprime, so the multiplication by $s$ induces a bijection on the congruences modulo $r$. Therefore, $\N \setminus A_{r,s}(\nu)$ is finite.

(3) $\Leftrightarrow$ (4) By definition, $A_{r,s}(\nu) \subseteq \N$ if and only if $r \kappa^{(i)}(\nu) + s(i-1) \geq 0$ for all $i \in [r]$. The inequality can be rewritten as $\kappa^{(i)}(\nu) \geq - \frac{(i-1)s}{r}$. But since $\kappa^{(i)}(\nu)$ is an integer, this is equivalent to $\kappa^{(i)}(\nu) \geq \lceil - \frac{(i-1)s}{r} \rceil = - \lfloor \frac{(i-1)s}{r} \rfloor$. 

Next, $A_{r,s}(\nu) = \N$ if and only if, for all $i\in [r]$, $0 \leq \Pi_{r,s}(i,\kappa^{(i)}(\nu)) \leq r - 1$, and $\Pi_{r,s}(i,\kappa^{(i)}(\nu)) $ spans all congruence classes as $i$ varies. This happens precisely when $\kappa^{(i)}(\nu) = - \lfloor \frac{(i-1)s}{r} \rfloor$.

(4) We first prove that $\kappa^{(i)}(\nu) = - \big\lfloor \frac{(i - 1)s}{r} \big\rfloor$ for $i \in [r]$ if and only if  $r = \pm 1\,\,\textnormal{mod}\,\,s$ and $\nu$ is the canonical partition associated to $(r,s)$. 
If $r = \pm 1\,\,\textnormal{mod}\,\,s$ and $\nu$ is the canonical partition associated to $(r,s)$, \cite[Proposition B.1]{BBCCN24} computes $\kappa^{(i)}(\nu) = - \big\lfloor \frac{(i - 1)s}{r} \big\rfloor$. Conversely, suppose that $\kappa^{(i)}(\nu) = - \big\lfloor \frac{(i - 1)s}{r} \big\rfloor$ for all $i \in [r]$. Then $\nu[i] = 1 + \big\lfloor \frac{(i - 1)s}{r} \big\rfloor$. In the proof of \cite[Proposition B.1]{BBCCN24} it is shown that if $s \in [r + 1]$, those values can be realized by a partition $\nu$ if and only if $r = \pm 1\,\,\textnormal{mod}\,\,s$ and $\nu$ is the canonical partition associated to $(r,s)$. If $s > r + 1$, we claim that there exists $j \in [r - 1]$ such that $\nu[j + 1] > \nu[j] + 1$, which is a contradiction since this can never happen for a partition $\nu$ (see \Cref{d:lambdaj}). 

To justify this claim, write $s = r  + s'$ with $s' \geq 2$. Then
\begin{equation}
\nu[j+1] - \nu[j] = 1 +  \bigg \lfloor \frac{j s'}{r} \bigg \rfloor - \bigg \lfloor \frac{(j-1) s'}{r} \bigg \rfloor  \geq 1.
\end{equation}
For this expression to be always equal to $1$, we need that
\begin{equation}
\forall j \in [r-1], \qquad \bigg \lfloor \frac{j s'}{r} \bigg \rfloor = \bigg \lfloor \frac{(j-1) s'}{r} \bigg \rfloor .
\end{equation}
But for $j=1$ the right-hand-side is equal to $0$, while for $j=r-1$ the left-hand-side is $\geq 1$ since $s' \geq 2$: this is the contradiction we were looking for.

We now prove the general statement. Let $\nu$ be any partition of $r$ of length $\ell(\nu) \leq s$. We want to show that $\kappa^{(i)}(\nu) \geq - \big\lfloor \frac{(i - 1)s}{r} \big\rfloor$ for all $i \in [r]$, that is, $\nu[i] \leq 1 + \big\lfloor \frac{(i - 1)s}{r} \big\rfloor$. The integer $s$ is independent of $\nu$ except that $\ell(\nu) \leq s$, so it is sufficient to prove
\begin{equation}
\label{lableunu}
\forall i \in [r],\qquad \nu[i] \leq 1 + \bigg\lfloor \frac{(i - 1)\ell(\nu)}{r} \bigg\rfloor.
\end{equation}
This property is an expression of the discrete convexity of the function $i \mapsto \nu[i]$.  To see this, we first upgrade the partition $\nu$ to the nonnegative weakly decreasing function $\overline{\nu}(x) := \nu_{1 + \lfloor x \rfloor}$ defined on the \emph{real} interval $[0,\ell(\nu))$. Its primitive $N(x) = \int_{0}^{x} \overline{\nu}(y)\dd y$ is a strictly increasing and concave function on the interval $[0,\ell(\nu)]$ with initial value $N(0) = 0$ and final value $N(\ell(\nu)) = r$. Then, the reciprocal $M$ of $N$ is a strictly increasing and convex function defined on the interval $[0,r]$ with initial value $M(0) = 0$ and $M(r) = \ell(\nu)$. In particular,
\begin{equation}
\label{convN}
\forall x \in [0,r],\qquad M(x) \leq \frac{x\ell(\nu)}{r}.
\end{equation}
For all $i \in [r]$ we have by definition
\[
N(\nu[i] - 1) = \sum_{j = 1}^{\nu[i] - 1} \nu_j \leq i - 1 = N(M(i - 1)).
\]
Since $N$ is strictly increasing, this gives $\nu[i] - 1 \leq M(i - 1) \leq \frac{(i - 1)\ell(\nu)}{r}$ due to \eqref{convN}, and further implies \eqref{lableunu} because $\nu[i]$ is an integer.
\end{proof}

\subsection{Coloured spectral curves and their partial Airy ideals}

The starting point of our construction is an enrichment of the notion of spectral curves.

\begin{definition}[Coloured spectral curve]\label{d:scc}
Let $\mathcal{S}=(C,x,\omega_{0,1})$ be a spectral curve. We say that it is \emph{coloured} by $(\bm{\nu},\bm{T})$ if the following additional data is provided:
\begin{itemize}
\item A family $\bm{\nu} = (\nu^{\alpha})_{\alpha}$ of partitions $\nu^{\alpha}$ of $r_\alpha$, indexed by $\alpha \in \Ram$;
\item A family $\bm{T} = (T_{\alpha;g}^{(i)})_{\alpha,i,g}$ of scalars, also called \emph{zero-mode shifts}, indexed by $\alpha \in \Ram$, $i \in [\nu_1^{\alpha} - \nu_2^{\alpha}]$ and $g \in \frac{1}{2}\N^*$. We denote
\[
T_{\alpha}^{(i)} \coloneqq \sum_{g} T_{\alpha;g}^{(i)}\hbar^{2g},
\]
and stress that $T_{\alpha}^{(i)} = O(\hbar)$. By convention, we set all $T_{\alpha}^{(i)}$ for $i  > \nu_1^{\alpha} - \nu_2^{\alpha}$ to zero, and if $\nu = (r)$ zero-mode shifts are allowed for all $i \in [r]$.
\end{itemize} 
\end{definition}
\begin{definition}[Coloured admissibility]
\label{d:cadmi}
We say that a coloured spectral curve is \emph{admissible} if, for all $\alpha \in \Ram$, $r_{\alpha}$ and $s_{\alpha}$ are coprime and we have $s_\alpha \geq \ell(\nu^{\alpha})$.
\end{definition}

Our main result is the construction of partial Airy ideals attached to admissible coloured spectral curves.
\begin{theorem}\label{t:PAISCc}
Let $\mathcal{S} = (C,x,\omega_{0,1},\bm{\nu},\bm{T})$ be an admissible coloured spectral curve. Define $I = \Ram \times \Z$ and
\begin{equation}
\label{colorA}
A \coloneqq \set{(\alpha,m)\,\,|\,\,\alpha \in \Ram \,\,\textnormal{and}\,\,m \in A_{r_{\alpha},s_{\alpha}}(\nu^{\alpha})},\qquad S \coloneqq I \setminus A.
\end{equation}
Let $\mathcal{I}_{\mathcal{S}} \subset \widehat{\mathcal{D}}_{\Ram \times \mathbb{Z}}^{\hbar}$ be the left ideal  generated by the 
set of shifted operators  $\mathbf{H} = \set{H_{\alpha,k}^{(i)} - T_{\alpha}^{(i)}\delta_{k,0}}$ (see \eqref{eq:HikN}), indexed by $\alpha \in \Ram$ and $(i,k) \in I_{r_\alpha}(\nu^{\alpha})$. Then $\mathcal{I}_{\mathcal{S}}$ is a partial Airy ideal on the set $I$ with actor set $A$ and spectator set $S$.
\end{theorem}
\begin{remark} According to \Cref{lem:N}, $S$ and $A$ differ from $\Ram \times \mathbb{Z}_{<0}$ and $\Ram \times \mathbb{N}$ by finite sets. Therefore, $\mathcal{D}_{I;\llbracket S,A \rrbracket}^{\hbar} = \widehat{\mathcal{D}}_{\Ram \times \Z}^{\hbar}$ is the same completion as in \Cref{d:WeylZcomplete} (see \Cref{rem:finites}).
\end{remark}

\Cref{t:PAISC} for (non-coloured) admissible spectral curves is a special case of \Cref{t:PAISCc}: a (non-coloured) admissible spectral curve can be turned into an admissible coloured spectral curve with a canonical choice of colours. Namely, we take vanishing zero-mode shifts, and for each $\alpha \in \Ram$ we choose $\nu^{\alpha}$ to be the canonical partition associated to $(r_{\alpha},s_{\alpha})$ (see \Cref{def:canopartition}). For this canonical colour, the actor variables for the partial Airy ideal in \Cref{t:PAISCc} are exactly the nonnegative ones (see statement (3) in \Cref{lem:N}), and the partial Airy ideal constructed there matches the one of \Cref{t:PAISC}. For other admissible coloured spectral curves the actor variables form a strict subset of the nonnegative ones, missing only finitely many of them (see \Cref{tables} for a few examples).

\begin{proof}[Proof of \Cref{t:PAISCc}]
It is clear that $\mathbf{H}$ is bounded. The condition $[\mathcal{I}_{\mathcal{S}},\mathcal{I}_{\mathcal{S}}] \subseteq \hbar^2 \mathcal{I}_{\mathcal{S}}$ is satisfied due to \Cref{l:partitionideal} and \cite[Sections 2.3.1 and Corollary 2.24]{BBKN25}. More precisely, the latter justifies that $\hbar^{-2} [\mathcal{J}_{r_{\alpha}}(\nu^{\alpha}),\mathcal{J}_{r_{\alpha}}(\nu^{\alpha})]$ is actually included in the smaller left ideal generated by the modes $\overline{W}^{(i)}_{\alpha,k}$ indexed by $(i,k) \in I_{r_{\alpha}}(\nu^{\alpha})$ but without including the modes with $k = 0$ for $i \leq  \nu_1^{\alpha} - \nu_2^{\alpha}$ (in the language of \cite{BBCC24}, the zero modes $\overline{W}^{(i)}_{\alpha,0}$ are \emph{extraneous} for $i \leq  \nu_1^{\alpha} - \nu_2^{\alpha}$). Hence, subtracting scalars $T_{\alpha}^{(i)} \delta_{k,0}$ from these modes does not perturb the ideal condition.\footnote{In fact \cite[Sections 2.3.1 and Corollary 2.24]{BBKN25} show that these are  the only extraneous modes for the ideals $\mathcal{J}_{r_\alpha}$. This is why we do not allow more general shifts in the definition of coloured spectral curves.}
 
What remains to be shown is that $\mathcal{I}_{\mathcal{S}}$ admits a generating set $\set{G_{\alpha,m} \,\,|\,\, (\alpha,m) \in A}$ such that 
\begin{equation}
G_{\alpha,m} = \hbar \partial_{\alpha,m} - \hbar h_{\alpha,m}^{(1)}(\bm{t}_I)+ O(\hbar^2).
\end{equation}

Consider first the case where $\omega_{0,1}(z) = -\tau_{\alpha} z^{s_\alpha-1}\, \dd z$ for all $z \in C_{\alpha}$ and $\alpha \in \Ram$, that is, the only non-zero coefficients in the $z$-expansion of $\omega_{0,1}$ are $F_{0,1} \begin{bsmallmatrix} \alpha \\ - s_\alpha \end{bsmallmatrix} = -\tau_{\alpha}$. Then the conjugation by $Z_{(0,1)}$ in \eqref{eq:HikN} amounts to doing the shifts
\[
J_{\alpha,-s_\alpha} \longrightarrow  J_{\alpha,-s_\alpha} - \frac{\tau_{\alpha}}{ \hbar}.
\]
Looking at \eqref{eq:overline}, we see that
\begin{equation}
H_{\alpha,k}^{(i)} = (-\tau_{\alpha})^i c_{\alpha,k}^{(i)}  \delta_{k+s_{\alpha},0} +  \hbar(-\tau_{\alpha})^{i - 1} d_{\alpha,k}^{(i)}  J_{\alpha,r_\alpha k + s_\alpha(i-1)} - \hbar \delta_{k,0} T_{\alpha;\frac{1}{2}}^{(i)}  + O(\hbar^2),
\end{equation}
The coefficients for the first two terms are
\[
c_{\alpha,k}^{(i)} = r_{\alpha}^{-i} \Psi^{(i,0)}_{r_{\alpha}}(\underbrace{-s_{\alpha},\ldots,-s_{\alpha}}_{i\,\,\textnormal{times}}),\qquad 
d_{\alpha,k}^{(i)} =  i r_{\alpha}^{-i} \Psi^{(i,0)}_{r_{\alpha}}\big(\underbrace{-s_{\alpha},\ldots,-s_{\alpha}}_{i - 1\,\,\textnormal{times}},r_{\alpha}k + s_{\alpha}(i-1)\big).
\]
They can be computed using \Cref{evallem} and the coprimality of $r_{\alpha}$ and $s_{\alpha}$:
\[
c_{\alpha,k}^{(i)} = r_{\alpha}^{-i} (-1)^{r_{\alpha} - 1} \delta_{i,r_{\alpha}},\qquad d_{\alpha,k}^{(i)} = (-r_{\alpha})^{1 -i}.
\]
Remember that the representation $\rho$ will replace $J_{\alpha,m}$ by $\partial_{m} + |m|t_{|m|}\delta_{m<0}$.

The term of order $\hbar^0$ is non-zero only for $i = r_{\alpha}$ and $k = -s_{\alpha}$. As we do not want $\hbar^0$-terms, we should make sure that it does not happen for $(r_{\alpha},k) \in I_{r_{\alpha}}(\nu^{\alpha})$. This translates into the condition $1 - \nu^{\alpha}[r_{\alpha}] = 1 - \ell(\nu^\alpha) > -s_{\alpha}$. This is precisely the admissibility condition $\ell(\nu^\alpha) \leq s_{\alpha}$ for coloured spectral curves from \Cref{d:cadmi}.

Next, to make sure that the $\hbar^1$-terms are of the right form, the restriction of $\Pi_{r_\alpha,s_\alpha}$ to $I_{r_{\alpha}}(\nu^{\alpha})$ must be injective. This is the case if and only if $r_{\alpha}$ and $s_{\alpha}$ are coprime. We conclude that the operators have the right form for a partial Airy ideal on $I$, and the actor and spectator sets $A$ and $S$ are as announced, because the $J_{\alpha,m}$ with $m \notin A_{r_{\alpha},s_{\alpha}}(\nu^{\alpha}) = \Pi_{r_{\alpha},s_{\alpha}}(I_{r_{\alpha}}(\nu^{\alpha}))$ are precisely those that do not appear in the $\hbar^1$-term. 
  
Conjugating by more general $Z_{(0,1)}$ does not change the conclusion; the $\hbar^1$-terms will now become linear combinations of $J$s, but the system is triangular and can be diagonalized to recover the same result as above (this part of the proof is the same as in \cite{BKS24,BCJ24}).
\end{proof}

\begin{table}
\centering
{\def\arraystretch{1,2}\begin{tabular}{|c|c||l|l|l|l|l|}
\hline
$\nu$ & $\begin{smallmatrix} \textnormal{Zero-mode} \\ \textnormal{shifts} \end{smallmatrix}$ & \multicolumn{4}{c|}{$\mathbf{s} : S \cap \N$} \\
\hline\hline
\begin{minipage}{0.14\textwidth}\centering\vspace{0.7em}\Scale[0.6]{\begin{ytableau} 1 & 2 \end{ytableau}}\vspace{0.7em}\end{minipage} & $1,2$ & $\mathbf{1} : \emptyset$ \cellcolor{yellow}  & $\mathbf{3} : 1$ & $\mathbf{5} : 1,3$ & $\mathbf{7} : 1,3,5$  \\
\hline
\begin{minipage}{0.14\textwidth}\centering \vspace{0.7em}\Scale[0.6]{\begin{ytableau} 1 \\ 2 \end{ytableau}} \vspace{0.7em}\end{minipage} & $\emptyset$ & $\mathbf{3} : \emptyset$  \cellcolor{yellow}  & $\mathbf{5} : 1$ & $\mathbf{7} : 1,3$ & $\mathbf{9} : 1,3,5$  \\
\hline
\hline
\begin{minipage}{0.14\textwidth}\centering\vspace{0.7em}\Scale[0.6]{\begin{ytableau} 1 & 2 & 3 \end{ytableau}}\vspace{0.7em}\end{minipage} & $1,2,3$ & $\mathbf{1} : \emptyset$ \cellcolor{yellow} & $\mathbf{2} : 1$ & $\mathbf{4} : 1,2,5$ & $\mathbf{5} : 1,2,4,7$  \\
\hline
\begin{minipage}{0.14\textwidth}\centering\vspace{0.7em}\Scale[0.6]{\begin{ytableau} 1 & 2 \\ 3 \end{ytableau}}\vspace{0.7em}\end{minipage} & $1$ & $\mathbf{2} : \emptyset$ \cellcolor{yellow} & $\mathbf{4} : 1,2$ & $\mathbf{5} : 1,2,4$ & $\mathbf{7} : 1,2,4,5,8$   \\
\hline
\begin{minipage}{0.14\textwidth}\centering\vspace{0.7em}\Scale[0.6]{\begin{ytableau} 1 \\ 2 \\ 3 \end{ytableau}}\vspace{0.7em}\end{minipage}& $\emptyset$ & $\mathbf{4} : \emptyset$ \cellcolor{yellow} & $\mathbf{5} : 1$ & $\mathbf{7} : 1,2,5$ & $\mathbf{8} : 1,2,4,7$  \\
\hline
\hline
\begin{minipage}{0.14\textwidth}\centering\vspace{0.7em}\Scale[0.6]{\begin{ytableau} 1 & 2 & 3 & 4 \end{ytableau}}\vspace{0.7em}\end{minipage} & $1,2,3,4$ & $\mathbf{1} : \emptyset$ \cellcolor{yellow} & $\mathbf{3} : 2,5$ & $\mathbf{5} :$ \!\!{\small \begin{tabular}{l} $1,2,3,6,7,$ \\[-1ex] $11$ \end{tabular}} & $\mathbf{7} :$ \!\!{\small \begin{tabular}{l} $1,2,3,5,6,9,$ \\[-1ex] $10,13,17$ \end{tabular}} \\
\hline
\begin{minipage}{0.14\textwidth}\centering\vspace{0.7em}\Scale[0.6]{\begin{ytableau} 1 & 2 & 3 \\ 4 \end{ytableau}}\vspace{0.7em}\end{minipage} & $1,2$ & $\mathbf{3} : 1,2$ & $\mathbf{5} : 1,2,3,6,7$ &  $\mathbf{7} :$ \!\!{\small \begin{tabular}{l} $1,2,3,5,6,9,$ \\[-1ex] $10,13$ \end{tabular}} & $\mathbf{9} :$ \!\!{\small \begin{tabular}{l} $1,2,3,5,6,7,$ \\[-1ex] $10,11,14,15,19$ \end{tabular}} \\
\hline
\begin{minipage}{0.14\textwidth}\centering\vspace{0.7em}\Scale[0.6]{\begin{ytableau} 1 & 2 \\ 3 & 4 \end{ytableau}}\vspace{0.7em}\end{minipage} & $\emptyset$ & $\mathbf{3} : 1$ & $\mathbf{5} : 1,2,3,7$ & $\mathbf{7} :$ \!\!{\small \begin{tabular}{l} $1,3,5,6,9,$ \\[-1ex] $10,13$\end{tabular}}  & $\mathbf{9} :$ \!\!{\small \begin{tabular}{l} $1,2,3,5,6,7,$ \\[-1ex] $10,11,15,19$\end{tabular}} \\
\hline
\begin{minipage}{0.14\textwidth}\centering\vspace{0.7em}\Scale[0.6]{\begin{ytableau} 1 & 2 \\ 3 \\ 4 \end{ytableau}}\vspace{0.7em}\end{minipage} & $1$ & $\mathbf{3} : \emptyset$ \cellcolor{yellow} & $\mathbf{5} : 1,2,3$ & $\mathbf{7} : 1,2,3,5,6,9$ & $\mathbf{9} :$ \!\!{\small \begin{tabular}{l} $1,2,3,5,6,7,$ \\[-1ex] $10,11,15$ \end{tabular}} \\
\hline
\begin{minipage}{0.14\textwidth}\centering\vspace{0.7em}\Scale[0.6]{\begin{ytableau} 1 \\ 2 \\ 3 \\ 4 \end{ytableau}}\vspace{0.7em}\end{minipage} & $\emptyset$ & $\mathbf{5} : \emptyset$ \cellcolor{yellow}& $\mathbf{7} : 1,2,5$ & $\mathbf{9} :$ \!\!{\small \begin{tabular}{l} $1,2,3,6,7,$ \\[-1ex] $11$\end{tabular}} & $\mathbf{11} :$ \!\!{\small \begin{tabular}{l} $1,2,3,5,6,9,$ \\[-1ex] $10,13,17$ \end{tabular}} \\
\hline
\end{tabular}}
\caption{\label{tables}Allowed zero-mode shifts and positive spectators for $\nu$ of size $r \leq 4$ and $s$ coprime to $r$ such that $\ell(\nu) \leq s$. In yellow, the canonical partitions: the associated partial Airy ideals (without shifts) were already known in the form of stable Airy ideals in \cite{BBCCN24}.}
\end{table}

\subsection{The poset of colouring partitions}

\label{secpose}

In this section, we consider the following question. Fix a spectral curve with one ramification point of order $r$. That is, we have a fixed choice of two coprime integers $r \geq 2$ and $s \geq 1$. Consider all possible admissible colourings on this spectral curve (i.e. choices of partitions $\nu$ of $r$ such that $\ell(\nu) \leq s$ and allowed zero-mode shifts). Can we relate the partition functions associated to the resulting partial Airy ideals?

\begin{definition}[Partial order]
We define a partial order on the set of partitions of size $r$ and length $\ell \leq s$ by
\[
\nu \preceq \nu' \qquad \Longleftrightarrow \qquad \forall i \in [r]\quad \nu[i] \leq \nu'[i].
\]
\end{definition}

\begin{figure}[h!]
\begin{center}
\includegraphics[width=0.3\textwidth]{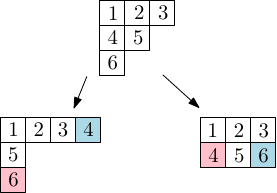}
\caption{\label{fig:partialorder} The partition on the top is larger than each of the partition on the bottom. The two partitions on the bottom are not comparable: in the left partition, the box numbered $4$ is above its position in the right partition, while the box numbered $6$ is below its position in the right partition.}
\end{center}
\end{figure}

If $\nu \preceq \nu'$, there is an inclusion of actor sets $A_{r,s}(\nu) \subseteq A_{r,s}(\nu')$. The larger the partition is, the larger the set of modes in the partial Airy ideals is. The poset has a minimum given by the single-row partition $\nu = (r)$. If $s \geq r + 1$ it has a maximum which is the partition $(1,\ldots,1)$. If $s < r$ and $r = \pm 1\,\,\textnormal{mod}\,\,s$, \Cref{lem:N} says that the canonical partition is the maximum. For $r \leq 5$ or $s \leq 2$ the order $\preceq$ is total, but in all other situations it is not (see \Cref{fig:partialorder} for the simplest example of non-comparable partitions).

The poset has a realization at the level of the sets of partition functions produced by the partial Airy ideals of \Cref{t:PAISCc}. First, if $\nu \preceq \nu'$ we have natural inclusions for actor and spectator sets
\[
A_{r,s}(\nu) \subseteq A_{r,s}(\nu'),\qquad S_{r,s}(\nu') \subseteq S_{r,s}(\nu),
\]
If there were no zero-mode shifts, larger partitions would give rise to smaller sets of partition functions. The story is more complicated, because the set of allowed zero-mode shifts can be larger or smaller independently of the relative order $\preceq$. To formalise it, let us call $\mathfrak{T}_{\nu}$ the set of zero-mode shifts supported by $\nu$, that is
\[
\mathfrak{T}_{\nu} = \hbar \mathbb{C}^{\nu_1 - \nu_2}\llbracket \hbar \rrbracket,
\]
and call $\hat{\mathfrak{Z}}_{\nu}$ the abelian group of partition functions in the spectator variables $\bm{t}_{S_{r,s}(\nu)}$. For a given admissible coloured spectral curve $\mathcal{S} = (C,x,\omega_{0,1})$ with a single ramification point of order $r$ and partition $\nu$ of $r$, the set of partition functions $\mathfrak{Z}_{\nu}^{\mathcal{S}}$ for the partial Airy ideals of  \Cref{t:PAISCc} is canonically identified  to $\hat{\mathfrak{Z}}_{\nu}\times \mathfrak{T}_{\nu}$.

Assume now $\nu \preceq \nu'$. If $\nu_1' - \nu'_2 \leq \nu_1 - \nu_2$, there is a natural inclusion map $\mathfrak{i} : \mathfrak{T}_{\nu'} \hookrightarrow \mathfrak{T}_{\nu}$, which takes the zero-mode shifts supported by $\nu'$ and add zeros for the zero-mode shifts supported by $\nu$ that were not supported by $\nu'$. For each $\bm{T}' \in \mathfrak{T}_{\nu'}$ and reduced partition function $\hat{Z}' \in \hat{\mathfrak{Z}}_{\nu'}$ as initial condition, we get a unique partition function $Z'(\bm{t}_{\Z})$ satisfying the partial Airy ideal constraints associated to $\nu'$ with zero-mode shifts $\bm{T}'$. As this partition function also obeys the smaller set of partial Airy ideal constraints associated to $\nu$ with zero-mode shifts $\mathfrak{i}(\bm{T}')$, it coincides with the partition function of the latter partial Airy ideal with initial condition $\mathfrak{j}(\hat{Z}') \in \hat{\mathfrak{Z}}_{\nu}$ defined by 
\[
\mathfrak{j}(\hat{Z}') :=  Z'(\bm{t}_I)_{|\bm{t}_{A_{r,s}(\nu)} = 0}.
\]
We stress that the map $\mathfrak{j}$ is non-linear and requires the construction of the full partition function $Z'$, before reducing it to the larger subset $S_{r,s}(\nu)$ of spectator variables. The map $\mathfrak{j}$ is injective, because the restriction of $\mathfrak{j}(\hat{Z}')$ to the smaller subset of spectator variables $S_{r,s}(\nu')$ is exactly $\hat{Z}'$. This gives rise to an injective map
\begin{equation}
\label{fleche}
\mathfrak{Z}_{\nu'}^{\mathcal{S}} \cong \mathfrak{T}_{\nu'} \times \hat{\mathfrak{Z}}_{\nu'} \,\, \mathop{\longrightarrow}^{(\mathfrak{i},\mathfrak{j})} \,\,\mathfrak{T}_{\nu} \times \hat{\mathfrak{Z}}_{\nu} \cong \mathfrak{Z}_{\nu}^{\mathcal{S}}.
\end{equation}

If $\nu_1 - \nu_2 > \nu_1' - \nu_2'$, the natural inclusion for spaces of zero-mode shifts goes in the opposite direction $\mathfrak{i} : \mathfrak{T}_{\nu} \rightarrow \mathfrak{T}_{\nu'}$. We can construct the map $\mathfrak{j}$ in the same way only for the zero-mode shifts of the form $\bm{T}' = \mathfrak{i}(\bm{T})$ for $\bm{T} \in \mathfrak{T}_{\nu}$. Instead of \eqref{fleche}, we get a zigzag of injective maps
\[
\mathfrak{Z}_{\nu'}^{\mathcal{S}} \cong \mathfrak{T}_{\nu'} \times \hat{\mathfrak{Z}}_{\nu'} \,\,\mathop{\longleftarrow}^{(\mathfrak{i},\textnormal{id})}\,\, \mathfrak{T}_{\nu} \times \hat{\mathfrak{Z}}_{\nu'} \mathop{\longrightarrow}^{(\textnormal{id},\mathfrak{j})}\,\,\mathfrak{T}_{\nu} \times \hat{\mathfrak{Z}}_{\nu} \cong \mathfrak{Z}_{\nu}^{\mathcal{S}}.
\]

\subsection{Elementary general properties}
\label{secprop}
We derive a few properties of the partial Airy ideals of \Cref{t:PAISCc}.

\begin{proposition}
\label{l69}
Let $\mathcal{S} = (C,x,\omega_{0,1})$ be an admissible coloured spectral curve, and $Z$ be the partition function of its partial Airy ideal with initial condition $\hat{Z}$.
\begin{enumerate}
\item The partition function $Z$ does not depend on the actor variables of the form $t_{\alpha, r_\alpha k}$ with $k > 0$, and it has the following simple dependence on the variables $t_{\alpha,0}$:
\begin{equation}
\label{eqtrs}
Z = \exp\left(\frac{T^{(1)}_{\alpha}}{\hbar} t_{\alpha,0}\right) Z_{|t_{\alpha,0}=0}.
\end{equation}
\item If $\hat{Z}$ does not depend on a given negative variable $t_{\alpha,m}$ with $m<0$, then $Z$ also does not depend on this $t_{\alpha,m}$.
\item Assume that we have a factorization $\hat{Z} = \prod_{\alpha \in \Ram} \hat{Z}_{\alpha}$ where $\hat{Z}_{\alpha}$ depends only on the spectator variables associated to $\alpha$. Then, we have a factorization  $Z = \prod_{\alpha \in \Ram} Z_{\alpha}$, where $Z_{\alpha}$ is the partition function for the Airy ideal generated by $\set{H_{\alpha,k}^{(i)}| (i,k) \in A_{r_{\alpha},s_{\alpha}}(\nu^{\alpha})}$ with initial condition $\hat{Z}_{\alpha}$.
\item 
Given $\lambda \in \mathbb{C}^*$, the partition function associated to the spectral curve $(C,x,\lambda \omega_{0,1})$ with initial condition $\lambda^{-\hbar\partial_{\hbar}}\hat{Z}$ is equal to $\lambda^{-\hbar \partial_{\hbar}}Z$ (note that $\lambda^{-\hbar \partial_{\hbar}}$ rescales $\hbar \rightarrow \lambda^{-1}\hbar$).
\item Let $\alpha \in \Ram$. If $\omega_{0,1}(z) = - \tau_{\alpha} z^{s_{\alpha} - 1}\dd z$, the partition function  $Z$ satisfies the Virasoro constraints: 
\[L_{\alpha,k} \cdot Z = 0 \qquad \text{for}\qquad  \begin{cases} k \geq -1 & \text{if } \nu^{\alpha} = (1,\ldots,1) \\ k \geq 0 &\text{for any other choices of partitions $\nu^\alpha$, } \end{cases}
\]
with the operators
 \begin{equation*}
 \begin{split}
 L_{\alpha,k}  & = \hbar \tau_{\alpha} \partial_{\alpha,r_{\alpha}k + s_{\alpha}} - \sum_{\substack{m_1 + m_2 = r_{\alpha}k \\ r \nmid m_c}} \frac{\hbar^2}{2} : (\partial_{\alpha,m_1} + |m_1|t_{\alpha,|m_1|}\delta_{m_1 < 0})(\partial_{\alpha,m_2}  + |m_2|t_{\alpha,|m_2|}\delta_{m_2 < 0}): \\
 & \quad - \left(\frac{\hbar^2(r_{\alpha}^2 - 1)}{24} +r_\alpha T^{(2)}_\alpha - \frac{r_{\alpha}-1}{2}(T_{\alpha}^{(1)})^2\right)\delta_{k,0}.
 \end{split}
 \end{equation*} 
 Note that $T_{\alpha}^{(1)}$ may be non-zero only if $\nu_1^{\alpha} > \nu_2^{\alpha}$, and $T_{\alpha}^{(2)}$ may be non-zero only if $\nu_1^{\alpha} > \nu_2^{\alpha} +1$.
 \end{enumerate}
\end{proposition}
\begin{proof}
(1) This is a consequence of the $\mathcal{W}$-constraints for $i = 1$: using $\Psi^{(1,0)}_{r}(m) = r\delta_{r|m}$ they read $H^{(1)}_{\alpha,k} = \hbar \partial_{\alpha,r_{\alpha}k} - T^{(1)}_{\alpha}\delta_{k,0}$, and hence property (1) follows.

(2) This property holds because the operators $H^{(i)}_{\alpha,k}$ from \eqref{eq:overline}-\eqref{theshift} do not explicitly depend on the negative variables $t_{\alpha,m}$ with $m<0$. The argument works by induction as in the proof of \Cref{t:PAISC1}. 

(3) This property expresses the fact that the operators $H_{\alpha,k}^{(i)}$ only involve the variables and derivatives associated to the ramification point $\alpha$. There is no mixing within the operators themselves, the mixing occurs in the reduced partition function.

(4) This property follows from the fact that conjugating $\lambda^{i} H_{\alpha,k}^{(i)}$ by $\lambda^{-\hbar \partial_{\hbar}}$ amounts to replacing $\omega_{0,1}$ with $\lambda \omega_{0,1}$.

(5) The operator $L_{\alpha,k}$ is $r_{\alpha} (H_{\alpha,k}^{(2)}- T^{(2)}_\alpha \delta_{k,0})$ after exploiting (1) to get rid of the variables and derivatives in the variables $t_{\alpha, r_\alpha k}$. That this is a representation of the Virasoro algebra follows from the definition of $\mathcal{W}(\mathfrak{gl}_r)$-algebras or can be checked directly.
\end{proof}

We know that in the free energy, each monomial in $\hbar$ and negative variables involves a finite number of positive variables, because this is how our completion is defined. In fact, we will show that a stronger property holds: the free energies $F_{g,n}$ are always polynomial in positive variables, and we can find explicit bounds for the indices of the positive variables involved.

\begin{definition}[Depth]
Denote $[x]_+ = \max(x,0)$. The \emph{depth} of a coloured spectral curve is
\begin{equation}
\label{varkap}
\varkappa \coloneqq  \Big[\max\set{\tfrac{m}{s_{\alpha}} | (\alpha,m) \in S}\Big]_+.
\end{equation}
We say that the zero-mode shifts $\bm{T}$ are \emph{regular} if $T_{\alpha}^{(i)} = O(\hbar^i)$ for all $\alpha \in \Ram$ and $i \in [r_{\alpha}]$. 
\end{definition}
By \Cref{lem:N}-(3), the depth is $0$ for canonically coloured spectral curves and positive for the others.

\begin{proposition}[Bound for free energies]
\label{p:ineqcase}
Let $\mathcal{S}$ be an admissible coloured spectral curve and $Z$ be the partition function of its partial Airy ideal with initial condition $\hat{Z}$. Assume that the zero-mode shifts are regular.
\begin{enumerate}
\item If $\varkappa = 0$ or if $\hat{Z}$ does not depend on positive variables, we have for any $g,n,\alpha_1,\ldots,\alpha_n,m_1,\ldots,m_n$
\begin{equation}
\label{vanishskap0}
\sum_{i = 1}^n \frac{[m_{i}]_+}{s_{\alpha_i}} > 2g - 2 + n \qquad \Longrightarrow \qquad F_{g,n}\begin{bsmallmatrix} \alpha_1 & \cdots & \alpha_n \\ m_1 & \cdots & m_n \end{bsmallmatrix} = 0.
\end{equation}
\item If $\varkappa > 0$, we have for any $g,n,\alpha_1,\ldots,\alpha_n,m_1,\ldots,m_n$
\begin{equation}
\label{vanishkap1}
\sum_{i = 1}^{n} \frac{[m_i]_+}{s_{\alpha_i}} > 2g - 2 + n + 2\varkappa(g -1 + n) \qquad \Longrightarrow \qquad F_{g,n}\begin{bsmallmatrix} \alpha_1 & \cdots & \alpha_n \\ m_1 & \cdots & m_n \end{bsmallmatrix} = 0.
\end{equation}
\end{enumerate}
Without assuming the regularity of zero-mode shifts, $F_{g,n}$ is a polynomial in positive variables for all $g,n$. 
\end{proposition}

In case (1), we can access the top coefficients (i.e. the ones for which the inequalities in \Cref{p:ineqcase} are equalities) by looking at a certain "top part" of the spectral curve and initial condition. We do not address this question in case (2) or when the zero-mode shifts are not regular, as the resulting description would not be as concise.

\begin{definition}[Top part of a spectral curve]
\label{Stopdef}Let $\mathcal{S}  = (C,x,\omega_{0,1},\bm{\nu},\bm{T})$ be a coloured spectral curve. We define the top part of $\mathcal{S}$ to be the coloured spectral curve $\mathcal{S}^{\textnormal{top}} = (C,x,\omega_{0,1}^{\textnormal{top}},\bm{\nu},\bm{T}^{\textnormal{top}})$ with
\[
\omega_{0,1}^{\textnormal{top}}(z) = -\tau_{\alpha} z^{s_{\alpha} - 1}\dd z,\qquad T^{(i),\textnormal{top}}_{\alpha} = T_{\alpha;\frac{i}{2}}^{(i)} \hbar^i
\]
for all $\alpha \in \Ram$, $z \in C_{\alpha}$ and $i \in [r_{\alpha}]$.
\end{definition}

\begin{proposition}[Top part of the free energies]
\label{p:eqcase}
Let $\mathcal{S}$ be an admissible coloured spectral curve and $Z$ be the partition function of its partial Airy ideal with initial condition $\hat{Z}$. Assume that the zero-mode shifts are regular. Assume that $\varkappa = 0$ or that $\hat{Z}$ does not depend on positive variables. Denote by $Z^{\textnormal{top}}$ the partition function with free energies
\[
F_{g,n}^{\textnormal{top}}\begin{bsmallmatrix} \alpha_1 & \cdots & \alpha_n \\ m_1 & \cdots & m_n \end{bsmallmatrix} = \left\{\begin{array}{lll} F_{g,n}\begin{bsmallmatrix} \alpha_1 & \cdots & \alpha_n \\ m_1 & \cdots & m_n \end{bsmallmatrix} & \quad & \textnormal{if}\,\,m_i \geq 0\,\,\textnormal{for all}\,\,i\,\,\textnormal{and}\,\,\sum_{i = 1}^{n} \frac{m_i}{s_{\alpha_i}} = 2g - 2 + n, \\ 0 & \quad & \textnormal{otherwise}. \end{array}\right.
\]
Then, $Z^{\textnormal{top}}$ is the partition function of the partial Airy ideal associated with $\mathcal{S}^{\textnormal{top}}$ and initial condition $\hat{Z}^{\textnormal{top}} \coloneqq 1$.
\end{proposition}
The results of \Cref{p:ineqcase,p:eqcase} are well-known for canonically coloured spectral curves (i.e. $\varkappa = 0$) and no zero-mode shifts.

\begin{proof}[Proof of \Cref{p:ineqcase}]
We introduce a grading in $\widehat{\mathcal{D}}^{\hbar}_{\Ram \times \Z}$, by declaring that
\begin{equation}
\label{degedef} \forall (\alpha,m) \in \Ram \times \mathbb{Z},\qquad  \deg^+ \partial_{\alpha,m} = \deg^+ t_{\alpha,-m} = -\frac{m}{s_{\alpha}}\delta_{m \geq 0} \qquad \textnormal{and}\qquad  \deg^+ \hbar = -1.
\end{equation}
We also introduce the increasing filtration $\set{\mathcal{F}_N^+ |N \in \mathbb{Q}}$ where the $N$-th piece is spanned by homogeneous terms of $\deg^+ \leq N$. Our starting observation is that
\begin{equation}
\label{linearfu} \Big(\hbar \partial_{\alpha,m} + \hbar |m|t_{\alpha,|m|}\delta_{m<0} + F_{0,1}\begin{bsmallmatrix} \alpha \\ m \end{bsmallmatrix} \delta_{m + s_{\alpha} \leq 0}\Big) \in \mathcal{F}^+_{-\frac{m}{s_{\alpha}} - 1}.
\end{equation}
These are the building blocks for the operators $H_{\alpha,k}^{(i)}$ of \eqref{eq:HikN}, which then satisfy
\[
\forall \alpha \in \Ram, \qquad \forall k \geq \kappa_{\alpha}^{(i)},\qquad H_{\alpha,k}^{(i)} \in \mathcal{F}^+_{-i - \frac{r_{\alpha}}{s_{\alpha}}k}.
\]

Let us assume regularity of the zero-mode shifts. This means that $T_{\alpha}^{(i)} \in \mathcal{F}_{-i}^+$ and thus the operators $H_{\alpha,k}^{(i)} - \delta_{k,0} T_{\alpha}^{(i)}$ generating the partial Airy ideal also belong to $\mathcal{F}_{-i - \frac{r_{\alpha}}{s_{\alpha}}k}^+$. The $\hbar$-degree $1$ term of highest $\deg^+$ degree is proportional to $\hbar \partial_{\alpha,m}$ with $m = \Pi_{r_{\alpha},s_{\alpha}}(i,k) = kr_{\alpha} + (i - 1)s_{\alpha}$: it has degree $-i-\frac{r_{\alpha}}{s_{\alpha}}k$. This term is "inverted" to convert the differential equation $(H_{\alpha,k}^{(i)} - T^{(i)}_{\alpha}\delta_{k,0})\cdot Z = 0$ into a recursion on $2g - 2 + n$ for the free energies $F_{g,n}$. Therefore, the recursion operators (the $H$-vertices in \Cref{Sec3}) preserve the filtration $\mathcal{F}^+$. Accordingly, if $\hat{F} \in \mathcal{F}^+_0$ then we must have $F \in \mathcal{F}^+_0$ as well. The latter property is equivalent to the vanishing property \eqref{vanishskap0}.

If $\varkappa = 0$ or if $\hat{Z}$ does not depend on positive variables, then we do have  $\hat{F} \in \mathcal{F}^+_0$, thus proving (1). If $\varkappa > 0$ and $\hat{Z}$ depends on positive variables, we only know a priori that $\hat{F}_{g,n} \in \mathcal{F}^+_{\varkappa n}$ for all $g,n$. Let us examine the implications for $F_{g,n}$, using its expression in \Cref{prop:gencolu}. The details of this expression do not matter, the only important features being that the $H$-vertices act as operations preserving the filtration $\mathcal{F}^+$, that each $\hat{F}_{h,p}$-vertex can increase $\deg^+$ by $\varkappa p$, and that $\hbar$-degrees of vertices sum up to $2g - 2 + n$. We distinguish three types of terms in the expression for $F_{g,n}$:
\begin{itemize}
\item The unique term $\hat{F}_{g,n}$ (no H-vertex): it belongs to $\mathcal{F}^+_{\varkappa n}$.
\item The terms without $\hat{F}$-vertices: they belong to $\mathcal{F}^+_0$.
\item The terms having at least one H-vertex and one $\hat{F}$-vertex: they belong to $\mathcal{F}^+_{\varkappa P}$ where $P$ is the total number of incoming half-edges on $\hat{F}$-vertices. The maximum value of $P$ is $2(g - 1 + n)$: it arises for instance when $H$-vertices form a graph of genus $0$ with one incoming leaf and $2g + n - 1$ outgoing leaves, and each of them is connected to $2g$ $\hat{F}_{\frac{1}{2},1}$ vertices and to $n - 1$ $\hat{F}_{0,2}$-vertices, so that each of the $\hat{F}_{0,2}$ vertices has one incoming leaf (not connected to anything).
\end{itemize}
Therefore $F_{g,n} \in \mathcal{F}^+_{N_{g,n}}$ with $N_{g,n} = \max\set{\varkappa n,0,2\varkappa(g - 1 + n)} = 2\varkappa(g - 1 + n)$: this is the vanishing property \eqref{vanishkap1}.

Last, let us review what happens if the assumption of regularity of $\bm{T}$ is dropped. We only know that $T_{\alpha}^{(i)} = O(\hbar)$,  so the zero-mode shift in the operators of the partial Airy ideal belongs to $\mathcal{F}^+_{-i + (i - 1)}$ instead of $\mathcal{F}^+_{-i}$. In the sum over graphs in \Cref{prop:gencolu}, this leads us to distinguish two types of H-vertices: the usual ones coming from $H_{\alpha,k}^{(i)}$, which preserve the filtration $\mathcal{F}^+$; and the ones coming from the zero-mode shifts ($\bm{T}$-vertices), which have a single incoming half-edge that is either a leaf or connected to an outgoing half-edge of a graph made of $H$-vertices, and can increase $\deg^+$ by an amount up to $r_{\max} \coloneqq \max\set{r_{\alpha} - 1|\alpha \in \Ram}$. In any case, the total number of $\hat{F}$-vertices and $\bm{T}$-vertices is bounded by a constant that depends only on $g$ and $n$. Hence $F_{g,n}$ must belong to $\mathcal{F}^+_{N_{g,n}}$ for some $N_{g,n}$. The optimal value of $N_{g,n}$ depends on the relative value of $r_{\max}$ and $\varkappa$. Nonetheless, $F_{g,n}$ can only involve a finite number (depending on $g$ and $n$) of positive variables, and thus is a polynomial in the positive variables.
\end{proof} 

\begin{proof}[Proof of \Cref{p:eqcase}] The homogeneous piece of maximal $\deg^+ = -\frac{m}{s_{\alpha}} - 1$ in the operator of \eqref{linearfu} is obtained by discarding the negative derivatives and keeping only the coefficient
\[
F_{0,1}\begin{bsmallmatrix} \alpha \\ -s_{\alpha} \end{bsmallmatrix} = -\tau_{\alpha}.
\]
Keeping only those homogeneous pieces in the formula \eqref{eq:overline}-\eqref{eq:HikN} for the operators $H_{\alpha,k}^{(i)} - \hbar^iT_{\alpha;\frac{i}{2}}^{(i)}\delta_{k,0}$ generating the partial Airy ideal associated to $\mathcal{S}$, we get a set of homogeneous operators that we call $\bm{H}^{\textnormal{top},+}$. If $\varkappa = 0$ or if $\hat{Z}$ does not depend on positive variables, we know that $\hat{F} \in \mathcal{F}_0^+$. It follows that the free energies $F$ such that $\exp(F)$ is annihilated by $\bm{H}^{\textnormal{top},+}$ are in $\mathcal{F}_0^+$. Then, if we compute modulo the terms of subleading $\deg^+$, the maximal $\deg^+ = 0$ part of $Z$ --- which is the definition of $Z^{\textnormal{top}}$ --- is also annihilated by $\bm{H}^{\textnormal{top},+}$ and corresponds to the initial condition $\hat{Z}^{\textnormal{top}} = 1$.

The operators in $\bm{H}^{\textnormal{top},+}$ coincide with the generators of the partial Airy ideal for $\mathcal{S}^{\textnormal{top}}$ (we call $\bm{H}^{\textnormal{top}}$ this generating set) in which one discards negative derivatives. Since there are no negative variables in the operators of $\bm{H}^{\textnormal{top}}$, the operators in $\bm{H}^{\textnormal{top},+}$ and $\bm{H}^{\textnormal{top}}$ annihilate the same partition function provided we take a trivial initial condition for the latter. This shows that $Z^{\textnormal{top}}$ is indeed the partition function for $\mathcal{S}^{\textnormal{top}}$ with trivial initial condition.
\end{proof}

We finish by computing the free energies in $\hbar$-degree $0$ and $1$ for the spectral curve $\mathcal{S}^{\textnormal{top}}$ of \Cref{Stopdef}.

\begin{definition}[$3$-point functions] Recall the map $\Pi_{r,s}$ from \Cref{d:map}.
If $m_l = \Pi_{r_{\alpha},s_{\alpha}}(i_l,k_l)$, we denote
\[
c_{\alpha}[m_1,m_2,m_3] = \tau_{\alpha}^{-1}(i - 1 - r_{\alpha} \delta_{i_1 + i_2,i})
\]
We denote $c_{\alpha}[m_1,m_2|m_3]$ the same expression where $m_3$ is rather written as $m_3 = - \Pi_{r_{\alpha},s_{\alpha}}(i_3,k_3)$. Likewise, $c_{\alpha}[m_1|m_2,m_3]$ is the same expression when we rather write $m_l = -\Pi_{r_{\alpha},s_{\alpha}}(i_l,k_l)$ for $l \in \set{2,3}$.
\end{definition}
When all indices are spectator, the free energies are simply given by the reduced free energies, which are initial conditions. Thus we only discuss the free energies with at least one actor index.
\begin{proposition}
Let $(m,\alpha) \in A$ written as $m = \Pi_{r_{\alpha},s_{\alpha}}(i,k)$. Let $(m_l,\alpha_l) \in I$ for $l \in \set{1,2}$. For the spectral curve $\mathcal{S}^{\textnormal{top}}$,  in $\hbar$-degree $0$ the only contributions come from the blobs and zero-mode shifts, i.e.
\begin{equation}
F_{0,2}\begin{bsmallmatrix} \alpha & \alpha_1 \\ m & m_1 \end{bsmallmatrix} = 0, \qquad
F_{\frac{1}{2},1}\begin{bsmallmatrix} \alpha \\ m \end{bsmallmatrix} = \delta_{m,0} T_{\alpha;\frac{1}{2}}^{(1)}.
\end{equation}
and in $\hbar$-degree $1$ we have
\begin{equation*}
\begin{split}
F_{0,3}\begin{bsmallmatrix} \alpha & \alpha_1 & \alpha_2 \\ m & m_1 & m_2 \end{bsmallmatrix} & = c_{\alpha}[m|m_1,m_2] \delta_{\alpha,\alpha_1,\alpha_2} m_1m_2 \delta_{m_1 > 0} \delta_{m_2 > 0} \delta_{m + m_1 + m_2,s_{\alpha}} \\
& \quad + c_{\alpha}[m,m + m_2 - s_{\alpha}|m_2] \hat{F}_{0,2}\begin{bsmallmatrix} \alpha_1 & \alpha \\ m_1 & m + m_2 - s_{\alpha}\end{bsmallmatrix} m_2 \delta_{m_2 > 0} \\
& \quad + c_{\alpha}[m,m + m_1 - s_{\alpha}|m_1]\hat{F}_{0,2}\begin{bsmallmatrix} \alpha_2 & \alpha \\ m_2 & m + m_1 - s_{\alpha} \end{bsmallmatrix} m_1 \delta_{m_1 > 0} \\
& \quad + \sum_{\substack{\hat{m}_1,\hat{m}_2 \in \mathbb{Z} \\ \hat{m}_1 + \hat{m}_2  + s_{\alpha} = m}} c_{\alpha}[m,\hat{m}_1,\hat{m}_2] \hat{F}_{0,2}\begin{bsmallmatrix}  \alpha_1 & \alpha \\ m_1 & \hat{m}_1 \end{bsmallmatrix} \hat{F}_{0,2}\begin{bsmallmatrix} \alpha_2 & \alpha \\ m_2 & \hat{m}_2 \end{bsmallmatrix}, \\
F_{\frac{1}{2},2}\begin{bsmallmatrix} \alpha & \alpha_1 \\ m & m_1 \end{bsmallmatrix} & = T_{\alpha;\frac{1}{2}}^{(1)} c_{\alpha}[m,0|m_1] \delta_{\alpha,\alpha_1},\\
F_{1,1}\begin{bsmallmatrix} \alpha \\ m \end{bsmallmatrix} & = \left(\frac{r_{\alpha}^2 - 1}{24} + r_{\alpha}T_{\alpha;1}^{(2)}- \frac{r_{\alpha} - 1}{2}(T_{\alpha;\frac{1}{2}}^{(1)})^2\right)\frac{\delta_{m,s_{\alpha}}}{\tau_{\alpha}} + \frac{1}{2} \sum_{\substack{\hat{m}_1, \hat{m}_2 \in \mathbb{Z} \\  \hat{m}_1 + \hat{m}_2 + s_{\alpha} = m}} c_{\alpha}[m,\hat{m}_1,\hat{m}_2]\hat{F}_{0,2}\begin{bsmallmatrix} \alpha & \alpha \\ \hat{m}_1 & \hat{m}_2 \end{bsmallmatrix} .
\end{split}
\end{equation*}
\end{proposition}
\begin{proof}
We need the expression of $H_{\alpha,k}^{(i)}$ up to $O(\hbar^3)$. We know that the $\hbar^0$-term is absent, and the $\hbar^1$-term was computed in the proof of \Cref{t:PAISCc}. What remains is to compute the $\hbar^2$-term. Using \eqref{eq:overline}-\eqref{eq:HikN} we find, with the coding $m = \Pi_{r_{\alpha},s_{\alpha}}(i,k)$, that $H_{\alpha,k}^{(i)}$ is equal to
\begin{equation}
\begin{split}
& \quad \bigg(\frac{\tau_{\alpha}}{r_{\alpha}}\bigg)^{i - 1} \hbar \partial_{\alpha,m} - \hbar^2 \frac{\tau_{\alpha}^{i - 2}}{r_{\alpha}^{i - 1}} \left(\frac{r_{\alpha}^2 - 1}{24} + r_{\alpha} T_{\alpha;1}^{(2)}\right) \delta_{m,s_{\alpha}} + O(\hbar^3) \\
& \quad + \hbar^2 (-\tau_{\alpha})^{i -2} \frac{i(i - 1)}{2r_{\alpha}^i} \sum_{m_1,m_2 \in \mathbb{Z}} \Psi_{r_{\alpha}}^{(i,0)}(\underbrace{-s_{\alpha},\ldots}_{i - 2\,\,\textnormal{times}},m_1,m_2) : \prod_{l = 1}^{2} \big(\partial_{\alpha,m_l} + \delta_{m_l < 0} |m_l|t_{\alpha,|m_l|}\big) : \delta_{-s_{\alpha}(i - 2) + m_1 + m_2,m}.
\end{split}
\end{equation}
Notice that for $m = s_{\alpha}$ we have $i = 2$.  In \Cref{evallem}, using the fact that $r_{\alpha}$ and $s_{\alpha}$ are coprime, we can get non-zero contributions only when each subset $L_a$ contain $m_1$ or $m_2$, that is, either $\mathbf{L} = \set{[i]}$ or $\mathbf{L}$ has two elements of cardinality $j_1 + 1$ and $j_2 + 1$, the first one containing $m_1$ and the second one containing $m_2$.
\begin{equation*}
\begin{split}
& \quad \Psi_{r_{\alpha}}^{(i,0)}(\underbrace{-s_{\alpha},\ldots}_{i - 2\,\,\textnormal{times}},m_1,m_2) \\
& = \frac{1}{i!} \left((-1)^{i - 1}(i - 1)! r_{\alpha} +  (-1)^{i - 2}  r_{\alpha}^2 \sum_{j_1 + j_2 = i - 2}  \frac{(i - 2)!}{j_1!j_2!}\cdot  j_1! j_2! \delta_{r_{\alpha}|-s_{\alpha}j_1 + m_1}\delta_{r_{\alpha}|-s_{\alpha}j_2 + m_2}\right) \\
& = (-1)^{i - 1}r_{\alpha} \frac{(i-2)!}{i!}\left(i - 1 - r_{\alpha} \sum_{j_1 + j_2 = i - 2} \delta_{r_{\alpha}|-s_{\alpha}j_1 + m_1} \delta_{r_{\alpha}|-s_{\alpha}j_2 + m_2}\right).
\end{split}
\end{equation*}
Since $r_{\alpha}$ and $s_{\alpha}$ are coprime, there is at most one non-zero term in the last sum, which is obtained for $j_l = i_l - 1$ when $m_l = \Pi_{r_{\alpha},s_{\alpha}}(i_l,k_l)$ for some $k_l \in \mathbb{Z}$. Setting
\[
c_{\alpha}[m,m_1,m_2] =  -\frac{i(i - 1)}{r_{\alpha}\tau_{\alpha}} \Psi_{r_{\alpha}}^{(i,0)}(\underbrace{-s_{\alpha},\ldots}_{i - 2\,\,\textnormal{times}},m_1,m_2)
\]
and
\[
c_{\alpha}[m,m_1|m_2] = c_{\alpha}[m,m_1,-m_2],\qquad c_{\alpha}[m|m_1,m_2] = c_{\alpha}[m,-m_1,-m_2]
\]
we then get
\begin{equation*}
\begin{split}
\bigg(\frac{r_{\alpha}}{\tau_{\alpha}}\bigg)^{i - 1} H_{\alpha,k}^{(i)} & = \hbar \partial_{\alpha,m} - \frac{\hbar^2}{2} \!\!\!\!\sum_{\substack{m_1,m_2 \in \mathbb{Z} \\ m_1 + m_2 + s_{\alpha} = m}} \!\!\! c_{\alpha}[m,m_1,m_2]\partial_{\alpha,m_1}\partial_{\alpha,m_2} -  \hbar^2\!\!\!\! \sum_{\substack{m_1 > 0 \\ m_2 \in \mathbb{Z} \\ m_2 + s_{\alpha} = m + m_1}} \!\!\!\! c_{\alpha}[m,m_2|m_1] m_1 t_{\alpha,m_1}\partial_{\alpha,m_2} \\
& \quad - \frac{\hbar^2}{2}\!\!\! \sum_{\substack{m_1,m_2 > 0 \\ s_{\alpha} = m + m_1 + m_2}} \!\!\! c_{\alpha}[m|m_1,m_2] m_1m_2 t_{\alpha,m_1}t_{\alpha,m_2} - \hbar^2\left(\frac{r_{\alpha}^2 - 1}{24} + r_{\alpha} T_{\alpha;1}^{(2)}\right) \frac{\delta_{m,s_{\alpha}}}{\tau_{\alpha}}.
\end{split}
\end{equation*}
The announced formulae then follow from \Cref{Sec3}, see in particular \Cref{fig:02graph,fig:recinicheck}.
\end{proof}

\vspace{0.5cm}

\section{Coloured loop equations and topological recursion}
\label{Sec7}

\subsection{Correspondence between coloured loop equations and partial Airy ideals}
\label{Sec71} 
We now define the notion of ``coloured loop equations'', which will be the equations satisfied by the system of correlators associated to the partition functions of the partial Airy ideals of \Cref{t:PAISCc}.

The proofs in \Cref{Sec51} were insensitive to the splitting of variables into actor and spectator (the splitting of fields in \Cref{d:fields} was coming from the normal ordering, not from the distinction actor/spectator). Therefore, proofs work in the same way for coloured loop equations (the only modification being the range of $k$ for the generators of the partial Airy ideal), so we will be brief.

We define systems of correlators $\bm{\omega}$ on $\mathcal{S}$ as in \Cref{d:sofc}; the colouring does not play a role, but it is now essential that this definition allowed poles for $\omega_{0,2}$ at the ramification divisor (outside the diagonal).

\begin{definition}[Coloured loop equations]\label{d:lec}
A system of correlators $\bm{\omega}$ on a coloured spectral curve $\mathcal{S}$ satisfies \emph{coloured loop equations} if, for all $\alpha \in \Ram$, $i \in [r_{\alpha}]$, and $2g-2+(1+n) > 0$
  \begin{equation}
    \mathcal{E}_{\alpha;g,n}^{(i)}(z;\bm{z}_{[n]}) - \delta_{n,0} T_{\alpha;g}^{(i)} \left(\frac{r_{\alpha} \dd z}{z}\right)^i = O\big( z^{r_\alpha \nu^{\alpha}[i]} \big) \left(\frac{\dd z}{z}\right)^i \qquad \text{as $z \rightarrow 0$}.
  \end{equation}
\end{definition}

Those are tailored to match with the partial Airy ideals constructed in the previous section. Indeed, we obtain the following theorem, which is analogous to \Cref{t:LEPAI}.

\begin{theorem}\label{t:LEPAIc}
Let $\mathcal{S}$ be a coloured spectral curve and $\bm{\omega}$ a system of correlators on $\mathcal{S}$. 
Using the expansion \eqref{eq:expcorr} for the correlators, we construct a partition function
\begin{equation}\label{eq:Zpfc}
  Z = \exp \left( \sum_{ \substack{g,n \\ 2g-2 + n \geq 0}} \frac{\hbar^{2g-2+n}}{n!} \sum_{\substack{\alpha_1, \ldots, \alpha_n \in \Ram \\ m_1, \ldots, m_n \in \mathbb{Z}}} F_{g,n} \begin{bsmallmatrix} \alpha_1 & \ldots & \alpha_n \\ m_1 & \ldots & m_n \end{bsmallmatrix} \prod_{j = 1}^{n} t_{\alpha_j,m_j}  \right).
\end{equation} 
Then $\bm{\omega}$ satisfies the coloured loop equations from \Cref{d:lec} if and only if $\mathcal{I}_{\mathcal{S}} \cdot Z = 0$, where $\mathcal{I}_{\mathcal{S}}$ is the partial Airy ideal attached to $\mathcal{S}$ via \Cref{t:PAISCc}.
\end{theorem}

\begin{proof}
The proof is essentially the same as for \Cref{t:LEPAI}. The coloured loop equations from \Cref{d:lec} are satisfied if and only if the following residue constraints hold: for all $\alpha \in \Ram$, $i \in [r_\alpha]$, $2g-2+(1+n) > 0 $ and  $k \geq 1- \nu^{\alpha}[i]$,
\begin{equation}
\label{Eresloop}
 \Res_{z=0} \frac{z^{r_\alpha k + i - 1}}{(\dd z)^{i-1}} \left(\mathcal{E}_{\alpha;g,n}^{(i)}(z;\bm{z}_{[n]})- \delta_{n,0}\delta_{k,0} r_{\alpha}^i T_{\alpha;g}^{(i)}\right) = 0.
 \end{equation}
Using \Cref{l:simplephi} and \Cref{d:projRWA}, we can rewrite these constraints as
\begin{equation}
 \Res_{z=0} \frac{z^{r_\alpha k + i - 1}}{(\dd z)^{i-1}} \Ev_{g,n}\left[  \overline{Z}^{-1} \Big(\overline{W}_{\alpha}^{(i)}(z) - T_{\alpha}^{(i)}\big(\tfrac{r_{\alpha} \dd z}{z}\big)^i\Big) \overline{Z} \cdot 1 \right](\bm{z}_{[n]}) = 0.
\end{equation}
where $T_{\alpha}^{(i)} \coloneqq \sum_{g} T_{\alpha;g}^{(i)}\hbar^{2g}$. A simple rewriting of the argument in \Cref{l:Zres} then concludes the proof.
\end{proof}

To state an existence and uniqueness theorem for solutions of coloured loop equations, we need to generalize the notion of blobs, as those may not be given by holomorphic differentials anymore. Instead, we want our blobs to be meromorphic  differentials whose Laurent series expansion contains terms associated to spectator  indices.

\begin{definition}[Coloured blobs]
Let $\mathcal{S}$ be a coloured spectral curve. A \emph{system of coloured blobs} on $\mathcal{S}$ is a collection
\[
\bm{\phi} = \Set{\phi_{g,n}\,\,|\,\,g  \in  \tfrac{1}{2} \N \,\,\textnormal{and}\,\,n \in \N\,\,\textnormal{such that}\,\,2g - 2 + n \geq 0},
\]
where each $\phi_{g,n}(\bm{z}_{[n]})$ is a meromorphic symmetric $n$-differential on $C^n$ which can have poles only at $z_i \in \Ram$ for each $i \in [n]$. We require that its Laurent series expansion  takes the form
\[
\phi_{g,n}(z_1,\ldots,z_n) = \sum_{(\alpha_1,m_1), \ldots, (\alpha_n,m_n) \in S}
\phi_{g,n}\begin{bsmallmatrix} \alpha_1 & \cdots & \alpha_n \\ m_1 & \cdots & m_n \end{bsmallmatrix} \prod_{i = 1}^{n} z_i^{-m_i - 1}\dd z_i 
\]
when $z_i$ approaches the ramification point in $C_{\alpha_i}$ for each $i \in [n]$. The spectator set $S$ was defined in \Cref{t:PAISCc}.
\end{definition}
\Cref{d:blobs} corresponds to blobs with the actor set $A = \Ram \times \mathbb{N}$ and spectator set $S = \Ram \times (- \mathbb{N}^*)$, making the blobs $\phi_{g,n}$ holomorphic. Clearly, our definition of blobs is tailored to match with the initial condition of the partial Airy ideals in \Cref{t:PAISCc}. We define compatibility of correlators and blobs as expected.

\begin{definition}[Coloured compatibility]\label{d:compatibilityc}
Let $\mathcal{S}$ be a coloured spectral curve. A system of coloured blobs $\bm{\phi}$ is compatible with a system of correlators $\bm{\omega}$ if $\omega_{g,0} = \phi_{g,0}$ for $g \geq 1$, and 
\[
\omega_{\frac{1}{2},1}(z) = \phi_{\frac{1}{2},1}(z) \qquad \textnormal{and}\qquad  \omega_{0,2}(z_1,z_2) = \omega_{0,2}^{\textnormal{std}}(z_1,z_2) + \phi_{0,2}(z_1,z_2),
\]
and for all $g \in \frac{1}{2}\N$ and $n \in \N^*$ such that $2g-2+n \geq 0$, for all $(\alpha_1,m_1),\ldots,(\alpha_n,m_n) \in S$, we have
\begin{equation}
F_{g,n} \begin{bsmallmatrix} \alpha_1 & \ldots & \alpha_n \\ m_1 & \ldots & m_n \end{bsmallmatrix} = \phi_{g,n} \begin{bsmallmatrix} \alpha_1 & \ldots & \alpha_n \\ m_1 & \ldots & m_n \end{bsmallmatrix}.
\end{equation}
\end{definition}
The following theorem is analogous to \Cref{t:EU}.
\begin{theorem}\label{t:EUc}
Let $\mathcal{S}$ be an admissible coloured spectral curve, and fix a system of coloured blobs $\bm{\phi}$ on $\mathcal{S}$. Then there exists a unique system of correlators $\bm{\omega}$ that satisfies the coloured loop equations and is compatible with $\bm{\phi}$. All solutions of the coloured loop equations are obtained in this way.
\end{theorem}
\begin{proof}
The proof is the same as for \Cref{t:EU}, with the spectator set $\Ram \times \Z_{< 0}$ being replaced by the set $S$ of \Cref{t:PAISCc}.
\end{proof}

\subsection{Coloured (blobbed) topological recursion}

We now derive a residue formula for the solutions of the coloured loop equations, as we did in \Cref{Sec:55} for the loop equations. We start by defining a coloured projection property, which singles out a particular solution of the coloured loop equations, and write an explicit topological recursion formula for this solution. We call the resulting formula "coloured topological recursion". We then formulate a topological recursion formula for general solutions of the coloured loop equations, which we call "coloured blobbed topological recursion".

\begin{definition}[Coloured standard bidifferential]
The \emph{coloured standard bidifferential} $\omega_{0,2}^{\textnormal{std},\bm{\nu}}$ is zero on $C_{\alpha_1} \times C_{\alpha_2}$ for distinct $\alpha_1,\alpha_2 \in \Ram$, while for $\alpha \in \Ram$ and $z_1,z_2 \in C_{\alpha}$ it is defined by the expansion for $|z_1| < |z_2|$
\[
\omega_{0,2}^{\textnormal{std},\bm{\nu}}(z_1,z_2)  = \sum_{m \in A_{r_{\alpha},s_{\alpha}}(\nu^{\alpha})} m z_1^{m - 1}z_2^{-(m + 1)}\dd z_1\dd z_2  =  \omega_{0,2}^{\std}(z_1,z_2) - \sum_{m \in S_{r_{\alpha},s_{\alpha}}(\nu^{\alpha}) \cap \N} m z_1^{m - 1} z_2^{-(m + 1)}\dd z_1\dd z_2
\]
where the second sum has finitely many terms (see \Cref{lem:N}). Unlike in $\omega_{0,2}^{\std}$, the variables $z_1$ and $z_2$ do not play a symmetric role in $\omega_{0,2}^{\std,\bm{\nu}}$.
\end{definition}
\begin{definition}[Coloured projection property]
A system of correlators $\bm{\omega}$ on a coloured spectral curve satisfies the \emph{coloured projection property} if it is compatible with zero blobs and has no residues at ramification points. Equivalently, the zero-mode shifts $T_{\alpha}^{(i)}$ vanish for all $\alpha,i$, we have
\[
\omega_{\frac{1}{2},1} = 0 \qquad \textnormal{and}\qquad \omega_{0,2} = \omega_{0,2}^{\textnormal{std}}
\]
and for $2g - 2 + (1+n) > 0$,
\begin{equation}
\label{colouredpro}
\omega_{g,1+n}(z_0,\bm{z}_{[n]}) = \sum_{\alpha \in \Ram} \Res_{z = \alpha} \bigg(\int_{\alpha}^{z} \omega_{0,2}^{\textnormal{std},\bm{\nu}}(\cdot,z_0)\bigg) \omega_{g,1+n}(z,\bm{z}_{[n]}).
\end{equation}
\end{definition}

\begin{theorem}[Coloured topological recursion]
\label{CTR0} Let $\mathcal{S}$ be an admissible coloured spectral curve. There exists a unique system of correlators $\bm{\omega}$ solving the coloured loop equations and satisfying the coloured projection property. It obeys the following recursion, for $2g - 2 + (1+ n) > 0$:
\[
\omega_{g,1+n}(z_0,\bm{z}_{[n]}) = \sum_{\alpha \in \Ram} \Res_{z = \alpha} \sum_{i = 2}^{r_{\alpha}} \,\sum_{\substack{\Xi \subseteq \fiber_{\alpha}'(z) \\ |\Xi| = i}} K_{\alpha}^{(i,\bm{\nu})}(z_0,z,\Xi) \mathcal{W}^{(i)\,\prime}_{\alpha;g,n}(z,\Xi;\bm{z}_{[n]}),
\]
with the coloured recursion kernels given for $(z_0,z) \in C \times C_{\alpha}$ and $\Xi \in \textnormal{Sym}^{i - 1}C_{\alpha}$ by
\[
K_{\alpha}^{(i,\bm{\nu})}(z_0,z,\Xi) =  \frac{-\int_{\alpha}^{z} \omega_{0,2}^{\textnormal{std},\bm{\nu}}(\cdot,z_0)}{\prod_{z' \in \Xi} (\omega_{0,1}(z') - \omega_{0,1}(z))} = \frac{ -\sum_{m \in A_{r_{\alpha},s_{\alpha}}(\nu^{\alpha}) \setminus \set{0}} z^{m} z_0^{-(m + 1)} \dd z_0}{\prod_{z' \in \Xi} (\omega_{0,1}(z') - \omega_{0,1}(z))}.
\]
\end{theorem}
 The only difference with the usual topological recursion of \Cref{t:BBCCN} is that we have to use the coloured standard differential $\omega_{0,2}^{\std,\bm{\nu}}$ (instead of $\omega_{0,2}$) in the recursion kernel.
\begin{proof} This is a simple adaptation of the proof of \Cref{t:BBCCN} given in \cite{BBCCN24}: we solve the coloured loop equation using \eqref{colouredpro} instead of the usual projection property \eqref{eq:pp}.
\end{proof}

Contrarily to the non-coloured case where \Cref{t:BS} is available, it does not seem possible to express the correlators of \Cref{t:EUc} solving coloured loop equations and compatible with arbitrary coloured blobs $\bm{\phi}$ as a function of the correlators solving loop equations and compatible with zero blobs (given by \Cref{CTR0}) and of the blobs themselves. Instead, we express the recursion described in \Cref{Sec31} in terms of correlators. First, we define the projectors on the spectator part. 
 
 \begin{definition}[Spectator projector]
\label{d:projector} Let $\mathcal{S}$ be a coloured spectral curve. Given a meromorphic $1$-form $\eta$ on $C$, we define new meromorphic $1$-form
 \[
 \mathcal{P}^{S}[\eta(z_1)] = \eta(z_1) -  \sum_{\alpha \in \Ram} \Res_{z = \alpha} \bigg(\int_{\alpha}^{z} \omega_{0,2}^{\textnormal{std},\bm{\nu}}(\cdot,z_1)\bigg) \eta(z)  - \big(\Res_{z = \alpha_1} \eta(z)\big) \frac{\dd z_1}{z_1},
 \]
when $z \in C_{\alpha_1}$ and $\alpha_1 \in \Ram$.

We write $\mathcal{P}^{S}_{z_j}$ if we apply this projector on the $j$-th variable of a meromorphic $n$-form on $C^n$ with poles only at the ramification divisor. To apply the projector on the second variable of a bidifferential on $C^2$ having a double pole on the diagonal (besides possible poles at the ramification divisor), it is sufficient to define for $z_1,z_2 \in C_{\alpha}$ with $|z_1|< |z_2|$.
\begin{equation}
\label{projnu02}
\mathcal{P}^{S}_{z_2}[\omega^{\textnormal{std}}_{0,2}(z_1,z_2)] = \omega_{0,2}^{\textnormal{std}}(z_1,z_2) - \omega_{0,2}^{\textnormal{std},\bm{\nu}}(z_1,z_2) = \sum_{m \in S_{r_{\alpha},s_{\alpha}}(\nu^{\alpha}) \cap \mathbb{N}} mz_1^{m - 1}z_2^{-(m + 1)} \dd z_1 \dd z_2.
\end{equation}
\end{definition}

If $\eta$ has the Laurent series expansion
\[
\eta(z) = \sum_{m \in \mathbb{Z}} \eta\begin{bsmallmatrix} \alpha \\ m  \end{bsmallmatrix} z^{-m-1}\dd z \qquad \textnormal{as}\,\,z \rightarrow \alpha,
\]
the projection $\mathcal{P}^{S}[\eta(z)]$ is characterized by the Laurent series expansion
 \[
 \mathcal{P}^{S}[\eta(z)] = \sum_{m \in S_{r_{\alpha},s_{\alpha}}(\nu^{\alpha})} \eta\begin{bsmallmatrix} \alpha_1 \\ m  \end{bsmallmatrix} z^{-m-1}\dd z \qquad \textnormal{as}\,\,z \rightarrow \alpha.
 \]

 \begin{theorem}[Coloured blobbed topological recursion]
\label{CTR1} Let $\mathcal{S}$ be an admissible coloured spectral curve and $\bm{\phi}$ a system of coloured blobs. The unique system of correlators $\bm{\omega}$ satisfying the coloured loop equations and compatible with $\bm{\phi}$ obeys the following recursion, for $2g - 2 + n > 0$:
 \begin{equation} 
 \label{colouredBTR}
 \begin{split}
 \omega_{g,n}(\bm{z}_{[n]}) & = \sum_{j = 1}^{n} \sum_{\alpha \in \textnormal{Ram}} \Res_{z = \alpha} \sum_{i = 2}^{r_{\alpha}} \sum_{\substack{\Xi \subseteq \fiber_{\alpha}'(z) \\ |\Xi| = i}} K_{\alpha}^{(i,\bm{\nu})}(z_j,z,\Xi) \mathcal{P}^{S}_{z_1} \cdots \mathcal{P}^{S}_{z_{j - 1}}\big[\mathcal{W}_{\alpha;g,n - 1}^{(i)\,\prime}(z,\Xi;\bm{z}_{[n] \setminus \set{j}})\big] \\
 & \quad + \left(T_{\alpha_1;g}^{(1)} \frac{\dd z_1}{z_1} - \sum_{\alpha \in \textnormal{Ram}} \sum_{i = 2}^{r_{\alpha}} T_{\alpha;g}^{(i)} \Res_{z = \alpha} K^{(r_{\alpha},\bm{\nu})}(z_1,\fiber_{\alpha}(z))   (-\omega_{0,1}(z))^{r_{\alpha} - i} \bigg(\frac{r_{\alpha} \dd z}{z}\bigg)^{i}\right)\delta_{n,1}  \\ 
 & \quad + \phi_{g,n}(z_1,\ldots,z_n),
 \end{split}
 \end{equation}
 where $z_1 \in C_{\alpha_1}$. 
 In $\mathcal{W}$, there are some terms involving $\omega_{0,2}(\xi,z_l)$ for some $\xi \in \Xi$: applying $\mathcal{P}^{S}_{z_l}$ is done using \eqref{projnu02} and linearity.
 \end{theorem} 
Formula \eqref{colouredBTR} can be taken as a definition of coloured blobbed topological recursion, the key property being that it defines correlators that are symmetric in all their variables.

\begin{proof}
Take $g,n$ such that $2g - 2 + n > 0$.  The compatibility of correlators with blobs means that the blobs are given by
\begin{equation}
\label{blobterm}
\phi_{g,n}(z_1,\ldots,z_n) = \mathcal{P}^{S}_{z_1}\cdots \mathcal{P}^{S}_{z_n}[\omega_{g,n}(z_1,\ldots,z_n)].
\end{equation}
Due to \Cref{l69}-(1), the multidifferential  $\omega_{g,n}(z_1,\ldots,z_n)$ can have a residue at the ramification divisor only if $n = 1$, in which case the residue at $\alpha$ is equal to $T_{\alpha;g}^{(1)}$. Then, we can write
\[
\omega_{g,n}(z_1,\ldots,z_n) = \delta_{n,1} T_{\alpha_1;g}^{(1)} \frac{\dd z_1}{z_1} + \sum_{\alpha \in \Ram} \Res_{z = \alpha} \bigg(\int_{\alpha}^{z} \omega_{0,2}^{\textnormal{std},\bm{\nu}}(\cdot,z_1)\bigg)\omega_{g,n}(z,z_2,\ldots,z_n) + \mathcal{P}^{S}_{z_1}[\omega_{g,n}(z_1,\ldots,z_n)].
\]
The middle term will be determined by the coloured loop equations in terms of $\omega_{g',n'}$ with $2g' - 2 + n' < 2g - 2 + n$. If $n = 1$, we can stop here. If $n > 1$ we apply the same decomposition successively in the variables $z_2,\ldots,z_n$, without having to worry about residue term, until we arrive to the blob term \eqref{blobterm}. The outcome is
\begin{equation}
\label{toucomePP}
\omega_{g,n}(\bm{z}_{[n]}) = \delta_{n,1} T_{\alpha_1;g}^{(1)} \frac{\dd z_1}{z_1} + \sum_{j = 1}^{r_{\alpha}} \mathcal{P}^{S}_{z_1} \cdots \mathcal{P}^{S}_{z_{j - 1}} \left[ \sum_{\alpha \in \Ram} \Res_{z = \alpha} \bigg(\int_{\alpha}^{z} \omega_{0,2}^{\textnormal{std},\bm{\nu}}(\cdot,z_j)\bigg)\omega_{g,n}(z,\bm{z}_{[n] \setminus \set{j}})\right] + \phi_{g,n}(\bm{z}_{[n]}). 
\end{equation}  

We now use the loop equations to compute the $j$-th term. As in  \cite[Proof of Proposition 5.10 and Equation 83]{BKS24} but using the variable $z_j$ instead of $z_1$ (this is possible because correlators are symmetric in all their variables) we can write
\begin{equation*}
\begin{split}
\Res_{z = \alpha} \bigg(\int_{\alpha}^{z} \omega_{0,2}^{\textnormal{std},\bm{\nu}}(\cdot,z_j)\bigg)\omega_{g,n}(z,\bm{z}_{[n]\setminus\set{j}}) & = \Res_{z = \alpha} K_{\alpha}^{(r_{\alpha},\bm{\nu})}(z_j,\fiber_{\alpha}(z))\bigg(-\sum_{i = 1}^{r_{\alpha}} \mathcal{E}^{(i)}_{\alpha;g,n - 1}(z;\bm{z}_{[n]\setminus \set{j}}) (-\omega_{0,1}(z))^{r_{\alpha} - i}  \\ \quad
& \quad + \sum_{i = 2}^{r_{\alpha}} \sum_{\substack{\Xi \subseteq \fiber_{\alpha}'(z) \\ |\Xi| = i}} K_{\alpha}^{(i,\bm{\nu})}(z_j,z,\Xi) \mathcal{W}_{\alpha;g,n - 1}^{(i)\,\prime}(z,\Xi;\bm{z}_{[n] \setminus \set{j}})\bigg).
\end{split} 
\end{equation*}
The coloured loop equations in the form \eqref{Eresloop} tells us that the only contribution of $\mathcal{E}^{(i)}$ to the residue comes from the $i$-th zero-mode shift if $n - 1 = 0$
\begin{equation}
\begin{split}
 \Res_{z = \alpha} \bigg(\int_{\alpha}^{z} \omega_{0,2}^{\textnormal{std},\bm{\nu}}(\cdot,z_j)\bigg)\omega_{g,n}(z,\bm{z}_{[n]\setminus \set{j}}) & = - \delta_{n,1}  \Res_{z = \alpha} K_{\alpha}^{(r_{\alpha},\bm{\nu})}(z_j,\fiber_{\alpha}(z)) \left( \sum_{i = 1}^{r_{\alpha}}  T_{\alpha;g}^{(i)} \frac{(r_{\alpha}\dd z)^i}{z^i} (-\omega_{0,1}(z))^{r_{\alpha} - i}\right) \\
& \quad + \Res_{z = \alpha}  \sum_{i = 2}^{r_{\alpha}} \sum_{\substack{\Xi \subseteq \fiber_{\alpha}'(z) \\ |\Xi| = i}} K_{\alpha}^{(i,\bm{\nu})}(z_j,z,\Xi) \mathcal{W}_{\alpha;g,n - 1}^{(i)\,\prime}(z,\Xi;\bm{z}_{[n]\setminus\set{j}}).
\end{split}
\end{equation}
Substituting in \eqref{toucomePP} and moving the operators $\mathcal{P}^{S}_{z_1}\cdots \mathcal{P}^{S}_{z_{j - 1}}$ inside the residue, we almost obtain \eqref{colouredBTR}. The only difference is the presence of the $i = 1$ term in the sum of residues involving $T_ {\alpha;g}^{(i)}$. However, using that $s_{\alpha}$ and $r_{\alpha}$ are coprime we can evaluate
\[
K_{\alpha}^{(r_{\alpha},\bm{\nu})}(z_1,\fiber_{\alpha}(z)) \frac{(r_{\alpha}\dd z)^i}{z^i} (-\omega_{0,1}(z))^{r_{\alpha} - i}  = O(z^{-s_{\alpha}(i - 1)})\qquad\textnormal{as}\,\,z \rightarrow \alpha,
\]
Therefore, the $i = 1$  term does not contribute to the residue. 
 \end{proof}

\section{Examples}
\label{Sec8}

\subsection{The simplest truly coloured spectral curves}

We consider the simplest coloured spectral curves that are out of the scope of the standard Eynard--Orantin topological recursion. We take $C = \mathbb{C}$, with $x=z^2$ (i.e. $r = 2$) and $\omega_{0,1} = -z^{s-1}\dd z$ (i.e. $\tau=1$). The usual cases considered (which correspond to the canonical partitions) are $\nu=(1,1)$ with $s=3$ (the Airy case) and $\nu=(2)$ with $s=1$ (the Bessel case). Therefore, the simplest new cases are the following:
\begin{itemize}
\item \textit{Quintic Airy:} $\nu = (1,1)$,  no zero-mode shift is allowed, and $s = 5$. Then $\omega_{0,1} = - z^4\dd z$ and $y = \frac{\omega_{0,1}}{\dd x} = -\frac{1}{2} z^{3}$. We note that $dy$ has a zero at the ramification point $z = 0$ (which is not allowed in the usual topological recursion). Thinking of it as an algebraic curve satisfied by $x$ and $y$, this would correspond to the cuspidal curve $4 y^2 + x^3 = 0$. If we tried applying the topological formula \eqref{usualTR} to this curve, we would get non-symmetric $\omega_{g,n}$ --- instead, we should use the  (blobbed) coloured topological recursion of \Cref{CTR0,CTR1}.
\item \textit{Queer Airy:} $\nu = (2)$, the zero-mode shifts $T^{(1)}$ and $T^{(2)}$ are allowed, and $s = 3$. Then $\omega_{0,1} = - z^2\dd z$ and  $y = \frac{\omega_{0,1}}{\dd x} = -\frac{1}{2} z$. This looks like the standard Airy curve $4 y^2 +  x = 0$, but with a non-canonical choice of partition and extra non-zero shifts.
\end{itemize}
In both cases the spectator set is $S = \mathbb{Z}_{<0} \cup \set{1}$, and the depth is $\varkappa = \frac{1}{s}$. 

For simplicity, we consider blobs which only depend on the single positive spectator variable
\begin{equation}
\label{positini}
\phi_{g,n}(z_1,\ldots,z_n) = \varphi_{g,n} \prod_{i = 1}^{n} \frac{\dd z_i}{z_i^2},
\end{equation}
and assume that they are zero when $g$ is not an integer. For the queer Airy case, we also assume that $T^{(1)} = 0$ while $T^{(2)} = c \hbar^2$. We set $u \coloneqq \varphi_{0,2}$. The outcome is a collection of correlators of the form
\begin{equation}
\label{egounsgsf}
\omega_{g,n}(z_1,\ldots,z_n) = \sum_{k_1,\ldots,k_n \geq 0} F_{g,n}[2k_1+1,\ldots,2k_n+1]\prod_{i = 1}^{n} \frac{\dd z_i}{z_i^{2k_i + 2}},
\end{equation}
with the correlators with non-integer $g$ being zero. The coefficients for $2g - 2 + n \leq 3$ are given in \Cref{s5colour} for quintic Airy and \Cref{s3colour} for queer Airy. These tables were obtained by solving order by order the Virasoro constraints of \Cref{l69}-(5), explicitly given by
\[
L_k = \hbar \partial_{2k + s} - \hbar^2\left(\frac{1}{2} \sum_{\ell = 0}^{k - 1}  \partial_{2\ell + 1}\partial_{2(k - \ell) - 1} + \sum_{\ell \geq 0} (2\ell + 1)t_{2\ell + 1}\partial_{2(\ell + k) + 1} + \frac{t_1^2}{2} \delta_{k,-1}  + \bigg(\frac{1}{8} + 2c\bigg)\delta_{k,0}\right),
\]
where $k \geq -1$, $c=0$ and $s = 5$ for quintic Airy, while $k \geq 0$ and $s = 3$ for queer Airy. We observe that the vanishing rules of \Cref{p:ineqcase} are respected. Interestingly, even if we take all blobs to be zero and all zero-mode shifts to be zero, we get non-zero coefficients in  \eqref{egounsgsf}, with $\sum_{i = 1}^{n} (2k_i + 1) = s(2g - 2 + n)$.

For queer Airy, taking the zero-mode shifts to zero and all blobs to zero except $\varphi_{0,3} = 1$, we retrieve the free energies of the non-coloured topological recursion for the Airy spectral curve $x(z) = z^2$, $\omega_{0,1}(z) = -z^2\dd z$.  This realizes the restriction map of \Cref{secpose} changing $\nu = (2)$ to $\nu = (1,1)$. It is well-known that for the Airy spectral curve, we have that, for all $g,n,k_1,\ldots,k_n \geq 0$,
\[
F_{g,n}[ 2k_1+1,\ldots,2k_n+1] = \int_{\overline{\mathcal{M}}_{g,n}} \prod_{i = 1}^{n} (2k_i + 1)!!\psi_i^{k_i}.
\]
Taking values $(\varphi_{0,3},c) \neq (1,0)$ while the other blobs are kept to zero yields a two-parameter deformation for which the free energy
\[
\sum_{i = 1}^{n} m_i \neq 3(2g - 2 + n) \qquad \Longrightarrow \qquad F_{g,n}[m_1,\ldots,m_n] = 0.
\]

{\tiny
\begin{table}
{\tiny
\centering
{\def\arraystretch{1,4}\begin{tabular}{|c|c||l|l|l|l|l|}
 \hline
$F_{0,2}$ & $\frac{2}{5}$ & $[1,1] : u$ & & & & \\
\hline
\hline 
$F_{0,3}$ & $\frac{3}{5}$ & $[1,1,1] : \varphi_{0,3}$ & & & & \\
&\cellcolor{yellow}$1$&\cellcolor{yellow}$[1,1,3] : 1$& & & & \\
& $\frac{7}{5}$ & $[1,1,5] :  2u$ & $[1,3,3] : 3u$ & & & \\
& $\frac{9}{5}$ & $[1,1,9] :  u^2$ & & & & \\
\hline
$F_{1,1}$ & $\frac{1}{5}$ & $[1] : \varphi_{1,1}$ & & & &  \\
&\cellcolor{yellow}$1$&\cellcolor{yellow}$[5] : \frac{1}{8}$& & & & \\
& $\frac{7}{5}$ & $[7] : \frac{u}{2}$ & & & & \\ 
\hline
\hline 
$F_{0,4}$ & $\frac{4}{5}$ & $[1,1,1,1] : \varphi_{0,4}$ & & & & \\
& $\frac{8}{5}$ & $[1,1,1,5] : 3\varphi_{0,3}$ & $[1,1,3,3] : 3\varphi_{0,3}$ & & & \\
&  \cellcolor{yellow} $2$ &\cellcolor{yellow}$[1,1,1,7] : 3u\varphi_{0,3}$ &\cellcolor{yellow}$[1,1,3,5] : 5$ &\cellcolor{yellow}$[1,3,3,3] : 6$ & & \\
& $\frac{12}{5}$ & $[1,1,1,9] : 9u$ & $[1,1,5,5] : 14u$ & $[1,1,3,7] : 14u$  & $[1,3,3,5] : 21u$ &  \\
& $\frac{14}{5}$ & $[1,1,1,11] : 9u^2$ & $[1,1,3,9] : 9u^2$ & $[1,1,5,7] : 9u^2$ & $[1,3,3,7] : 3u^2$ & \\
& $\frac{16}{5}$ & $[1,1,1,13] : \frac{3u^3}{2}$ & $[1,1,7,7] : 2u^3$ & & & \\
\hline
$F_{1,2}$ & $\frac{2}{5}$ & $[1,1] : \varphi_{1,2}$ & & & & \\
& $\frac{6}{5}$ & $[1,5] : \varphi_{1,1}$ & $[3,3] : 3\varphi_{1,1}$ & & & \\
& $\frac{8}{5}$ & $[1,7] : \frac{\varphi_{0,3}}{2} + u\varphi_{1,1}$ & & & & \\
&\cellcolor{yellow}$2$&\cellcolor{yellow}$[1,9] : \frac{9}{8}$ &\cellcolor{yellow}$[3,7] : \frac{7}{8}$ &\cellcolor{yellow}$[5,5] : \frac{5}{8}$ & & \\
& $\frac{12}{5}$ & $[1,11] : \frac{33u}{8}$ & $[3,9] : \frac{9u}{2}$ & $[5,7] : \frac{7u}{2}$ & & \\
& $\frac{14}{5}$ & $[1,13] : \frac{3u^2}{2}$ & $[7,7] : \frac{u^2}{2}$ & & & \\
\hline\hline 
$F_{0,5}$ & $1$ & $[1,1,1,1,1] : \varphi_{0,5}$ & & & & \\
& $\frac{9}{5}$ & $[1,1,1,1,5]  : 4\varphi_{0,4}$ & $[1,1,1,3,3] :  6\varphi_{0,4}$ & & &\\
& $\frac{11}{5}$ & $[1,1,1,1,7] : 3\varphi_{0,3}^2 + 4u\varphi_{0,4}$  & & & & \\
& $\frac{13}{5}$ & $[1,1,1,1,9] : 18\varphi_{0,3}$ & $[1,1,1,3,7] : 21\varphi_{0,3}$ & $[1,1,1,5,5] : 24\varphi_{0,3}$ & $[1,1,3,3,5] : 24\varphi_{0,3}$ & $[1,3,3,3,3] : 27\varphi_{0,3}$ \\
&\cellcolor{yellow}$3$&\cellcolor{yellow}$[1,1,1,1,11] : 15 + 36u\varphi_{0,3}$&\cellcolor{yellow}$[1,1,1,3,9] : 27(1+u\varphi_{0,3})$&\cellcolor{yellow}$[1,1,3,3,7] : 44+9u\varphi_{0,3}$\cellcolor{yellow}&\cellcolor{yellow}$[1,3,3,3,5] : 60$\cellcolor{yellow}&\cellcolor{yellow}$[3,3,3,3,3] : 72$ \\ 
& $\frac{17}{5}$ & $[1,1,1,1,13] : 60u + 18u^2\varphi_{0,3}$ & $[1,1,1,3,11] : 99u$ & $[1,1,1,5,9] : 108u$ & $[1,1,1,7,7] : 114u + 12u^2\varphi_{0,3}$ & $[1,1,3,3,9] : 153u$ \\
&& $[1,1,3,5,7] : 168u$ & $[1,1,5,5,5] : 168u$ & $[1,3,3,3,7] :  231u$ & $[1,3,3,5,5] : 252u$ & $[3,3,3,3,5] : 324u$ \\
& $\frac{19}{5}$ & $[1,1,1,1,15] : 90u^2$ & $[1,1,1,3,13] : 117u^2$ & $[1,1,1,5,11] : 126u^2$ & $[1,1,1,7,9] : 135u^2$ & $[1,1,3,3,11] : 126u^2$ \\ && $[1,1,3,5,9] : 126u^2$ &  $[1,1,5,5,7] : 126u^2$ & $[1,3,3,3,9] : 81u^2$ & $[1,3,3,3,9] : 81u^2$  &  $[1,1,5,5,7] : 126u^2$ \\
&&  $[3,3,3,3,7] : 27u^2$  & & & & \\
& $\frac{21}{5}$ & $[1,1,1,1,17] : 60u^3$ & $[1,1,1,3,15] : 45u^3$ & $[1,1,1,5,13] : 48u^3$ &  $[1,1,1,7,11] : 60u^3$ & $[1,1,1,9,9] : 54u^3$  \\
& &  $[1,1,3,3,13] : 9u^3$ & $[1,1,3,7,9] : 18u^3$ & $[1,1,5,7,7] : 32u^3$ & $[1,3,3,7,7] : 6u^3$ & \\
& $\frac{23}{5}$ & $[1,1,1,1,19] : 15u^4$ & $[1,1,1,7,13] : 9u^4$ & $[1,1,7,7,7] : 6u^4$ & & \\
\hline 
$F_{1,3}$ & $\frac{3}{5}$ & $[1,1,1] : \varphi_{1,3}$ & & & & \\
& $\frac{7}{5}$ & $[1,3,3] : 3\varphi_{1,2}$ & & & & \\
& $\frac{9}{5}$ & $[1,1,7] : \varphi_{1,1}\varphi_{0,3} + 2u\varphi_{1,2} + \frac{\varphi_{0,4}}{2}$ & & & & \\
& $\frac{11}{5}$ & $[1,1,9] : 3\varphi_{1,1}$ &  $[1,3,7] : 7\varphi_{1,1}$ & $[1,5,5] : 6\varphi_{1,1}$ &  $[3,3,5] : 18\varphi_{1,1}$ &  \\
& $\frac{13}{5}$ & $[1,1,11] :  6u\varphi_{1,1} + \frac{45}{8}\varphi_{0,3}$ & $[1,3,9] : 9u\varphi_{1,1} + \frac{9\varphi_{0,3}}{2}$ &  $[1,5,7] : 8u\varphi_{1,1} + 4\varphi_{0,3}$  & $[3,3,7] : 3u\varphi_{1,1} + \frac{3}{2}\varphi_{0,3}$ & \\ 
&\cellcolor{yellow}$3$&\cellcolor{yellow}$[1,1,13] : \frac{83}{3} + 3u^2\varphi_{1,1} + \frac{9u\varphi_{0,3}}{2}$&\cellcolor{yellow}$[1,3,11] : \frac{99}{8}$&\cellcolor{yellow} $[1,5,9] : \frac{45}{4}$&\cellcolor{yellow}$[1,7,7] : \frac{41}{4} + 2u\varphi_{0,3} + 2u^2\varphi_{1,1}$&\cellcolor{yellow}$[3,3,9] : \frac{45}{4}$  \\
& &  $[3,5,7] : \frac{35}{4}$ & $[5,5,5] : \frac{25}{4}$ & & & \\
& $\frac{17}{5}$ & $[1,1,15] : \frac{165u}{4}$ & $[1,3,13] : \frac{429u}{8}$  & $[1,5,11] : \frac{99u}{2}$ & $[1,7,9] : \frac{189u}{4}$ & $[3,3,11] : \frac{495u}{8}$ \\
&& $[3,5,9] : 54u$ & $[5,5,7] : 42u$ & & &  \\
& $\frac{19}{5}$ & $[1,1,17] : \frac{315u^2}{8}$ &  $[1,3,15] : \frac{45u^2}{2}$  &  $[1,5,13] : 21u^2$ & $[1,7,11] : \frac{201u^2}{8}$ & $[1,9,9] : 27u^2$ \\
&& $[3,3,13] : \frac{9u^2}{2}$ & $[3,7,9] : \frac{9u^2}{2}$ & $[5,7,7] : 7u^2$ & $\frac{21}{5}$ & $[1,1,19] : \frac{15u^3}{2}$ \\
&& $[1,7,13] : 3u^3$ & $[7,7,7] : u^3$ & && \\
\hline
$F_{2,1}$ & $\frac{1}{5}$ & $[1] : \varphi_{2,1}$ & & & & \\
& $\frac{7}{5}$ & $[7] : \frac{\varphi_{1,1}^2}{2} + \frac{\varphi_{1,2}}{2}$ & & & & \\
& $\frac{11}{5}$ & $[11] : \frac{21\varphi_{1,1}}{8}$ & & & & \\
& $\frac{13}{5}$ & $[13] : \frac{3u\varphi_{1,1}}{2}$ & & & & \\
&\cellcolor{yellow}$3$&\cellcolor{yellow}$[15] :  \frac{297}{128}$& & & & \\
& $\frac{17}{5}$ & $[17] : \frac{195u}{16}$ & & & & \\
& $\frac{19}{5}$ & $[19] : \frac{15u^2}{8}$ & & & & \\
\hline
\end{tabular}}
\vspace{-0.1cm}
\caption{\label{s5colour} Non-zero coefficients $F_{g,n}[m_1,\ldots,m_n]$ for quintic Airy with blobs \eqref{positini}, sorted by $\deg^+ = \frac{1}{5}\sum_{i = 1}^n [m_i]_+$ ($2{}^{\textnormal{nd}}$ column) and $\deg^+ = 2g - 2 + n$ highlighted.}}
\end{table}}

{\tiny
\begin{table}
{\tiny
\centering
{\def\arraystretch{1,4}\begin{tabular}{|c|c||l|l|l|l|}
 \hline
$F_{0,2}$ & $\frac{2}{3}$ & $[1,1] : u$ & & & \\
\hline
\hline 
$F_{0,3}$ &\cellcolor{yellow}$1$&\cellcolor{yellow}$[1,1,1] : \varphi_{0,3}$ & & & \\
& $\frac{5}{3}$ & $[1,1,3] : 2u$ & & & \\
& $\frac{7}{3}$ & $[1,1,5] :  u^2$ & & &  \\
\hline
$F_{1,1}$ & $\frac{1}{3}$ & $[1] : \varphi_{1,1}$ & & &  \\
&\cellcolor{yellow}$1$&\cellcolor{yellow}$[3] : \frac{1}{8} + 2c$ & & & \\
& $\frac{5}{3}$ & $[5] : \frac{u}{2}$ & & & \\ 
\hline
\hline 
$F_{0,4}$ & $\frac{4}{3}$ & $[1,1,1,1] : \varphi_{0,4}$ & & & \\
&  \cellcolor{yellow} $2$ &\cellcolor{yellow}$[1,1,1,3] : 3\varphi_{0,3}$ & & & \\
& $\frac{8}{3}$ & $[1,1,1,5] : 3u(2+ \varphi_{0,3})$ & $[1,1,3,3] : 10u$ & &  \\
& $\frac{10}{3}$ & $[1,1,1,7] : 9u^2$ & $[1,1,3,5] : 7u^2$ & & \\
& $4$ & $[1,1,1,9] : 3u^3$ & $[1,1,5,5] : 2u^3$ & & \\
\hline
$F_{1,2}$ & $\frac{2}{3}$ & $[1,1] : \varphi_{1,2}$ & & & \\
& $\frac{4}{3}$ & $[1,3] : \varphi_{1,1}$ & & & \\
&\cellcolor{yellow}$2$&\cellcolor{yellow}$[1,5] : \frac{1}{8} + \frac{\varphi_{0,3}}{2} + u\varphi_{1,1} + 2c$&\cellcolor{yellow}$[3,3] : \frac{3}{8} + 24c^2$ & & \\
& $\frac{8}{3}$ & $[1,7] : \frac{21u}{8} + 2cu$ & $[3,5] : \frac{5u}{2}$ & & \\
& $\frac{10}{3}$ & $[1,9] : \frac{3u^2}{2}$ & $[5,5] : \frac{u^2}{2}$ & & \\
\hline\hline 
$F_{0,5}$ & $\frac{5}{3}$ & $[1,1,1,1,1] : \varphi_{0,5}$ & & & \\
& $\frac{7}{3}$ & $[1,1,1,1,3]  : 4\varphi_{0,4}$ & & &\\
&\cellcolor{yellow}$3$&\cellcolor{yellow}$[1,1,1,1,5] : 12\varphi_{0,3} + 3\varphi_{0,3}^2 + 4u\varphi_{0,4}$&\cellcolor{yellow}$[1,1,1,3,3] : 18\varphi_{0,3}$& & \\
& $\frac{11}{3}$ & $[1,1,1,1,7] : 24u + 36u\varphi_{0,3}$ & $[1,1,1,3,5] : 48u + 24u\varphi_{0,3}$ & $[1,1,3,3,3] : 80u$ & \\
& $\frac{13}{3}$ & $[1,1,1,1,9] : 72u^2$ & $[1,1,1,3,7] : 90u^2$ & $[1,1,1,5,5] : 84u^2 + 12u^2\varphi_{0,3}$ & $[1,1,3,3,5] : 70u^2$ \\ 
& $5$ & $[1,1,1,1,11] : 60u^3$ & $[1,1,1,3,9] : 36u^3$ & $[1,1,1,5,7] : 48u^3$ & $[1,1,3,5,5] : 24u^3$ \\
& $\frac{17}{3}$ & $[1,1,1,1,13] : 15u^4$ & $[1,1,1,5,9] : 9u^4$ & $[1,1,5,5,5] : 6u^4$ &\\
\hline 
$F_{1,3}$ & $1$ & $[1,1,1] : \varphi_{1,3}$ & & & \\
& $\frac{5}{3}$ & $[1,1,3] : 2\varphi_{1,2}$ & & & \\
& $\frac{7}{3}$ & $[1,1,5] : 2\varphi_{1,1} + \varphi_{1,1}\varphi_{0,3} + \frac{\varphi_{0,4}}{2} +2u\varphi_{1,2}$ & $[1,3,3] : 4\varphi_{1,1}$ & & \\
&\cellcolor{yellow}$3$&\cellcolor{yellow}$[1,1,7] : \frac{1}{4} + \frac{33}{8} \varphi_{0,3} + 2c\varphi_{0,3} + 4c + 6u\varphi_{1,1}$ &\cellcolor{yellow}$[3,3,3] : \frac{9}{4} + 9c$&\cellcolor{yellow}$[1,3,5] : \frac{3}{4} + 3\varphi_{0,3} + 12c + 6u\varphi_{1,1}$& \\
& $\frac{11}{3}$ & $[1,1,9]  : \frac{67}{4}u + \frac{9\varphi_{0,3}u}{2} + 12cu + 3u^2\varphi_{1,1}$ & $[3,3,5] : 20u$ & $[1,3,7] : 21u + 16cu$ &  \\ 
& $\frac{13}{3}$ & $[1,1,11] : \frac{203}{8}u^2$ & $[1,3,9] : 15u^2$ & $[1,5,7] : \frac{141u^2}{8} + 2cu^2$ & $[3,5,5] : 5u^2$  \\
& $5$ & $[1,1,13] : \frac{15u^3}{2}$ & $[5,5,5] : u^3$ & $[1,5,9] : 3u^3$ &  \\
\hline
$F_{2,1}$ & $\frac{1}{3}$ & $[1] : \varphi_{2,1}$ & & & \\
& $\frac{5}{3}$ & $[5] : \frac{\varphi_{1,1}^2}{2} + \frac{\varphi_{1,2}}{2}$ & & & \\
& $\frac{7}{3}$ & $[7] : \frac{9\varphi_{1,1}}{8} + \frac{c\varphi_{1,1}}{2}$ & & & \\
&\cellcolor{yellow}$3$&\cellcolor{yellow}$[9] : \frac{41}{128} + \frac{\varphi_{0,3}}{2} + \frac{21c}{4} + 2c^2 + \frac{3u\varphi_{1,1}}{2}$& & & \\
& $\frac{11}{3}$ & $[11] :  \frac{83u}{16} + 3cu$ & & & \\
& $\frac{13}{3}$ & $[13] : \frac{15u^2}{8}$ & & & \\
\hline
\end{tabular}}
\vspace{-0.1cm}
\caption{\label{s3colour} Non-zero coefficients $F_{g,n}[m_1,\ldots, m_n]$ for queer Airy with blobs \eqref{positini} and zero-mode shifts $T^{(1)} = 0$ and $T^{(2)} = c\hbar^2$, sorted by $\deg^+ = \frac{1}{3}\sum_{i = 1}^n [m_i]_+$ ($2{}^{\textnormal{nd}}$ column) and $\deg^+ = 2g - 2 + n$ highlighted.}}
\end{table}}

\subsection{The \texorpdfstring{$\Theta^{(r,s)}$}{Thetars}-cohomological field theories}

\label{s:theta}

Let $r \geq 2$ be an integer and $s \in [r - 1]$ without assuming that it is coprime to $s$. In \cite[Sections 4 and 5]{BCGS}, $\mathcal{W}$-constraints for the descendant potential of the $\Theta^{(r,s)}$-cohomological field theories are constructed. We will not define $\Theta^{(r,s)}$ here. We only mention that it is a family of cohomology classes $\Theta^{(r,s)}_{g,n}$ on $\overline{\mathcal{M}}_{g,n}$ forming a cohomological field theory which is nowhere semi-simple. It comes from the geometry of the moduli of $\frac{s-r}{r}$-th roots of the canonical bundle over $\overline{\mathcal{M}}_{g,n}$ and the cohomological classes are obtained as top Chern classes of Chiodo bundles \cite{Chi08-twisted,Chi08-class}. We refer the reader to \cite[Section 2]{BCGS} for details.

To construct the associated $\mathcal{W}$-constraints, we consider the following coloured spectral curves with a single ramification point.
\begin{itemize}
\item $C = \mathbb{C}$, $x = z^r$, and $\omega_{0,1} = - z^{s-1}\dd z$ (i.e. $\tau = 1$);
\item $\nu = (r-s+1, 1, \ldots, 1)$;
\item $T^{(i)} = \hbar^i T^{(i)}_{\frac{i}{2}}$ for $i \in [r-s]$, where
\[
T^{(i)}_{\frac{i}{2}} = \mathsf{e}_i\left( \frac{2+s-r-1}{2(r-s)}, \frac{4+s-r-1}{2(r-s)}, \ldots, \frac{2(r-s)+s-r-1}{2(r-s)} \right),
\]
and $\mathsf{e}_i$ is the $i$-th elementary symmetric polynomial. Note that $T^{(i)}_{\frac{i}{2}}$ vanishes when $i$ is odd.
\end{itemize}
The $\mathcal{W}$-constraints of  \cite[Theorem 5.3]{BCGS} coincide with the set of operators $\bm{H}$ from \Cref{t:PAISCc}. The set of spectator variables is described in \cite[Lemma 5.15]{BCGS}, it is $S = \mathbb{Z}_{< 0} \cup S_{r,s}^+$ with
\[
S_{r,s}^+ = \set{m \in \N^*| \forall a,b \in \N \quad m \neq a(r-s) + bs}.
\]
We note that $S^+_{r,s} = \emptyset$ if and only if $s=1$ or $s=r-1$. If $s \notin \set{1,r-1}$, the reduced partition function playing the role of initial condition is the restriction of the descendant potential  of $\Theta^{(r,s)}$ to the variables $\bm{t}_{S^+_{r,s}}$, see \cite[Definition 5.17]{BCGS}. It was proved in \cite[Proposition 5.18]{BCGS} that the $\mathcal{W}$-constraints determine uniquely the partition function given the reduced partition function.

This coloured spectral curve is admissible if and only if $r$ and $s$ are coprime. In that case, $\bm{H}$ forms a partial Airy structure and the existence and uniqueness of the partition function with given reduced partition function appears as a consequence of our general result \Cref{t:PAISCc}. The two cases $s = r-1$ and $s = 1$ are special. For $s=r-1$, the partition is $\nu = (2,1,\ldots,1)$, which is a canonical partition. Moreover, in this case the zero-mode shifts are all zero. Thus we recover Airy ideals as in \cite{BBCCN24}. This case was already studied in \cite{CGG} using the framework of Airy ideals and the usual topological recursion. For $s=1$, the partition is $\nu=(r)$, which is again a canonical partition. Thus there are no positive spectator variable, as $S^+_{r,1} = \emptyset$. However, the zero-mode shifts $T^{(i)}$ are non-zero for even $i$. Thus this is a non-trivial example of a coloured spectral curve, but only because of the zero-mode shifts. Such examples were already studied in \cite{BBKN25} (the corresponding coloured loop equations were called "shifted loop equations" there). For $s \notin \set{1,r-1}$, $\nu = (r-s+1, 1,\ldots, 1)$ is not a canonical partition. In that case, the positive spectator set $S^+_{r,s}$ is not empty, and we have a truly non-trivial example of admissible coloured spectral curves, including both non-zero shifts and non-canonical partitions. 

More surprising is the case where $r$ and $s$ are not coprime. Then $\bm{H}$ does not form a partial Airy ideal, because there are several operators containing the same active derivative in their $\hbar^1$-term. It is therefore puzzling that these differential equations are compatible, i.e. that there exists a partition function annihilated by the $\mathcal{W}$-constraints. Nevertheless, geometry (via the $\Theta^{(r,s)}$-classes) gives us a solution. We hope to return to this puzzle in the future.

\newpage

\appendix

\section{The $(V,D)$-completed Rees Weyl algebra}

\label{a:completion}

In this appendix we provide a detailed construction of the $(V,D)$-completed Rees Weyl algebra introduced in \Cref{d:comprees}.

Recall that we can define the completion of an algebraic structure as the limit of an inverse system of subspaces and homomorphisms. We will define the \((V,D)\)-completed Rees Weyl algebra by first completing the subspaces $\mathcal{F}^d\mathcal{D}_I$ in the Bernstein filtration from \Cref{d:bernstein}, and then use the Rees construction on the completed subspaces to obtain our desired algebra. Throughout this appendix, $I$ is an arbitrary fixed set (finite or infinite), and we write $\mathcal{F}^d := \mathcal{F}^d\mathcal{D}_I$. We will take two subsets $V,D \subseteq I$ which can be finite or countably infinite --- the construction gives new operators only in the latter case.

We will achieve the completion of $\mathcal{F}^d$ first with respect to the set of derivatives indexed by $D$ followed by the completion in the set of variables indexed by $V$. We start by defining the first of the two inverse systems. 

\begin{lemma}
  [Inverse system for derivatives] \label{l:inverse_sys_D} Let $D\subseteq I$ together with a bijection $f : N \rightarrow D$ for some $N \subseteq \N$. For each $u \in N$ define
\[
    \mathcal{F}^{d,(u)}_{D} := \mathcal{F}^d_{|\partial_{f(v)}=0 \,\,\forall v>u}= \Set{\sum_{m = 0}^{d}  \sum_{\mathbf{v} \in [u]^m} p_{\mathbf{v}}^{(d-m)}(\bm{t}_I, \bm{\partial}_{I\setminus D}) \partial_{f(v_1)} \cdots \partial_{f(v_m)}},
\]
where $ p_{\mathbf{v}}^{(d-m)}(\bm{t}_I,\bm{\partial}_{I\setminus D})$ are polynomials of degree $\leq d-m$ and the sum contains finitely many terms. 
For $u \leq u'$ define the linear map $R^{d,(u,u')} : \mathcal{F}^{d,(u')}_{D} \rightarrow \mathcal{F}^{d,(u)}_{D}$ by setting $\partial_{f(v)}$ to zero for $v > u$.  The family $(\mathcal{F}^{d,(u)}_D)_{u \in N}$ equipped with these morphisms forms an inverse system of vector spaces.
\end{lemma}
\begin{proof}
     Note that the operators are normal-ordered and we set derivatives to zero only in the normal-ordered form. The linear maps $R$ are therefore well-defined. The only thing we need to check is $R^{d,(u,u)} =\text{id}$ (it is obvious), and that for $u \leq u' \leq u''$ we have
\[
R^{d,(u,u')} \circ R^{d,(u',u'')} (H) =R^{d,(u,u')} \big(H_{|\partial_{f(v)}=0\,\,\forall v> u'}\big)=H_{|\partial_{f(v)}=0\,\,\forall v>u} =R^{d,(u,u'')}(H).
\]
\end{proof}

We now define the completion in derivatives of $D$ as the inverse limit.
\begin{definition}
    [\((\emptyset, D)\)-completed subspace] We define the \((\emptyset, D)\)-\emph{completed subspace} \( \mathcal{F}^{d}_{\llbracket \emptyset,D\rrbracket}\) as the inverse limit
\begin{equation}
\label{eq: completed_derivative_sqnce}
        \mathcal{F}^{d}_{\llbracket \emptyset,D\rrbracket}:=\lim_{\longleftarrow} \mathcal{F}^{d,(u)}_D=\Set{ \bm{A} \in \prod_{u \in N} \mathcal{F}^{d,(u)}_D\,\, |\,\, \forall u \leq u' \quad A^{(u)}=R^{d,(u,u')}(A^{(u')})}.
\end{equation}
\end{definition}
Formally, elements of $\mathcal{F}^{d}_{\llbracket \emptyset,D\rrbracket}$ are sequences, by definition of the inverse limit. In practice, however, we can think of them as operators where possibly infinite formal sums in the derivatives are allowed. Let us make this precise.
\begin{definition}[Alternative description of $(\emptyset,D)$-completed subspace]
Define the vector space $\widehat{\mathcal{F}}^d_D$ made of elements of the form
\[
\sum_{m = 0}^{d}  \sum_{\mathbf{i} \in D^m} p_{\mathbf{i}}^{(d-m)}(\bm{t}_I, \bm{\partial}_{I\setminus D}) \partial_{i_1} \cdots \partial_{i_m}.
\]
\end{definition}
Before proving that $\widehat{\mathcal{F}}^d_D$ is isomorphic to $\mathcal{F}^{d}_{\llbracket \emptyset,D\rrbracket}$, we need to define a  operator which will help us build sequences from operators of $\widehat{\mathcal{F}}^d_D$. 
\begin{definition}[Restriction]
We call $R^{d,(u)} : \widehat{\mathcal{F}}^d_D \rightarrow \mathcal{F}^{d,(u)}_D$ the operation of sending $\partial_{f(v)}$ to $0$ for $v > u$. 
\end{definition}

\begin{lemma}
 The vector spaces $\widehat{\mathcal{F}}^d_D$ and  $\mathcal{F}^{d}_{\llbracket \emptyset,D\rrbracket}$ are isomorphic.
\end{lemma}

\begin{proof}
We define a linear map $\widehat{R}^d : \widehat{\mathcal{F}}^d_{D} \rightarrow \mathcal{F}^{d}_{\llbracket \emptyset,D\rrbracket}$  by setting $\widehat{R}^d(H) = (R^{d,(u)}(H))_{u \in N}$. By construction, this sequence belongs to $\mathcal{F}^{d}_{\llbracket \emptyset,D \rrbracket}$. Conversely, if $\bm{A} \in \mathcal{F}^{d}_{\llbracket \emptyset,D \rrbracket}$, let us define $A^{(u)}$ for all $u \in \N^*$.
\begin{itemize}
\item  for $u \in N$, $A^{(u)}$ is the corresponding component of $\bm{A}$;
\item for $u < \min N$, we set $A^{(u)} = 0$;
\item for $u \in \N^* \setminus N$, we set $A^{(u)} = A^{(\overline{u})}$, where $\overline{u} = \max\set{v \in N|v \leq u}$.
\end{itemize}
Then, we decompose, for $u > 1$,
\[
A^{(u)} - A^{(u - 1)} = \sum_{m = 1}^{d} \sum_{v_1 \leq \cdots \leq v_m = u} p_{\mathbf{v}}^{(d - m)}(\bm{t}_I,\bm{\partial}_{I \setminus D}) \prod_{\ell = 1}^{m} \partial_{f(v_\ell)},
\]
where the constraint $v_m = u$ guarantees that every term contains at least one copy of $\partial_{f(u)}$. Sending sequences $\bm{A} \in \mathcal{F}^{d}_{\llbracket \emptyset,D \rrbracket}$ to $A^{(1)} + \sum_{u > 1} (A^{(u)}- A^{(u - 1)})$ provides the inverse map of $\widehat{R}^d$. 
\end{proof}

Thinking of the elements of $\mathcal{F}^{d}_{\llbracket \emptyset,D\rrbracket}$ as possibly infinite sums instead of sequences will help us define the next completion in terms of variables of some subset $V\subseteq I$. It will be important here to assume that $V \cap D$ is finite since otherwise we would get divergent sums.

Similar to what we did before, we start by defining an inverse system and then we take the inverse limit.

\begin{lemma}
  [Inverse system for variables] \label{l:inverse_sys_V} Let $D,V \subseteq I$ and choose bijections $f : N \rightarrow D$ and $g : M \rightarrow V$ with subsets $M,N \subseteq \N$. For each $u \in M$ define the subspace $\mathcal{G}^{d,(u)}_V:= \widehat{\mathcal{F}}^{d}_{\llbracket \emptyset,D\rrbracket}{}_{|t_{g(v)}=0\,\,\forall v>u}$, made of elements of the form
  \[
 \sum_{m = 0}^{d}  \sum_{m'+n'=m}\sum_{\substack{\mathbf{v} \in[u]^{n'} \\ \mathbf{i} \in D^{m'}}} \left(\prod_{\ell=1}^{n'} t_{g(v_\ell)}\right)p_{\mathbf{v},\mathbf{i}}^{(d-m)}(\bm{t}_{I\setminus V}, \bm{\partial}_{I\setminus D}) \left(\prod_{\ell=1}^{m'}\partial_{i_\ell}\right),
\]
where $p_{\mathbf{v},\mathbf{i}}^{(d-m)}(\bm{t}_{I\setminus V}, \bm{\partial}_{I\setminus D})$ are polynomials of degree $\leq d-m$. For $u \leq u'$ we define a linear map $S^{d,(u,u')} : \mathcal{G}^{d,(u')}_{V} \rightarrow \mathcal{G}^{d,(u)}_{V}$ by setting $t_{g(v)}$ to zero  for $v > u$. The family $(\mathcal{G}^{d,(u)}_{V})_{u \in M}$ equipped with these morphisms forms an inverse system.
\end{lemma}

It is important to keep the product of variables on the left so that the operator remains normal-ordered and the maps $S$ are well-defined. The proof of the lemma is the same as that of \Cref{l:inverse_sys_D}.
\begin{definition}\label{d:VDsub}
    [\((V, D)\)-completed subspace] We define the $(V, D)$-\emph{completed subspace} $\mathcal{F}^{d}_{\llbracket V,D\rrbracket}$ as the inverse limit
\begin{equation}
\label{eq: completed_variables_sqnce}
        \mathcal{F}^{d}_{\llbracket V,D\rrbracket}:=\lim_{\longleftarrow} \mathcal{G}^{d,(u)}_V=\Set{\bm{A} \in\prod_{u \in M} \mathcal{G}^{d,(u)}_V\,\,|\,\,\forall u \leq u' \quad A^{(u)} =S^{d,(u,u')}(A^{(u')})}.
\end{equation}
\end{definition}
Following the same steps as for \((\emptyset, D)\)-completed subspace, elements of $\mathcal{F}^{d}_{\llbracket V,D\rrbracket}$ can also be thought of as possibly infinite sums.
\begin{definition}[Alternative description of $(V,D)$-completed subspace]
Define the vector space $\widehat{\mathcal{F}}^d_{V,D}$ made of elements of the form
\begin{equation}
    \label{eq: infinite_series_form_II} \sum_{m = 0}^{d}  \sum_{m'+n'=m}\sum_{\substack{\mathbf{j} \in V^{n'} \\ \mathbf{i} \in D^{m'}}} \left(\prod_{\ell = 1}^{n'} t_{j_\ell}\right) p^{(d-m)}_{\mathbf{j},\mathbf{i}}(\bm{t}_{I\setminus V},\bm{\partial}_{I\setminus D}) \left(\prod_{\ell = 1}^{m'} \partial_{i_\ell}\right).
\end{equation}
\end{definition}
\begin{lemma}
  The vector spaces $\widehat{\mathcal{F}}^d_{V,D}$ and $\mathcal{F}^{d}_{\llbracket V,D\rrbracket}$ are isomorphic.
\end{lemma}
 
We make two important observations. First, it is clear in the definition of $\widehat{\mathcal{F}}^d_{V,D}$ that the bijections $f$ and $g$ do not affect the final form of the operators. In other words, the completion is independent of the choice of these maps. The next lemma shows that removing finite sets from $V$ and $D$ does not affect the completion.

\begin{lemma}
\label{lem:removfinit}
  Let $D'$ be a finite subset of $D$ and $V'$ be a finite subset of $V$. Then $\mathcal{F}^{d}_{\llbracket V,D \rrbracket} = \mathcal{F}^{d}_{\llbracket V \setminus V',D \setminus D'\rrbracket}$.
\end{lemma}

\begin{proof}
Set $V'' = V \setminus V'$ and $D'' = D \setminus D'$. Observe that we could write the elements of $\mathcal{F}^{d}_{\llbracket V,D\rrbracket}$ as
\begin{equation}
\label{eqgdgsfgum}
 \sum_{m = 0}^{d}  \sum_{m' + m'' + n' + n'' =m} \sum_{\substack{\mathbf{j}' \in (V')^{n'},\,\,\mathbf{j}'' \in (V'')^{n''} \\ \mathbf{i}' \in (D')^{m'},\,\,\mathbf{i}'' \in (D'')^{n''}}}  \left(\prod_{\ell = 1}^{n'} t_{j'_\ell} \prod_{\ell = 1}^{n''} t_{j''_\ell}\right) p^{(d-m)}_{\mathbf{j}',\mathbf{j}'',\mathbf{i}',\mathbf{i}''}(\bm{t}_{I\setminus V},\bm{\partial}_{I\setminus D}) \left(\prod_{\ell = 1}^{m'} \partial_{i'_\ell} \prod_{\ell = 1}^{m''} \partial_{i_\ell''}\right).
\end{equation}
What we have done here is simply separating variables in $V'$ and derivatives in $D'$. But since $V'$ and $D'$ are finite, we can absorb these into the polynomials $p^{(d-m)}$, defining polynomials of higher degree $\tilde{p}^{(d - k)}$ with $k = m'' + n''$, so that \eqref{eqgdgsfgum} can be rewritten as
\[
\sum_{k = 0}^{d} \sum_{m' + n' = k} \sum_{\substack{\mathbf{j}' \in (V')^{n'} \\ \mathbf{i}' \in (D')^{m'}}} \left(\prod_{\ell = 1}^{n'} t_{j_\ell'}\right) \tilde{p}^{(d - k)}_{\mathbf{j}',\mathbf{i}'}(\bm{t}_{I \setminus V''},\bm{\partial}_{I \setminus D''})  \left(\prod_{\ell = 1}^{m'} \partial_{i_{\ell}}\right).
\]
Hence, we have $\mathcal{F}^{d}_{\llbracket V,D\rrbracket} =\mathcal{F}^{d}_{\llbracket V'',D''\rrbracket}$.
\end{proof}

So far we only considered the vector space structure. If we assume that $V \cap D$ is finite, the inverse limits $(\mathcal{F}^{d}_{\llbracket V,D \rrbracket})_{d \geq 0}$ inherit the structure of a filtered associative algebra from the product of differential operators. The finiteness condition is essential. For instance, if $I = D = V = \N$ the product
\[
\left(\sum_{n \in \N} \partial_{n}\right)\left(\sum_{n \in \N} t_n\right) = \sum_{n \in \N} t_n\partial_n + \sum_{n \in \N} 1.
\]
would be ill-defined. 

In view of \Cref{lem:removfinit}, there is no harm in assuming that each of the sets $V$ and $D$ are either empty or countably infinite, and that they are disjoint. We are ready to define the \((V,D)\)-completed Rees Weyl algebra.

\begin{definition}[\((V,D)\)-\emph{completed Rees Weyl algebra}]\label{d:rees_comp}
Let $V,D\subseteq I$ such that $V$ and $D$ are empty or countably infinite, and $V \cap D = \emptyset$. Then the \((V,D)\)-\emph{completed Rees Weyl algebra} $\mathcal{D}^\hbar_{I;\llbracket V,D\rrbracket}$ is the associative algebra defined by
\begin{equation}
\mathcal{D}^\hbar_{I;\llbracket V,D\rrbracket} = \prod_{d \geq 0} \hbar^d \mathcal{F}^{d}_{\llbracket V,D\rrbracket}.
\end{equation}
and the operators are exactly those of the form \eqref{eq: operator final form}.  It is a graded algebra for the $\hbar$-degree.
\end{definition}

This is equivalent to \Cref{d:comprees}, where the operators of the algebra were written in explicit form.

\begin{remark}\label{r:aboutcomp}
The strategy employed in this appendix was to first construct the $(V,D)$-completion of the subspaces $\mathcal{F}^d\mathcal{D}_I$ of the Bernstein filtration (see \Cref{d:VDsub}) and then construct the $(V,D)$-completed Rees Weyl algebra $\mathcal{D}^\hbar_{I;\llbracket V,D\rrbracket}$ using the Rees construction (see \Cref{d:rees_comp}). We could have instead constructed the $(V,D)$-completion of the Weyl algebra $\mathcal{D}_{I;\llbracket V,D\rrbracket}$ directly (this is the completion referred to in \Cref{d:WeylZcomplete}), and then introduce $\hbar$ using the Bernstein filtration on this completed Weyl algebra. Both approaches work and give the same resulting algebra $\mathcal{D}^\hbar_{I;\llbracket V,D\rrbracket}$.

The reason we decided to complete the subspaces of the Bernstein filtration first is technical. To proceed with the second approach, the first step is to construct the completed Weyl algebra $\mathcal{D}_{I;\llbracket V,D\rrbracket}$ in exactly the same way as in this appendix. However, $\mathcal{D}_{I;\llbracket V,D\rrbracket}$ includes certain infinite sums (as an example, if $V = \mathbb{Z}_{<0}$ and $D = \mathbb{N}$ as in \Cref{d:rees_comp}, the operator $\sum_{n \in \mathbb{N}} \partial_{n}^{n+1} = \partial_0 + \partial_1^2 + \partial_2^3 + \ldots$ is in the completion $\mathcal{D}_{\mathbb{Z};\llbracket \mathbb{Z}_{<0},\mathbb{N}\rrbracket}$ ) which makes an explicit presentation as in \Cref{d:comprees} a little more involved. Nevertheless, the next step is to consider the Bernstein filtration on this completed algebra $\mathcal{D}_{I;\llbracket V,D\rrbracket}$, which is not exhaustive anymore as these infinite sums are never included in the subspaces of the Bernstein filtration. As a result, using the Rees construction we ultimately get the same $(V,D)$-completed Rees Weyl algebra $\mathcal{D}^\hbar_{I;\llbracket V,D\rrbracket}$.
\end{remark}

\newpage
\bibliographystyle{amsalpha}
\bibliography{PASblobbed2}
\end{document}